\documentclass[aps,pre,twocolumn,superscriptaddress,floats,floatfix]{revtex4-2}
\usepackage{amssymb,amsmath,amsthm}
\usepackage{graphicx}
\usepackage{epstopdf}
\usepackage{color}
\usepackage{subfigure}
\usepackage{epsfig}
\usepackage{rotating}
\usepackage{mwe}
\usepackage{float}
\usepackage{dcolumn,bm}
\usepackage{verbatim}
\usepackage{hyperref}
\usepackage{pgf,tikz}
\usepackage{comment}
\usepackage[normalem]{ulem}
\hypersetup{
  colorlinks   = true, %Colours links instead of ugly boxes
  urlcolor     = blue, %Colour for external hyperlinks
  linkcolor    = blue, %Colour of internal links
  citecolor   = red %Colour of citations
}
\begin{document}

\begin{abstract}

We show how to use the cubic-quintic Gross-Pitaevskii-Poisson equation (cq-GPPE) and the cubic-quintic Stochastic Ginzburg-Landau-Poisson equation (cq-SGLPE) to investigate the gravitational collapse of a tenuous axionic gas into a collapsed axionic condensate for both zero and finite temperature $T$. At $T=0$, we use a Gaussian Ansatz for a spherically symmetric density to obtain parameter regimes in which we might expect to find compact axionic condensates. We then go beyond this Ansatz, by using the cq-SGLPE to investigate the dependence of the axionic condensate on the gravitational strength $G$ at $T = 0$. We demonstrate that, as $G$ increases, the equilibrium configuration goes from a tenuous axionic gas, to flat sheets or \textit{Zeldovich pancakes}, cylindrical structures, and finally a spherical axionic condensate. By varying $G$, we show that there are first-order phase transitions, as the system goes from one of these structures to the next one; we find hysteresis loops that are associated with these transitions. We examine these states and the transitions between these states via the Fourier truncated cq-GPPE; and we also obtain the thermalized $T > 0$ states from the cq-SGLPE; the transitions between these states yield thermally driven first-order phase transitions and their associated hysteresis loops. Finally, we discuss how our cq-GPPE approach can be used to follow the spatiotemporal evolution of a rotating axionic condensate and also a rotating binary-axionic-condensate system; in particular, we demonstrate, in the former, the emergence of vortices at large angular speeds $\Omega$ and, in the latter, the rich dynamics of the mergers of the components of this binary system, which can yield vortices in the process of merging. 

%We study the evolution and rotational dynamics of axion stars in the cubic-quintic potential, $\mathcal{V} = \frac{1}{2}g | \psi|^4+ \frac{1}{3}g_2| \psi|^6$, by using the Gross-Pitaevskii-Poisson equation (GPPE). We obtain the equilibrium configurations at $T = 0$ and then investigate finite-temperature effects ($T>0$) using the Fourier-truncated GPPE. We show that, beyond a critical angular velocity, quantum vortices thread the axionic star; this critical velocity depends on the coefficient of the quintic nonlinearity. We also find the first-order phase transition while heating and cooling of the spherically collapsed star, which transforms to a noncollapsed density distribution at high temperatures.

\end{abstract}
\title{Gravity- and temperature-driven phase transitions in a model for collapsed axionic condensates}
%with qubic-quintic self-interaction}
\author{Sanjay Shukla }%$^1$}
\email{shuklasanjay771@gmail.com}
\affiliation{Centre for Condensed Matter Theory, Department of Physics, Indian Institute of Science, Bangalore 560012, India}
\author{Akhilesh Kumar Verma }%$^1$}
\email{akvermajnusps@gmail.com}
\affiliation{Civil and Architectural Engineering, University of Miami, Coral Gables, Florida FL 33146, USA}
\author{Marc E. Brachet }
\email{brachet@phys.ens.fr}
\affiliation{Laboratoire de Physique de l’École Normale Supérieure, ENS, Université PSL,
CNRS, Sorbonne Université Université de Paris, 24 Rue Lhomond, 75005 Paris, France}

\author{Rahul Pandit $^1$}
\email{rahul@.iisc.ac.in}

\maketitle
\section{ Introduction}
Dark matter has a rich history~\cite{bertone2018history}; it includes the \textit{dark bodies} discussed by Kelvin~\cite{kelvin1904baltimore}, the suggestion of \textit{mati\`ere obscure} by Poincar\'e~\cite{hp1906pa}, Zwicky's proposal~\cite{zwicky6redshift,zwicky2009republication} of \textit{Dunkler Materie}, inferred via the virial theorem, and the path-breaking studies of
Rubin and Ford on the rotation curves of spiral galaxies that provided evidence for dark-matter haloes (DMH)~\cite{rubin1970rotation,binney2008galactic,Rotation_curve}. Dark matter now plays a central role in cosmology~\cite{bertone2018history}, e.g., in the simple $\Lambda$-cold-dark-matter ($\Lambda$CDM) model, where $\Lambda$ is the cosmological constant, and its generalizations~\cite{perivolaropoulos2022challenges,mnras/stx3304_collision_less,Harko_2011,rotation_curves_observation,Moore_excess_satellite}. Experiments indicate that $\simeq 85\%$ of the matter in the Universe is nonbaryonic cold dark matter~\cite{spergel2003first,komatsu2009five,collaboration2015planck,lisanti2017lectures,abdallah2018search}.  Weakly interacting massive particles (WIMPS)~\cite{abdallah2018search,Schumann_2019} are among the leading dark-matter candidates. Several experiments have been carried out to establish the nature of dark matter;
unfortunately, there is still no unambiguous dark-matter candidate~\cite{rottpos,bertone2018history,collaboration2018dark,lhaaso2022constraints}.
While such experimental studies continue, it is important to explore theoretically the properties of other dark-matter candidates, such as self-gravitating assemblies of bosons, also called ultra-light dark matter (ULDM) (see, e.g., Refs.~\cite{ruffini_1969,chavanis_2011,Suarez,madarassy2015evolution,chavanis_2016,hui2017ultralight,chavanis2021jeans,PRR_akverma2021,verma2022rotating}) or axions~\cite{PRD_cq_chavanis}. We have studied the former, at temperature $T \geq 0$,  by using the Galerkin-truncated Gross-Pitaevskii-Poisson equation~\cite{PRR_akverma2021,verma2022rotating}; here, we generalize this to study a cubic-quintic Gross-Pitaevskii-Poisson equation~\cite{PRD_cq_chavanis} that is of relevance to axion stars~\cite{braaten_axion} and axion cosmology~\cite{sikivie2008axion}.

%Several models exist to study the properties of DMH, one of which is cold dark matter (CDM). In this %model, we consider DM to be made of weakly interacting massive particles %(WIMPS)\cite{abdallah2018search,Schumann_2019}. The CDM model accounts for the large-scale (a few %megaparsecs)\cite{mnras/stx3304_collision_less} properties but fails to explain the observed data at %galactic scales. A prediction from the theory that disagrees with observations is the presence of a %cuspy central density profile\cite{Harko_2011,rotation_curves_observation}, favoring the flat core %density. Another discrepancy between theory and observation arises when N-body simulations (assuming %CDM) predict an excess of satellite galaxy numbers \cite{Moore_excess_satellite}. 

%These discrepancies motivate the hunt for other dark matter candidates, one of which is the self-gravitating Bose-Einstein %condensates (BECs)\cite{chavanis_2011,ruffini_1969, PRL_boson}, also called the ultra-light dark matter (ULDM). It suggests the %presence of a scalar field $\psi$ with a scalar potential $\mathcal{V}(\psi)$ and ultra-light mass of the order of $\sim %10^{-23}eV$.

A three-dimensional (3D) system of non-interacting mass-$m$ bosons exhibits a Bose-Einstein condensate (BEC) for $T < T_c = [(2\pi \hbar^2 n^{2/3})/(mk_B)]$, the critical temperature at which the thermal de Broglie wavelength $\lambda_{dB} = [2\pi \hbar^2/mk_BT]^{1/2}$ becomes comparable to the mean interparticle spacing $\sim n^{-1/3}$, where $n$ is the number density of bosons and $k_B$ is the Boltzmann constant. To study a system of weakly interacting bosons we use the Gross-Pitaevskii equation (GPE), in which the BEC is a superfluid. To account for a non-relativistic gravitational interaction between such bosons, we couple the GPE with the Poisson equation, i.e., we employ the Gross-Pitaevskii-Poisson equation (GPPE). The scattering length 
$a$ of the bosonic atoms leads to a self-interaction between the bosons that can either be repulsive ($a > 0$) or attractive ($a < 0$). In the repulsive case, we use the GPPE with a cubic nonlinearity; the $T=0$ equilibrium state follows by balancing the gravitational interaction with the repulsive self-interaction and the quantum pressure~\cite{chavanis_2011}. For 
$T > 0$ we have carried out an extensive study of this GPPE, by a Galerkin-truncated pseudospectral method~\cite{PRR_akverma2021,verma2022rotating}, to obtain compact objects, which can be threaded by vortices if we include rotation.

For the case of attractive self-interactions $a < 0$, which is directly relevant to axionic systems~\cite{braaten_axion,sikivie2008axion}, the equilibrium $T=0$ state is unstable above an extremely low critical mass~\cite{chavanis_2016}, because the repulsive quantum pressure cannot overcome attractive gravitational and self-interactions. Self-gravitating bosonic systems with $a < 0$ are also interesting because it has been hypothesized~\cite{Suarez} that they can accelerate the formation of structures if the system starts from a homogeneous distribution of bosons~\footnote{This is
based on the assumption~\cite{Suarez} that the scattering length could be negative initially, to accelerate the structure formation, and later become positive to prevent complete collapse. However, the mechanism by which the sign changes remains unknown.}. It behooves us, therefore, to study such systems theoretically.

%When the system cools down to low temperatures $T$, such that the de Broglie wavelength $\lambda_{dB}$ becomes comparable to the %inter-particle distance, the system enters into a  quantum degenerate state. For a system of bosons, this degenerate state is %called Bose-Einstein condensate (BEC) which happens below a critical temperature $T_c$. The critical temperature $T_c$ can be %determined by comparing de Broglie wavelength with the mean interparticle spacing $d$. If $n$ is the number density of bosons, %$d\sim n^{-1/3}$. The de Broglie wavelength for a system of bosons with mass $m$ and temperature $T$ is
%\begin{eqnarray}
%\lambda_{dB} = \bigg(\frac{2\pi \hbar^2}{mk_BT}\bigg)^{1/2}
%\end{eqnarray}
%so the transition temperature, $\lambda_{dB} \sim n^{-1/3}$, is 
%\begin{eqnarray}
%T_c = \frac{2\pi \hbar^2 n^{2/3}}{mk_B}
%\end{eqnarray}

To study the spatiotemporal evolution of axion stars, we must use the GPPE with both cubic and quintic nonlinearities;
the former is negative (because $a<0$) and favors the collapse instability mentioned above; the quintic nonlinearity, with a coefficient $g_2 >  0$, controls this instability. The resulting cubic-quintic GPPE (cq-GPPE), has been used, in the absence of gravity, to study (a) the evolution and merging dynamics of bright solitons~\cite{cq_GPPE_prl,PhysRevA_cq_gppe} in one dimension (1D) and (b) in three dimensions (3D), without the quintic term, for the collision of bright solitons~\cite{Parker_cq_gppe} and their collapse times during collisions; these studies use harmonic traps. With self-gravitation, Ref.~\cite{PRD_cq_chavanis} has employed the cq-GPPE to obtain a phase transition between dilute and dense axion matter by using a Gaussian Ansatz at temperature $T=0$; aside from this study, there are very few investigations of the cq-GPPE, in the context of axionic stars, with negative scattering lengths. 

Our study of phases and transitions in the cq-GPPE is the first to go beyond the Gaussian Ansatz and $T=0$. It leads to 
important insights into structure formation in self-gravitating axionic matter. We give a qualitative summary of our principal results before we present the details of our work. We first obtain the equilibrium configurations at $T = 0$ and then investigate finite-temperature effects ($T>0$) by using the Fourier-truncated cq-GPPE and building on our boson-star studies with the GPPE~\cite{PRR_akverma2021,verma2022rotating} that generalise Fourier-truncated investigations of the GPE~\cite{Berloff4675,krstulovic2011energy,shukla2013turbulence}. Furthermore, we obtain the equilibrium configuration, for $T \geq 0$, by using an auxiliary cubic-quintic stochastic Ginzburg-Landau-Poisson equation (cq-SGLPE), the imaginary time ($t \to -it$) version of the cq-GPPE.

If we start with a nearly uniform density, the Fourier-truncated $T=0$ cq-SGLPE collapses, first along one direction, leading to a structure that is reminiscent of a stack of \textit{Zeldovich pancakes}~\cite{shandarin1989large}. As we increase the  gravitational strength $G$, these pancakes transform into cylindrical layers, which finally collapse into a spherical 
axion star. If we cycle $G$ from low to high values and back, this system displays hysteresis loops, associated with first-order phase transitions between these pancake, cylindrical, and spherical states; these loops are examples of \textit{cosmological hysteresis}~\cite{sahni2012cosmological}. We next investigate finite-temperature ($T>0$): we demonstrate that these collapsed axionic states transform to tenuous, non-collapsed states at high $T$. Finally, we impose an angular velocity $\Omega$ 
on these states; we show that, beyond a critical angular velocity, quantum vortices  thread the axionic star; this critical angular velocity depends on the coefficient $g_2$ of the quintic nonlinearity. Finally, we employ our cq-GPPE approach to follow the spatiotemporal evolution of a rotating axionic condensate and also a rotating binary-axionic-condensate system; in the former, we show the emergence of vortices at large angular speeds $\Omega$; and, in the latter, we elucidate the rich dynamics of the mergers of the components of this binary system.

The remainder of this paper is organised as follows: In Section~\ref{sec:model} we describe the cq-GPPE and cq-SGLPE models and the pseudospectral methods that we use to study these. We present our results in Section~\ref{sec:result}. 
Section~\ref{sec:conclusion} contains our conclusions and a discussion of the significance of our results.

\section{Model and Numerical simulation}
\label{sec:model}

We define below the models we use and the numerical methods that we employ to study them.
\subsection{The cq-GPPE and cq-SGLPE}
\label{subsec:ZeroOmega}

At low temperatures, 3D bosonic systems form a Bose-Einstein condensate (BEC), which can be described by a macroscopic complex wavefunction $\psi({\bf x},t)$. For weakly interacting bosons, we can use the GPE; and, in the presence of Newtonian gravity, this can be generalised to the GPPE (see, e.g., Refs.~\cite{PRR_akverma2021,verma2022rotating}). For attractive self interactions 
between the bosons, as in axionic systems~\cite{braaten_axion,sikivie2008axion}, we must include a quintic nonlinearity for stability and employ the following cq-GPPE:
\begin{eqnarray}
i\hbar \frac{\partial \psi}{\partial t} &=& -\frac{\hbar^2}{2m} \nabla^2 \psi + \left[ G \Phi + g |\psi|^2 + g_2 |\psi|^4 \right] \psi\,; \nonumber \\
\nabla^2 \Phi  &=& |\psi|^2 - \left\langle |\psi|^2 \right\rangle\,;
\label{eq:GPPE}
\end{eqnarray}
\noindent $m$ and $ n = |\psi|^2 $ are, respectively, the mass and number density of bosons, $\Phi$ is the gravitational potential, $G \equiv 4\pi G_Nm^2$, and $g \equiv 4\pi a\hbar^2/m$, with $a < 0$ the $s$-wave scattering length, and $g_2 > 0$  the coefficient of the quintic term. If we linearise Eq.~(\ref{eq:GPPE}) about the constant density $|\psi|^2 = n_0$, we get the dispersion relation, between the frequency $\omega$ and the wave number $k$,
\begin{eqnarray}
    \omega(k) = \sqrt{\frac{\hbar^2k^4}{(2m)^2} -\frac{k^2}{m} \left(\frac{Gn_0}{k^2} -gn_0 -2g_2n_0^2\right)}\,,
\label{eq:disprel}
\end{eqnarray}
whence we define the wave number
\begin{equation}
\begin{aligned}
    k_J^2 &=& \frac{2m(gn_0+2g_2n_0^2)}{\hbar^2} \bigg[-1+ \sqrt{1+ \frac{G\hbar^2 n_0}{m(gn_0+2g_2n_0^2)^2}} 
 \bigg]
 \end{aligned}
\end{equation}
 %\begin{eqnarray}
 %    k_J = \sqrt{\frac{G}{g+2g_2n_0}}\left[\frac{1}{2} %\left( 1 + \sqrt{1+ \frac{G\hbar^2}{m(g+2g_2n_0)^2 %n_0}}\right)\right]^{-1/2}
 %\label{eq:kJeans}
 %\end{eqnarray}
below which the low-$k$ Jeans instability occurs.

Equation~(\ref{eq:GPPE}) conserves the number of particles $ N \equiv \int |\psi|^2 d^3 {\bf x} $ and the total energy $E \equiv E_{k} + E_{int} + E_G $:
\begin{eqnarray}
E &=& \int \bigg[\frac{\hbar^2}{2m} |\nabla \psi|^2 + \mathcal{V}(\psi) + \frac{G}{2} |\psi|^2 \nabla^{-2} |\psi|^2 \bigg] d^3 {\bf x}\,; \nonumber \\
\mathcal{V}(\psi) &\equiv& \frac{g}{2} |\psi|^4 + \frac{g_2}{3} |\psi|^6\,.
\label{eq:total_energy}
\end{eqnarray}

In the absence of gravity [$G=0$], the stationary solution of Eq.~(\ref{eq:GPPE}) in a volume $V$, has a constant density $|\psi_0|^2 = n_0 = \rho_0/m$, total energy $E_0$ [Eq.~(\ref{eq:total_energy})], pressure $P_0$, and speed of sound 
$v$:
\begin{eqnarray}
E_0 &=& \frac{1}{2}g|\psi_0|^4 V+ \frac{1}{3}g_2|\psi_0|^6V
= \frac{1}{2}\frac{gN_0^2}{V}+ \frac{1}{3}\frac{g_2N_0^3}{V^2}\,; \nonumber \\
P_0 &\equiv& -\frac{\partial E_0}{\partial V} = \frac{1}{2}\frac{g\rho_0^2}{m^2} + \frac{2}{3} \frac{g_2\rho_0^3}{m^3}\,; \nonumber\\
%\label{eq:E0}
%\end{eqnarray}
%whence we obtain the pressure $P_0 = -\frac{\partial E_0}{\partial V}$' and the speed of sound 
%$v = \sqrt{\frac{\partial P_0}{\partial \rho_0}}$:
%\begin{eqnarray}
%P_0 = \frac{gN_0^2}{2V^2} + \frac{2}{3}\frac{g_2N_0^3}{V^3}\nonumber \\
%P_0 = \frac{1}{2}gn_0^2 + \frac{2}{3}g_2n_0^3\\
%P_0 &=& \frac{1}{2}\frac{g\rho_0^2}{m^2} + \frac{2}{3} \frac{g_2\rho_0^3}{m^3}\,; \nonumber\\
v &\equiv& \sqrt{\frac{\partial P_0}{\partial \rho_0}} = \sqrt{\frac{gn_0 + 2g_2n_0^2}{m}}\,.
\label{eq:E0P0v}
\end{eqnarray}
If we make the approximation $|\nabla \psi|^2 \simeq \psi/\xi^2$, then we can estimate the coherence length $\xi$ by equating the kinetic and interaction terms as follows:
\begin{eqnarray}
\frac{\hbar^2}{2m}\frac{\psi}{\xi^2} &=& g|\psi|^2 \psi + g_2|\psi|^4 \psi\,, {\rm whence} \nonumber\\
\xi &=& \frac{\hbar}{\sqrt{2m(gn_0+g_2n_0^2)}}\,.
\end{eqnarray}
We use $\tau \equiv \xi/v$ to non-dimensionalise the time $t$ in our direct numerical simulations (DNSs). The time scale based on $\tau$ and length $L$ of the simulation box are not comparable to the scales used in astrophysics. We define length, time, and mass scales relevant to astrophysics in the Appendix. We use pseudospectral DNSs to solve the 3D cq-GPPE and cq-SGLPE,
%For the GPPE~\ref{eq:GPPE} 
in a cubical domain, with side $L = 2\pi$ and $N^3$ collocation points, and periodic boundary conditions in all three spatial directions. We employ the Fourier expansion
\begin{eqnarray}
\psi ({\bf x}) = \sum_{ {\bf k} } \hat{\psi}_{{\bf k}} \exp (i {\bf k} \cdot {\bf x}) \,, 
\label{eq:Fourier}
\end{eqnarray}
and the $2/3$-rule for dealiasing, i.e., we truncate the Fourier modes by setting $\hat{\psi} \equiv 0$ for $ |{\bf k}| > k_{max}$\cite{giorgio_2011,HOU_dealiasing}, with $k_{max} = [N/3]$. The Fourier-truncated cq-GPPE is
\begin{eqnarray}
i\hbar \frac{\partial \psi}{\partial t} &=& P_G \bigg[-\frac{\hbar^2}{2m} \nabla^2 \psi + P_G \biggl\{ \bigg( G \nabla^{-2} + g \nonumber \\
&+& g_2 P_G (|\psi|^2)\bigg)  |\psi|^2 \biggr\} \psi\bigg] \,,
\label{eq:TGPPE}
\end{eqnarray}
where $P_G$ is the Galerkin projector [with $ P_G[\hat{\psi}_{{\bf k}}] = \theta (k_{max} - |{\bf k}|) \hat{\psi}_{{\bf k}} $ ]. For time marching we use the fourth-order Runge-Kutta scheme RK4.\\

%Equation \ref{eq:TGPPE} conserves the number of particles $ N = \int |\psi|^2 d^3 {\bf x} $ and the total energy $E = E_{kq} + E_{int} + E_G $, where

The total energy $E \equiv E_{k} + E_{int} + E_G $ now becomes
\begin{eqnarray}
E_{k}  &=&  \frac{\hbar^2}{2m} \int d^3 {\bf x} |\nabla \psi |^2 \,;\nonumber \\
E_{int} &=&  \int d^3 {\bf x} \left[  \frac{g}{2} (P_G |\psi|^2)^2 + \frac{g_2}{3} \left[  P_G \left\lbrace (P_G |\psi|^2)^2\right\rbrace \right] |\psi|^2  \right] \,; \nonumber \\
E_G &=& \frac{G}{2} \int d^3 {\bf x} \left[ P_G |\psi|^2 \right] \nabla^{-2}  \left[ P_G |\psi|^2 \right] \,.
\label{eq:energy}
\end{eqnarray} 
Equilibrium configurations of the cq-GPPE can be obtained efficiently by using the following Galerkin-truncated cq-SGLPE, which follows from Eq.~(\ref{eq:TGPPE}) via the Wick rotation $t \to -it$:
%We study the formation of structures, starting from a nearly uniform density distribution at $T = 0$ directly using the SGLPE. We study the finite temperature effects using spectrally truncated GPPE or by using the truncated SGLPE. 
%in we obtain the spectrally truncated SGLPE.
%\onecolumngrid
%\begin{multline}
\begin{eqnarray}
\hbar \frac{\partial \psi}{\partial t} &=& P_G \bigg[ \frac{\hbar^2}{2m} \nabla^2 \psi +\mu \psi - P_G \bigg[  \biggl\{  G \nabla^{-2} + g \nonumber \\
&+&  g_2 P_G (|\psi|^2)\biggr\}  |\psi|^2 \bigg]  \psi \Bigg]
+ \sqrt{\frac{2\hbar}{\beta}} P_G\left[ \xi({\bf x},t)\right]\,, 
\label{eq:TSGLPE}
\end{eqnarray}
%\end{multline}
%\twocolumngrid
\noindent where $\beta = 1/(k_B T)$, $k_B$ is the Boltzmann constant,
$T$ is the temperature, and $\xi({\bf x},t)$ is a zero-mean Gaussian white noise with $\left\langle \xi({\bf x},t) \xi^*({\bf x'},t') \right\rangle = \delta(t -t') \delta({\bf x} - {\bf x'})$. Although the truncated cq-SGLPE Eq.~(\ref{eq:TSGLPE}) 
does not conserve the total energy, its DNS converges more rapidly to the long-time
solution of the truncated cq-GPPE~(\ref{eq:TGPPE}) than does a direct DNS of the latter  (cf., Refs.\cite{PRR_akverma2021,marc_giorgio_particle} for the GPPE and the GPE).

\subsection{cq-GPPE with rotation ($\Omega\neq 0$)}
\label{subsec:NonzeroOmega}

One of the most remarkable features of superfluids, rotating with an angular
frequency $\Omega$, is the formation of quantized vortices when $\Omega > \Omega_c $, a critical angular frequency. The circulation around the  vortex line is quantized:
\begin{eqnarray}
\oint_C {\bf v}_s \cdot {\bf dl} = \kappa \,,
\label{eq:circulation}
\end{eqnarray}
where ${\bf v}_s$ is the superfluid velocity and $\kappa \equiv h/m$. We investigate the formation of quantized vortices in gravitationally collapsed axionic condensates
 %. To study the rotational dynamics of BEC halos and axion stars, 
 by introducing the rotation term $-\Omega L_z\psi$ into the cq-GPPE~(\ref{eq:GPPE}),
 where $L_z = -i\hbar (x\partial_y - y \partial_x)$ is the $z$-component of the angular momentum $\bf L = x\times P$. The equilibrium configuration can then be obtained 
 by using the following cq-SGLPE with the rotation term:
\begin{eqnarray}
\hbar \frac{\partial \psi}{\partial t} &=& \frac{\hbar^2}{2m} \nabla^2 \psi - \big[ G \Phi \nonumber \\ 
&+& g |\psi|^2 + g_2 |\psi|^4 - \Omega L_z\big] \psi \,. 
\label{eq_SGLPE_rot}
\end{eqnarray}
We first obtain a spherical collapsed condensate by using Eq.~(\ref{eq_SGLPE_rot}) for $\Omega=0$; we then increase $\Omega$ slowly up until the critical angular speed $\Omega_c$, beyond which vortices thread the system.
%The final equations that we solve using direct numerical simulations are Eq.\ref{eq:TGPPE} for GPPE and Eq.\ref{eq:SGLPE}, \ref{eq_SGLPE_rot} for SGLPE. 
In our DNSs with $\Omega\neq 0$, we use the pseudospectral methods that we have described above.
%in a cubic periodic box of dimension $L=2\pi$. 

In the remaining part of this paper, we work with the dimensionless forms of these equations with $\hbar = 1$ and $m = 1$. [In the Appendix, we give the length and times scales that we should use for different astrophysical systems.] We characterise the equilibrium configuration by the scaled radius of gyration:
\begin{eqnarray}
\frac{R}{L} = \frac{1}{L}\sqrt{\frac{\int \rho(r)r^2 d{\bf r}}{\int \rho(r) d{\bf r}}}\,.
\label{eq:gyration}
\end{eqnarray}

\subsection{Initial conditions}
\label{subsec:IC}

We use the following initial conditions in our DNSs:
%set for the simulations of Eq. \ref{eq:TGPPE},\ref{eq:SGLPE},\ref{eq_SGLPE_rot},
\begin{itemize}
    \item {\bf IC1}: To study the formation of different structures, we solve the cq-SGLPE~(\ref{eq:TSGLPE}) at $T=0$, with an initially uniform density on which we superimpose a small nonuniform perturbation.
    \item {\bf IC2}: After we obtain a stable axionic condensate, we study the collision dynamics of two such condensates by using Eq.~(\ref{eq:TGPPE}) and the following initial condition for this binary system [cf., Ref.~\cite{PRL_newton_method}]: We first obtain a radially symmetric solution $\psi(r,t)$ via the expansion 
\begin{equation}
\psi(r,t) = \sum_{n=0}^{N_R/2} \hat{\psi}_{2n}(t) T_{2n}(r/R)\,,
\end{equation}
where $T_n$ is the order-$n$ Chebyshev polynomial (of the first kind) and $\hat{\psi}_{N_R}$ is chosen to satisfy the boundary condition $\psi(R,t) = 0$. We then use the following
relaxation method to obtain the stationary state of Eq.~(\ref{eq:GPPE}):
\begin{eqnarray}
\psi(r,t+dt) &=& \Theta^{-1} \bigg[ \psi(r,t) - dt \biggl\{ (G\Phi \nonumber \\ 
&+& g |\psi|^2 + g_2|\psi|^4) \psi \biggr\} \bigg]\,, 
\label{eq:relaxation}    
\end{eqnarray}
where $\Theta = 1 -dt \nabla^2|_r/2$ and $\nabla^2|_r \equiv \frac{1}{r^2} \frac{\partial}{\partial r}  \biggl(r^2 \frac{\partial}{\partial r} \biggr)$. For rapid convergence to the stationary state, we use the following Newton method:
%starting from a stable state obtained by Eq.~\ref{eq:relaxation}. 
We define $f_j(\psi) \equiv \psi_j(t+dt) - \psi_j(t)$ and look for the root $\psi_{j*}$ at which $f_j(\psi_{j*}) = 0$; here, $\psi_j(t)$ is the value of $\psi(t)$ at the collocation point $j$. At every Newton iteration step, we solve (numerically) $\sum_k [( df_j/d\psi_k) \delta \psi_k]  = -f_j(\psi_j)$, to find $\delta \psi_k$~\footnote{We can also use the SGLPE to prepare the initial condition for our binary-star study, but we use the Newton method because of it yields much faster convergence to the stationary solution of the GPPE.}.
\end{itemize}

\section{Results}
\label{sec:result}

We present our results as follows: In Subsection~\ref{subsec:Gaussian} we present analytical results that use a Gaussian Ansatz for a spherically symmetric density profile. Subsections~\ref{subsec:ZeroT} and \ref{subsec:NonzeroT} are devoted, respectively, to our studies at temperature $T = 0$ and $T > 0$. In Subsections~\ref{subsec:Rotation} and \ref{subsec:Rotbinary} we discuss, respectively, a rotating axionic condensate and a rotating binary-axionic-condensate system.

\subsection{The Gaussian Ansatz }
\label{subsec:Gaussian}

\begin{figure*}[!hbt]
  \resizebox{\linewidth}{!}{
    \includegraphics[width=0.5\linewidth]{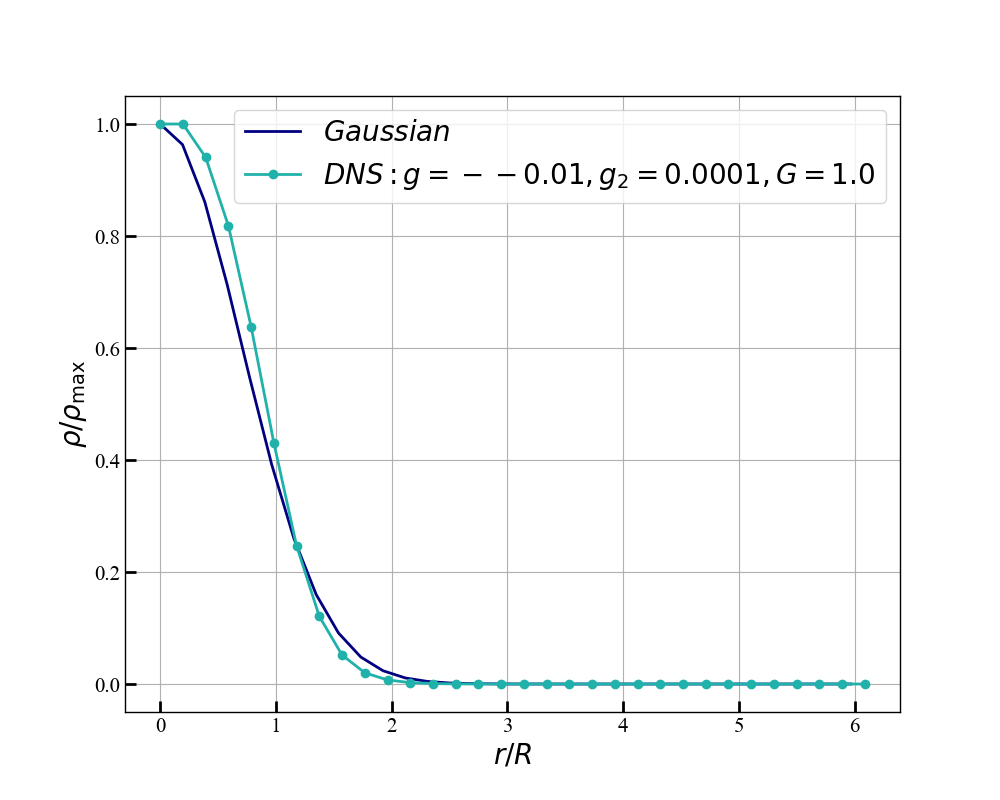}
    \includegraphics[width=0.5\linewidth]{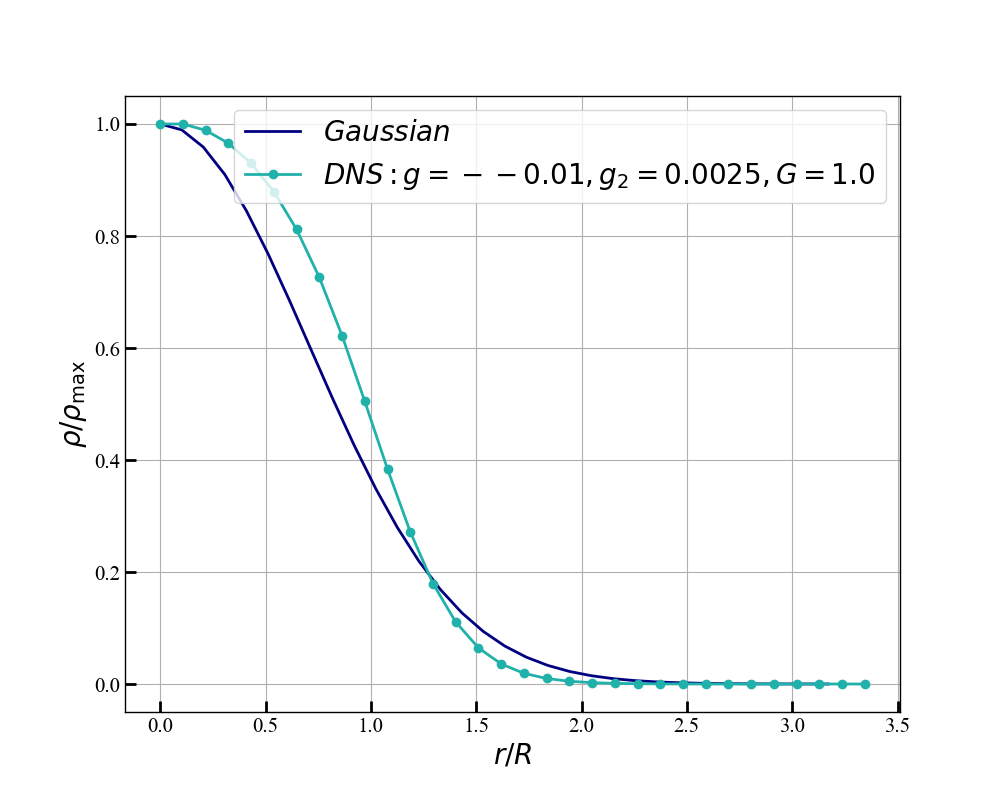}
    \includegraphics[width=0.5\linewidth]{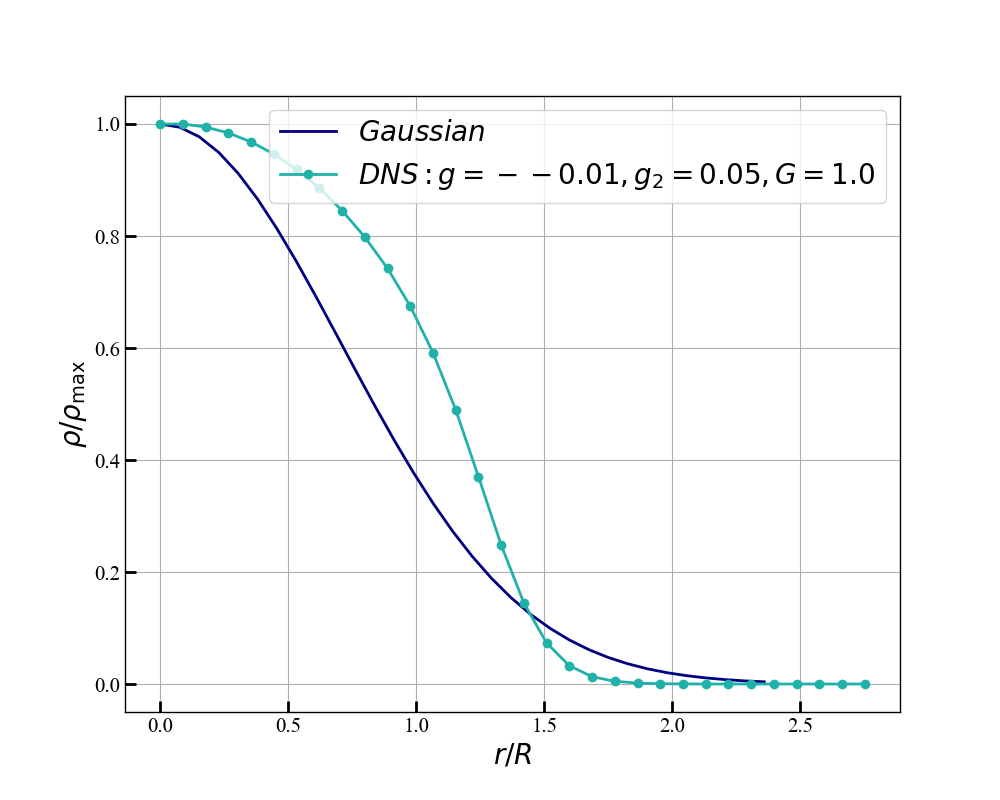}
    }
    \caption{Plots of the density $\rho(r)/\rho_{max}$ versus  $r/R$, the scaled distance from the center of the axion star, for different values of the parameters $g$, $g_2$, and $G$. The light blue curve is from our DNS and the dark blue curve follows from the Guassian Ansatz [see the text and Eq.~(\ref{eq:gaussian_ansatz})]. $R$ is the radius of gyration given in Eq.~(\ref{eq:gyration}).}
    \label{fig:gaussian_density}
\end{figure*}

Most analytical treatments of the cq-GPPE make the following Gaussian approximation for a spherically symmetric density profile (see, e.g., Ref.~\cite{chavanis_2011}):
\begin{eqnarray}
\rho(r) = \rho(0) e^{-r^2/R^2} \,,
\label{eq:gaussian_ansatz}
\end{eqnarray}
where $\rho(0) = M/(\pi^{3/2} R^3)$ is the central density and $R$ is the radius of gyration given by Eq.~(\ref{eq:gyration}). We contrast, in Fig.~\ref{fig:gaussian_density}, illustrative density profiles of spherically collapsed axion stars, which we obtain from this Gaussian Ansatz and our DNSs of Eq.~(\ref{eq:TSGLPE}), for various values of $g_2$, but with fixed $g=-0.01$ and $G=1$. We find that this Ansatz approximates the density profiles very well for small $g_2$ [see Fig.~\ref{fig:gaussian_density}(a)]. As $g_2$ increases
[see Figs.~\ref{fig:gaussian_density}(b) and (c)], the DNS density profile approaches that of a polytrope of index $n=1$, which has a compact support~\cite{chavanis_2011}, and the Gaussian Ansatz becomes a poor approximation.

\begin{figure}[!hbt]
    \centering
    \includegraphics[width=1\linewidth]{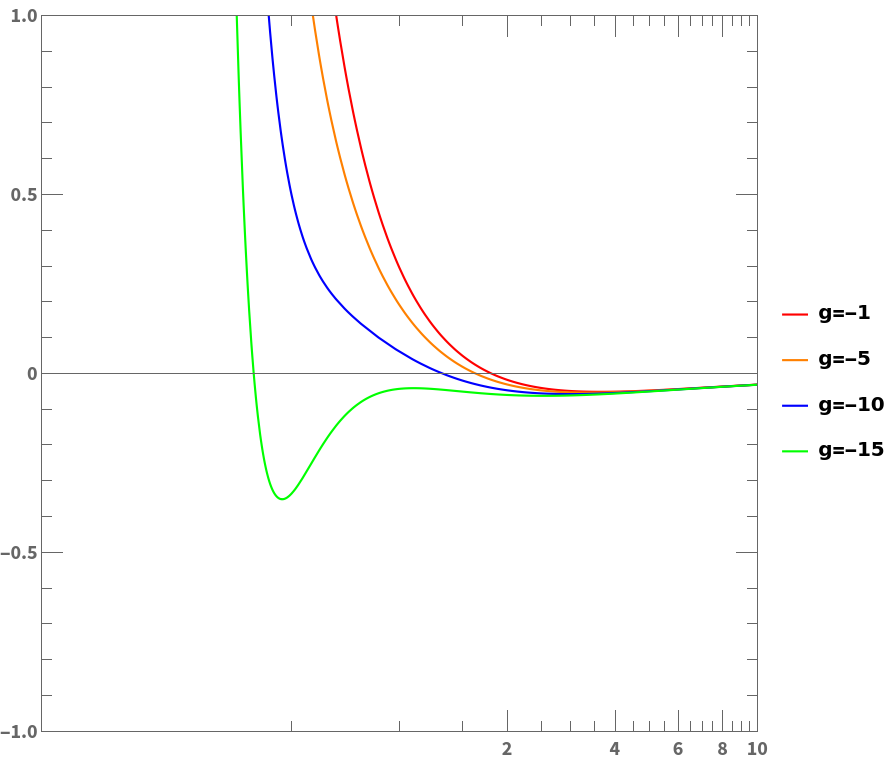}
    \put(-100,90){{\bf LM}}
    \put(-175,65){{\bf GM}}
    \caption{Plots of the effective potential $\mathcal{V}_{eff}$ versus the axionic condensate's radius of gyration $R$ [see  Eq.~(\ref{eq:gyration})], for $g_2=8$ and $G=0.8$. The minima labelled ${\bf LM}$ and ${\bf GM}$ correspond, respectively, to low- and high-density phases of axionic condensates. }
    \label{fig:eff_pot}
\end{figure}

If we continue with this simple Gaussian Ansatz, we can calculate the effective-potential-energy curve, whose minimum yields the collapsed axion star, as follows. By using the Madelung transformation
\begin{eqnarray}
\psi({\bf r},t) = \sqrt{\frac{\rho}{m}} e^{i\vartheta({\bf r},t)}
\end{eqnarray}
we rewrite the different parts of the total energy Eq.~(\ref{eq:total_energy}) as
\begin{eqnarray}
E_{k} &=&\frac{1}{2} \int \rho {\bf v}_s^2d^3{\bf x} + \frac{\hbar^2}{8m^2} \int \frac{(\nabla \rho)^2}{\rho}d^3{\bf x}\,, \nonumber\\
E_{int} &=& \int \bigg( \frac{g}{2m^2}\rho^2 + \frac{g_2}{3m^3}\rho^3 \bigg)d^3{\bf x}\,,\nonumber\\
E_G &=& \frac{G}{2}\int \rho \Phi d^3{\bf x}\,, 
\end{eqnarray}
where ${\bf v}_s = \frac{\hbar}{m} \nabla \vartheta$ and the gravitational potential is calculated via 
$\Phi = \int \frac{\rho({\bf x'})}{|{\bf x} - {\bf x'}|} d^3{\bf x'}$. The kinetic energy $E_k$ is a sum of the classical $E_{kc}$ and quantum $E_{kq}$ kinetic energies:
\begin{eqnarray}
E_{kc} &=& \frac{1}{2} \int \rho {\bf v}_s^2d^3{\bf x}\,;\nonumber\\
E_{kq} &=& \frac{\hbar^2}{8m^2} \int \frac{(\nabla \rho)^2}{\rho}d^3{\bf x}\,. 
\end{eqnarray}
Given the Gaussian Ansatz~(\ref{eq:gaussian_ansatz}), we can calculate these energies (by performing different integrals via Mathematica) to obtain
\begin{eqnarray}
E_{kq} &=& \frac{3\hbar^2 M}{4m^2R^2}\,,\nonumber\\
E_{int}&=& \frac{g_2M^3}{9\sqrt{3}m^3\pi^3 R^6} + \frac{gM^2}{4\sqrt{2}m^2 \pi^{3/2}R^3}\,,\nonumber\\
E_G &=& -\frac{GM^2}{\sqrt{\pi}R}\,,
\end{eqnarray}
whence we get
\begin{eqnarray}
E = E_{kc} + \mathcal{V}_{eff}\,,
\end{eqnarray}
where the effective potential is
\begin{eqnarray}
 \mathcal{V}_{eff}=E_{kq}+E_{int}+E_{G}\,.
\end{eqnarray}
In Fig.~\ref{fig:eff_pot} we present plots of $\mathcal{V}_{eff}$ versus $R$, for $g_2=8$ and $G=0.8$ and different values of the attraction parameter $g$, to show that there is a single, low-density minimum at small negative values of $g$. As we increase the attraction, a new minimum appears: At large negative values of $g$, there are two minima, labelled ${\bf LM}$ and ${\bf GM}$, which correspond, respectively, to low- and high-density phases of the axionic system; e.g., at $g=-15$, the dense axionic condensate is the global minimum ${\bf GM}$
of $\mathcal{V}_{eff}$.

\subsection{Collapsed Axionic Condensates: $T=0$}
\label{subsec:ZeroT}

\begin{figure*}[!hbt]
	%\resizebox{\linewidth}{!}{
		\includegraphics[width= 0.25\linewidth]{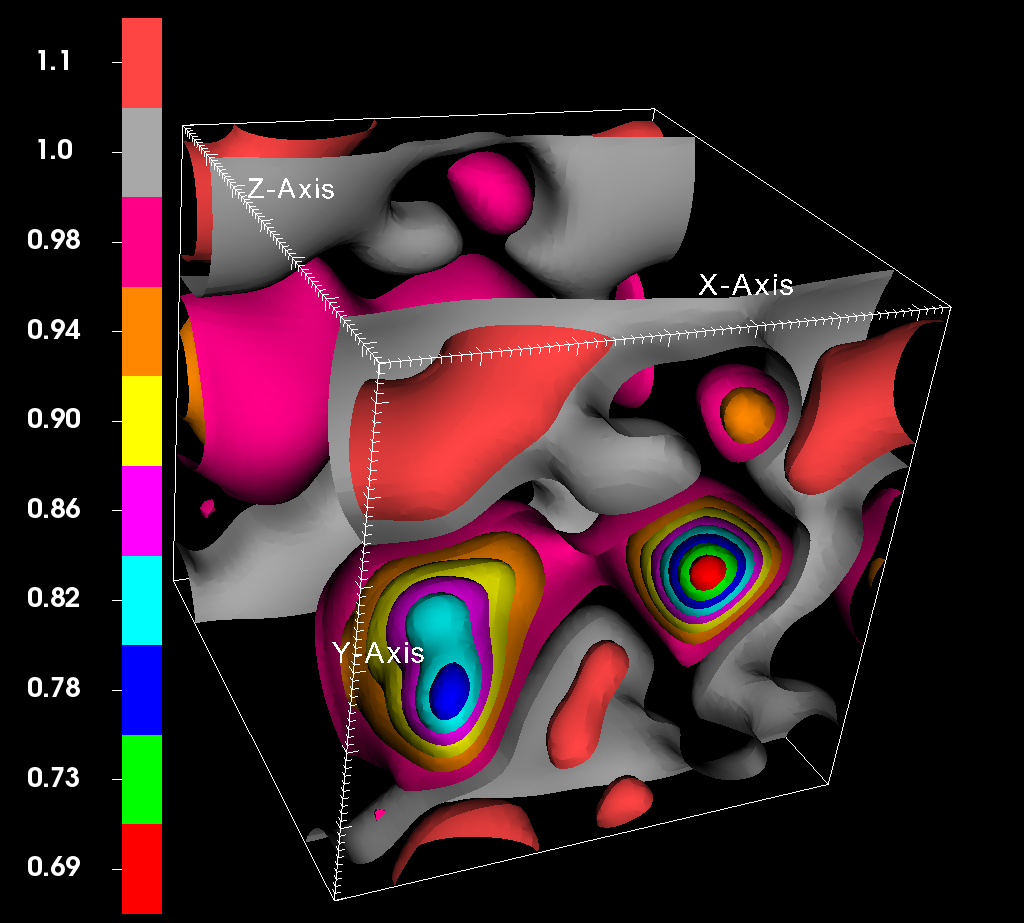} 
		%\put(-80,70){(a)}
		\includegraphics[width= 0.25\linewidth]{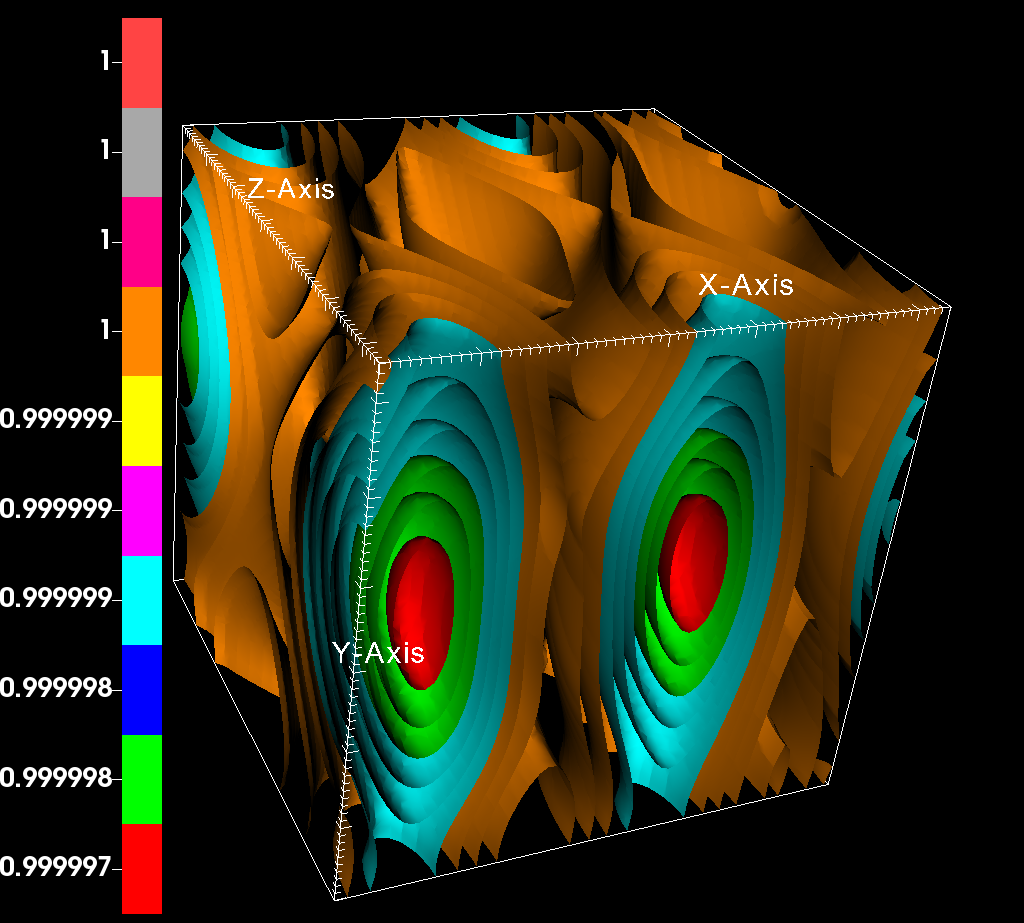} 
		\includegraphics[width= 0.25\linewidth]{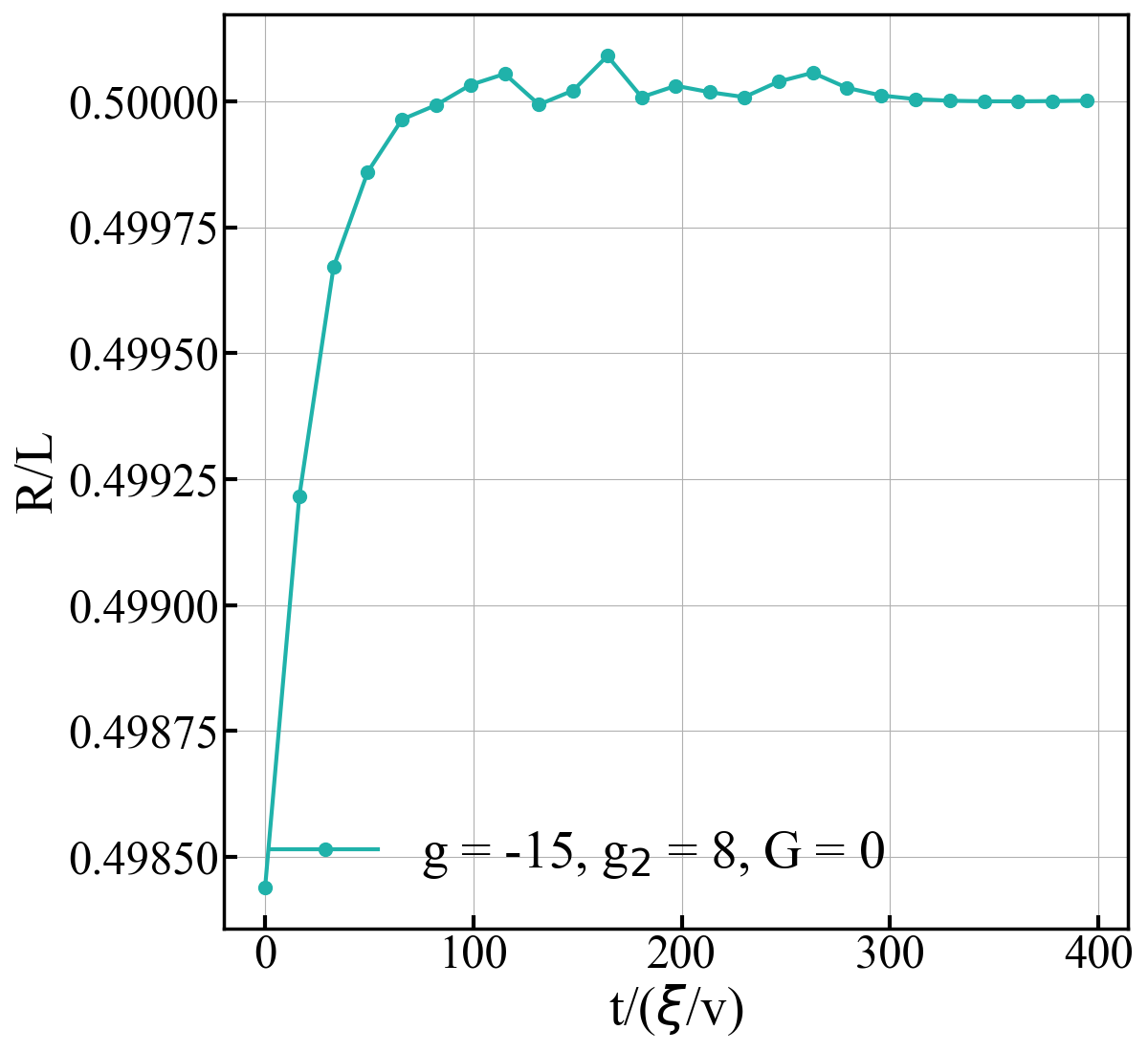} \put(-100,95){{\bf (a)}} %}	
	
	%\resizebox{\linewidth}{!}{
		\includegraphics[width= 0.25\linewidth]{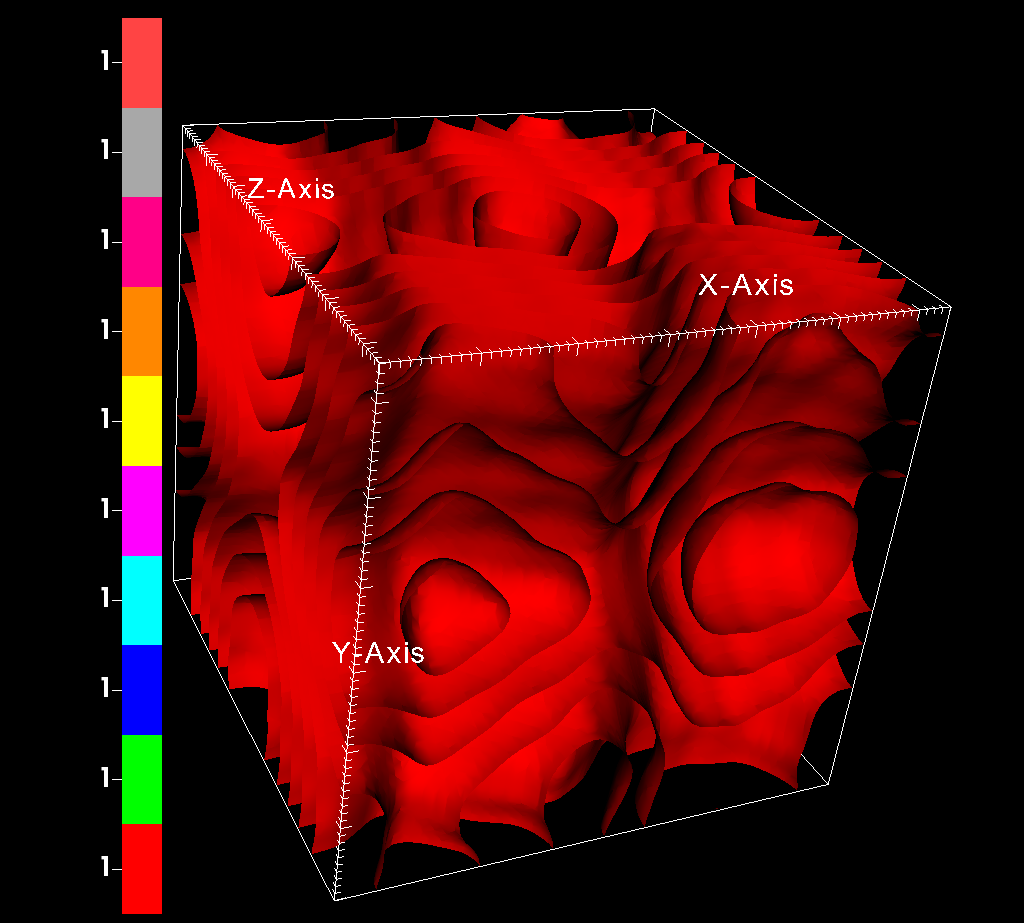} 
		%\put(-80,70){(a)}
		\includegraphics[width= 0.25\linewidth]{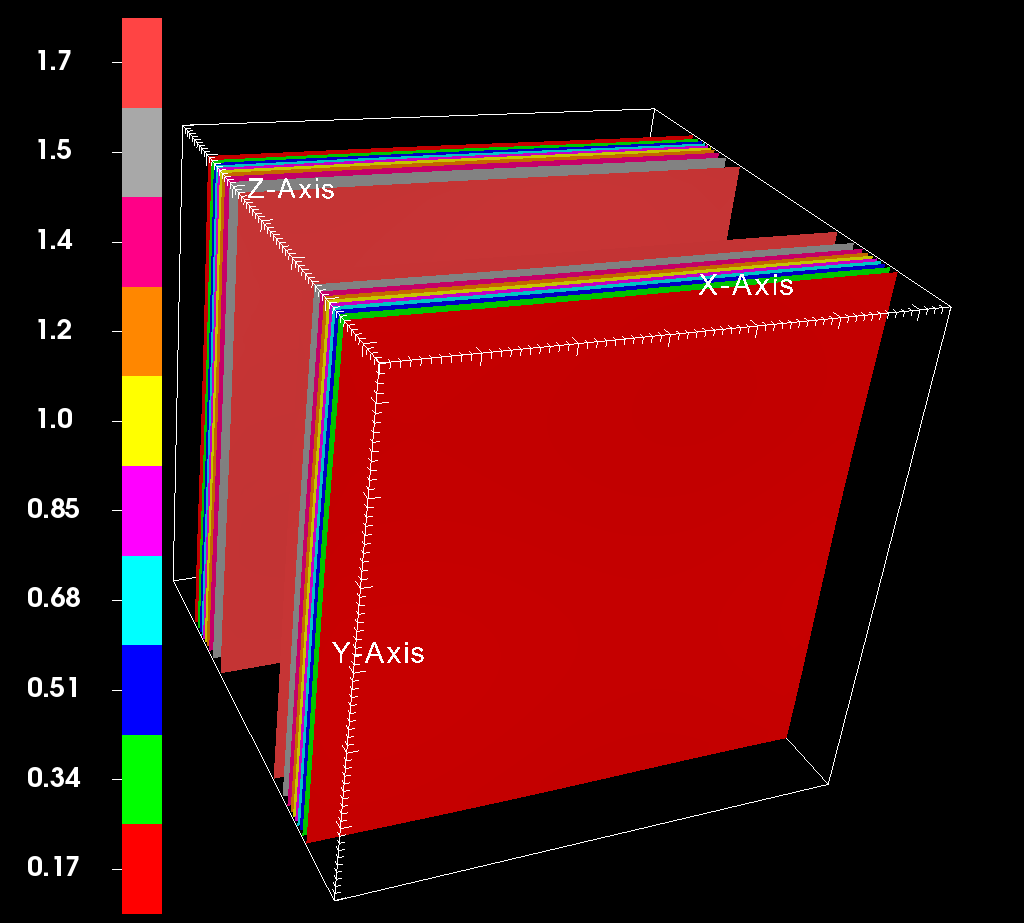}
		\includegraphics[width= 0.25\linewidth]{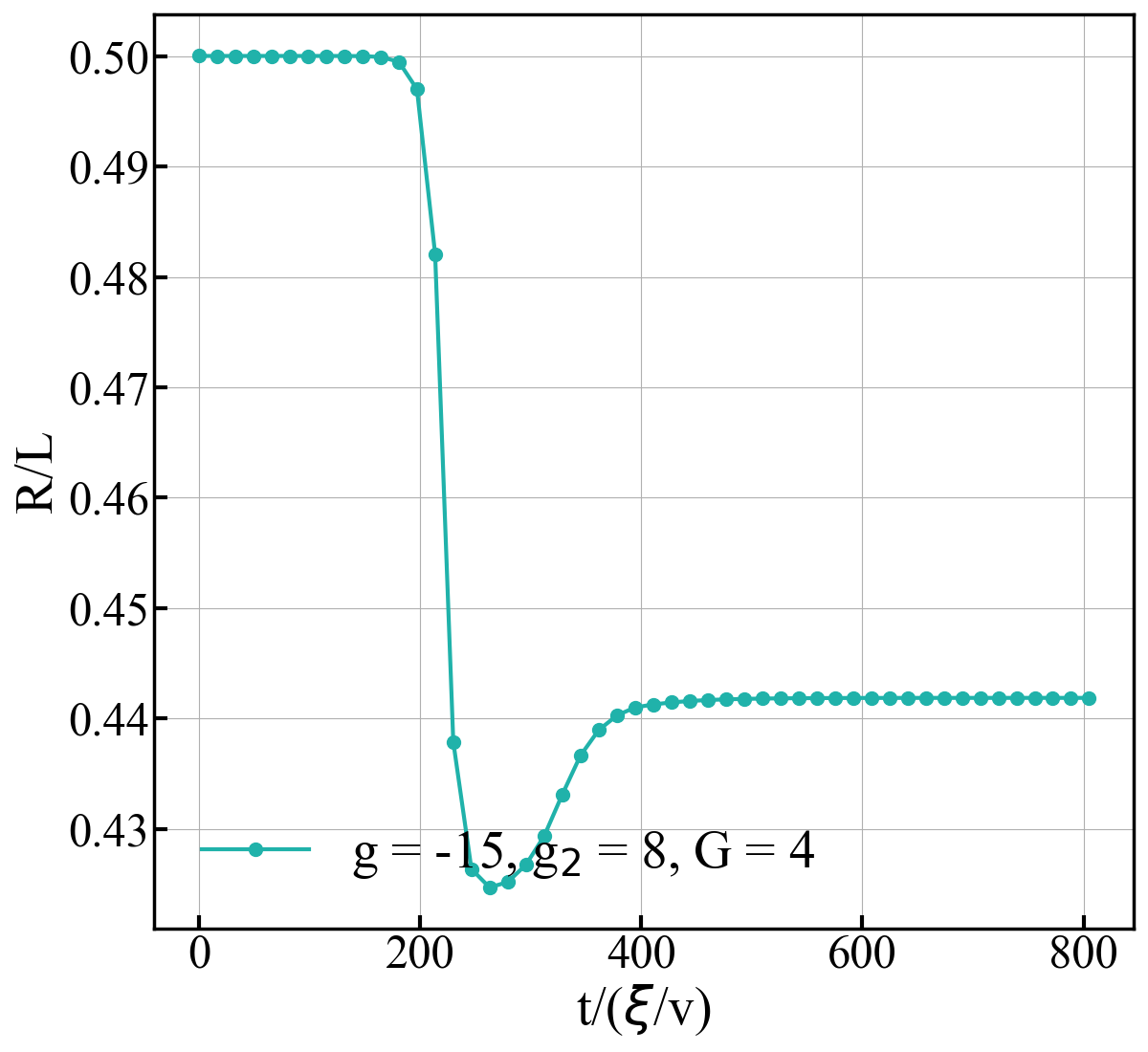} \put(-100,95){{\bf (b)}} %}	
	
	%\resizebox{\linewidth}{!}{
		\includegraphics[width= 0.25\linewidth]{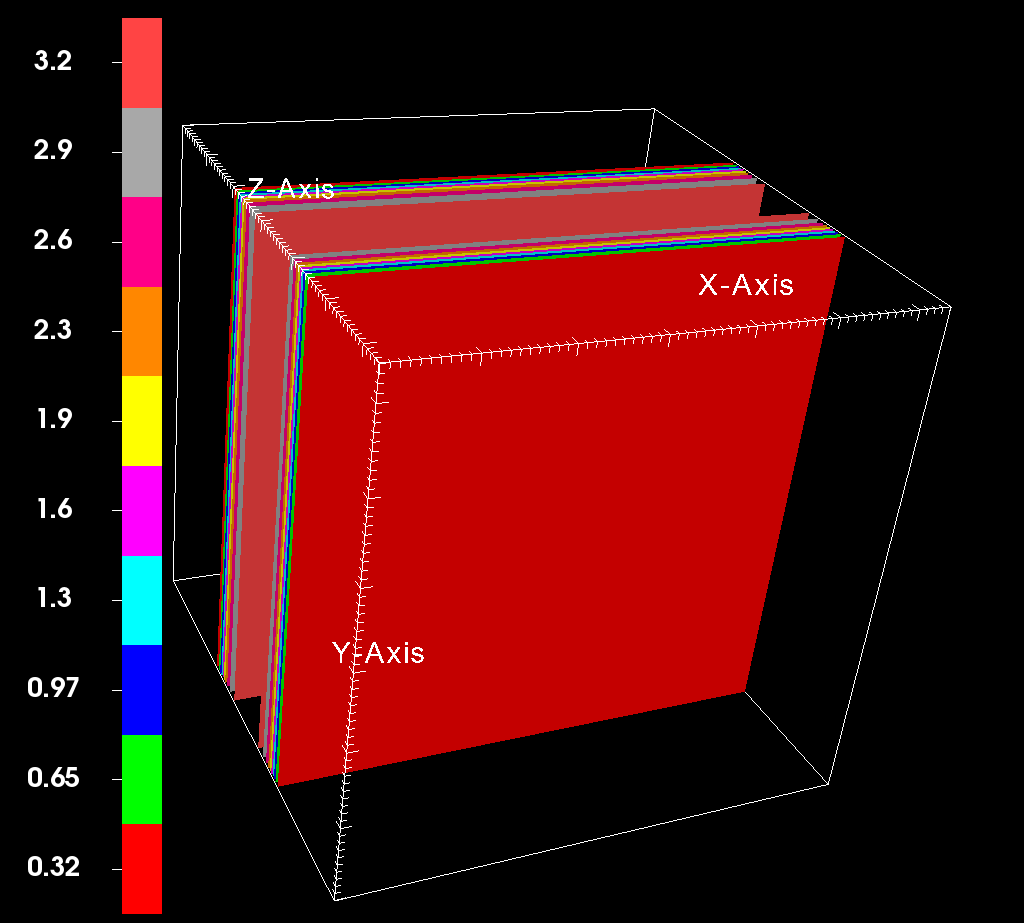} 
		%\put(-80,70){(a)}
		\includegraphics[width= 0.25\linewidth]{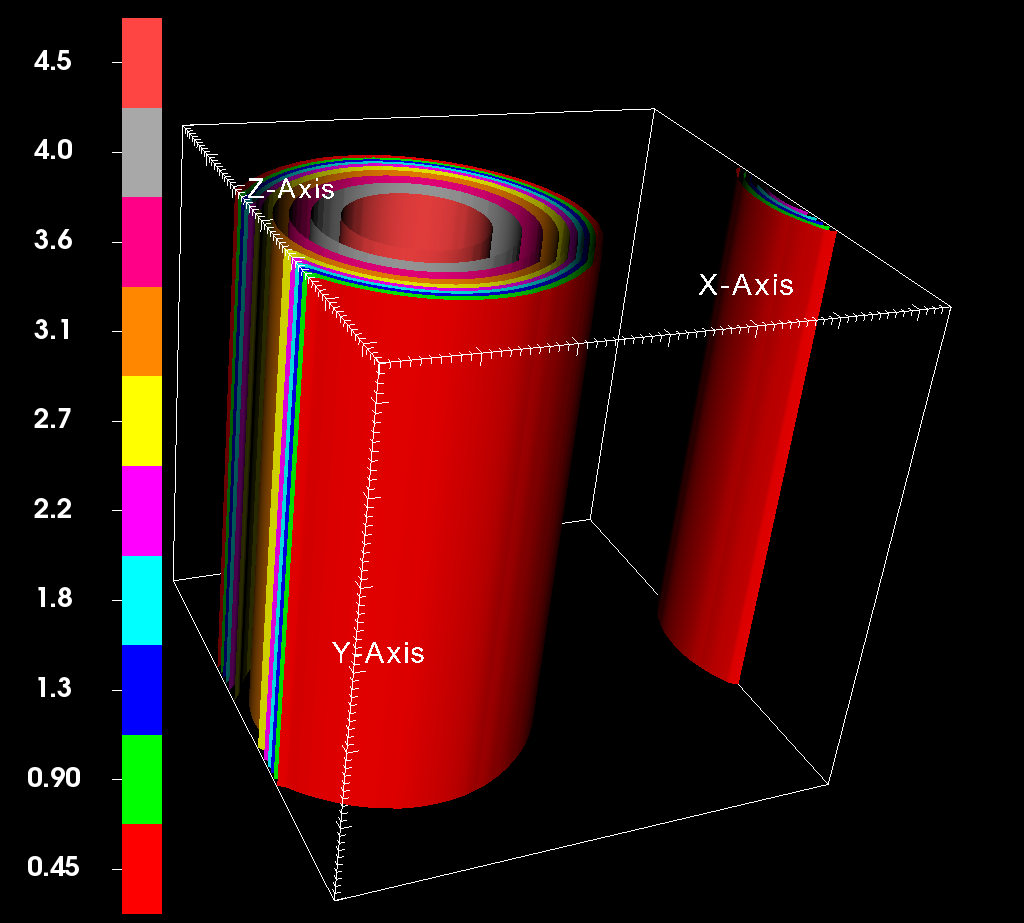}
		\includegraphics[width= 0.25\linewidth]{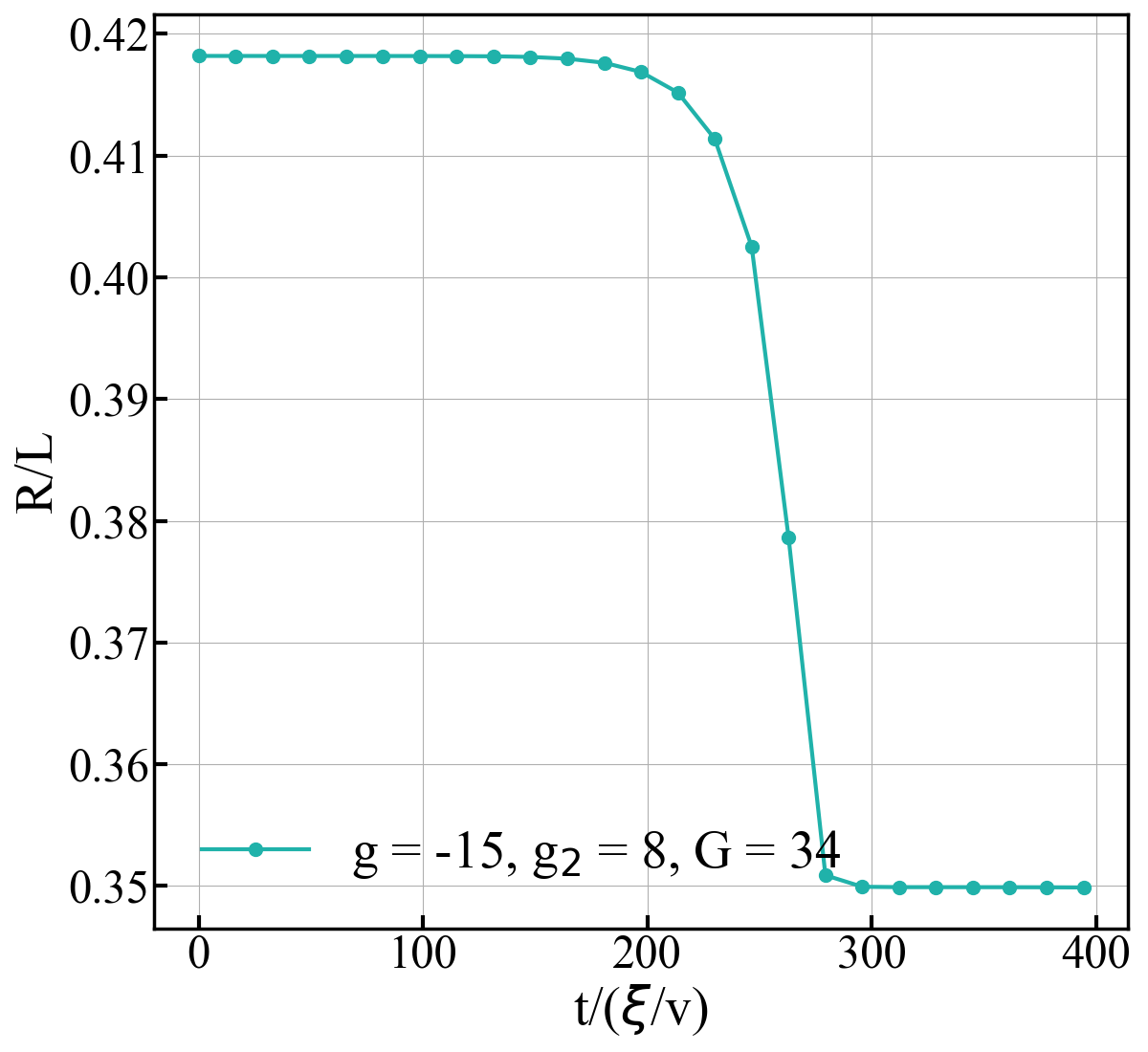} \put(-100,95){{\bf (c)}} %} 
	
	%\resizebox{\linewidth}{!}{
		\includegraphics[width= 0.25\linewidth]{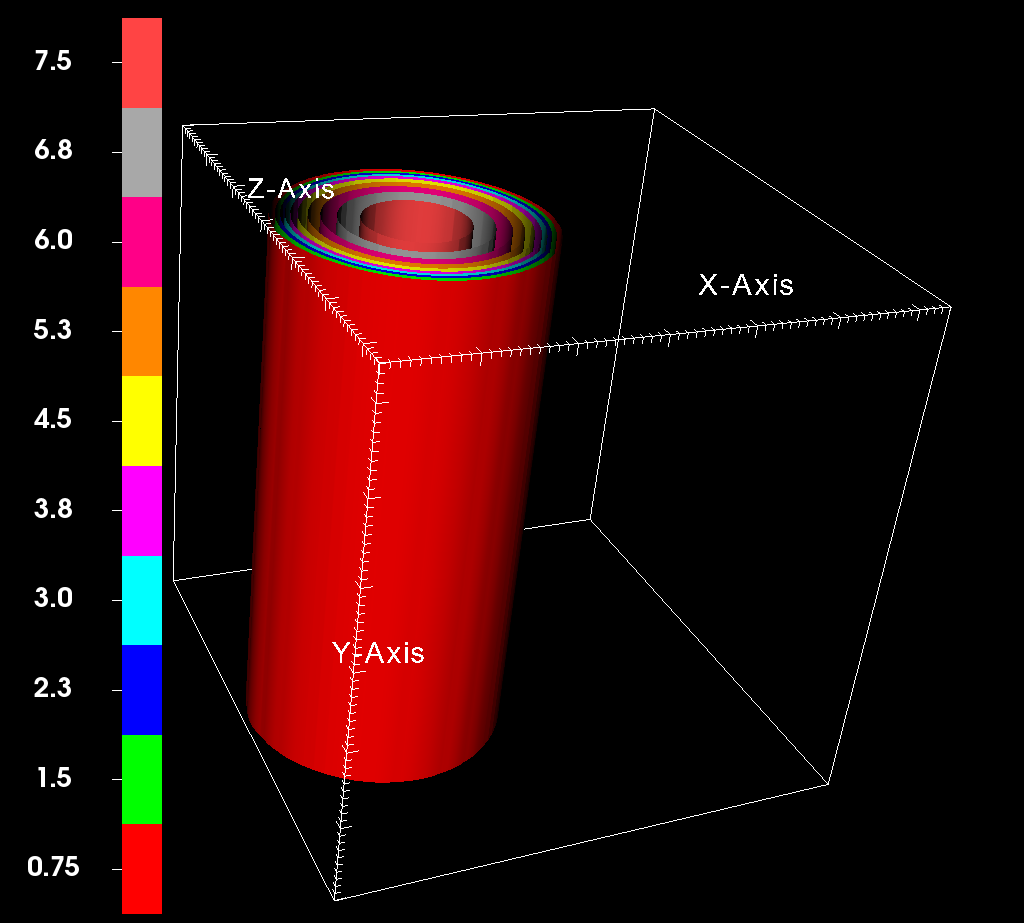} 
		%\put(-80,70){(a)}
		\includegraphics[width= 0.25\linewidth]{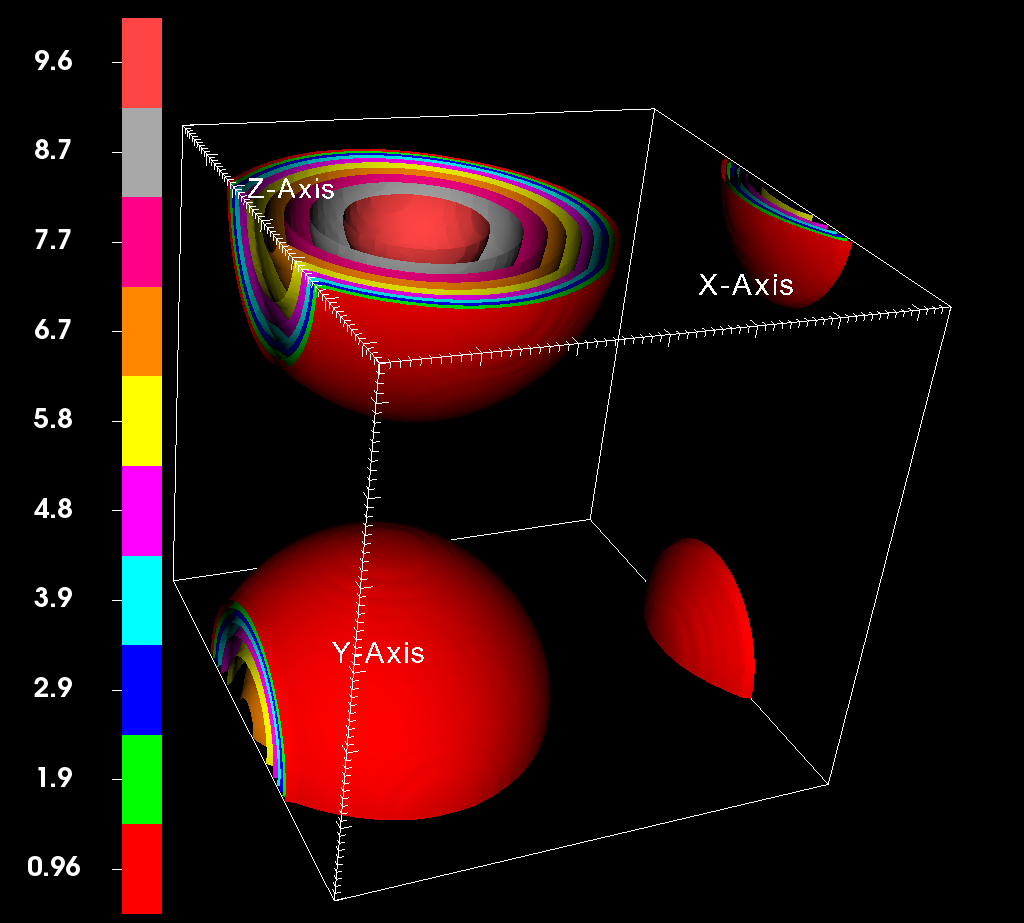}
		\includegraphics[width= 0.25\linewidth]{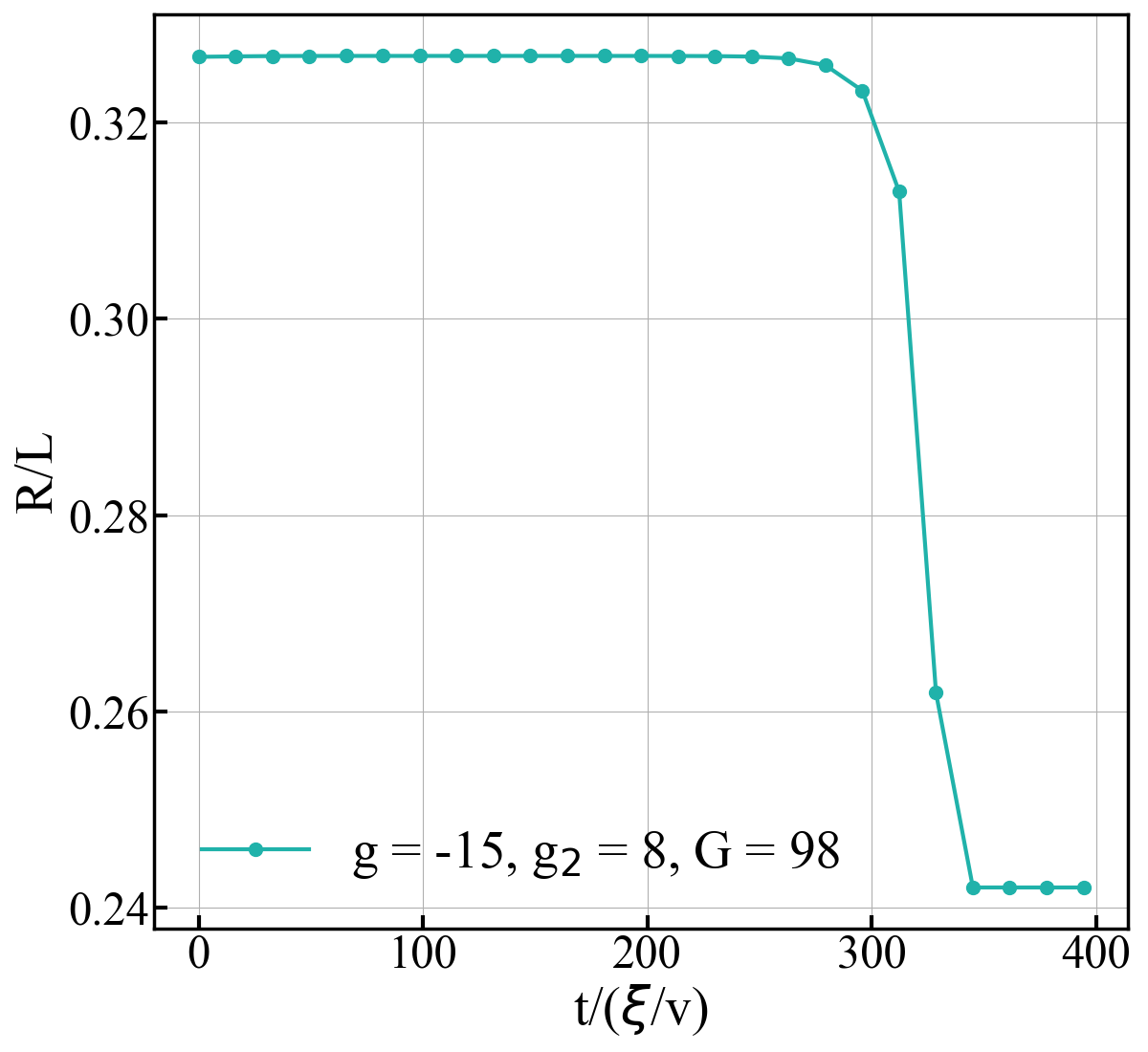} \put(-100,95){{\bf (d)}} %}
  
        \includegraphics[width= 0.25\linewidth]{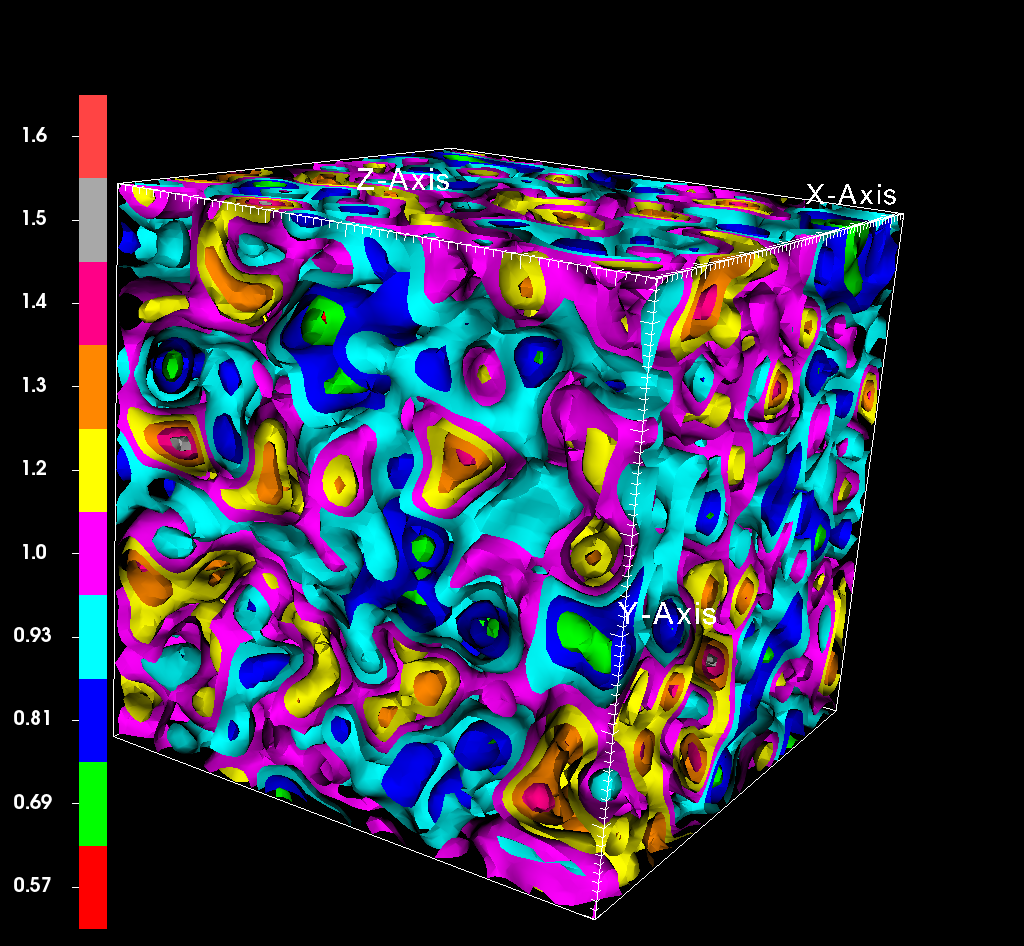} 
		%\put(-80,70){(a)}
		\includegraphics[width= 0.25\linewidth]{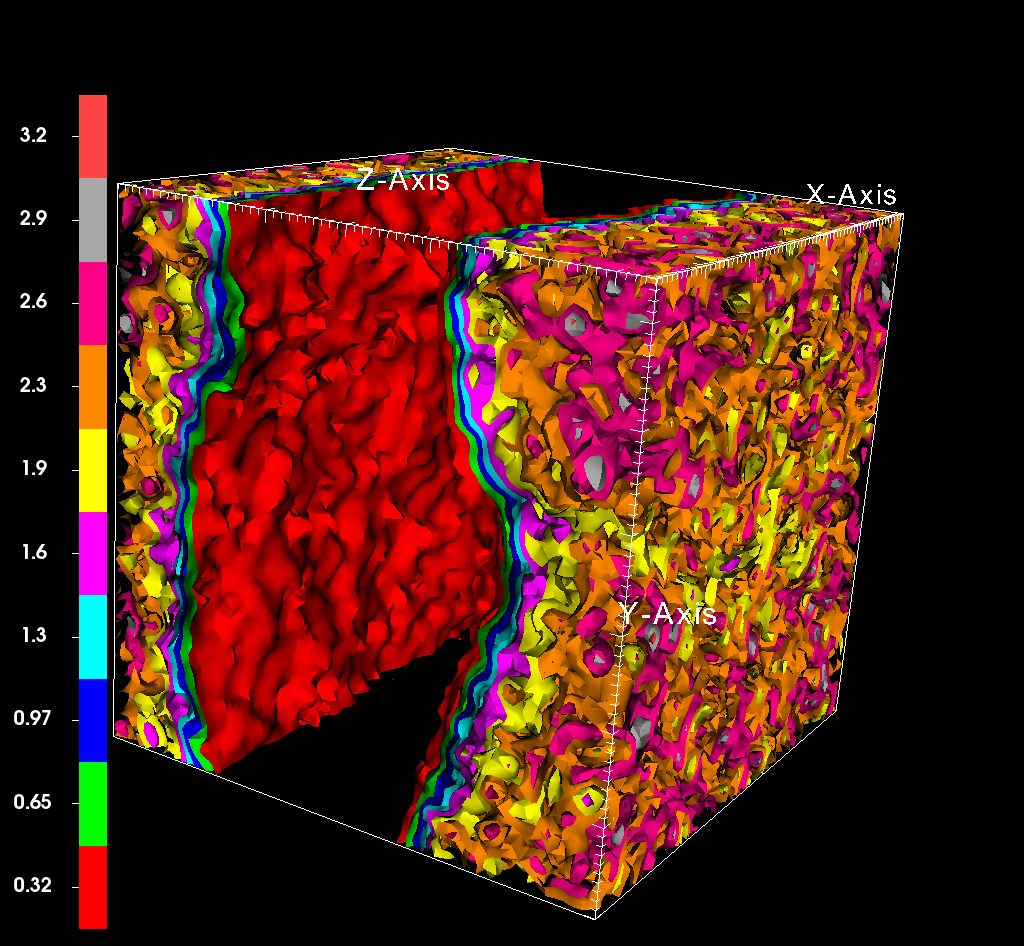}
		\includegraphics[width= 0.25\linewidth]{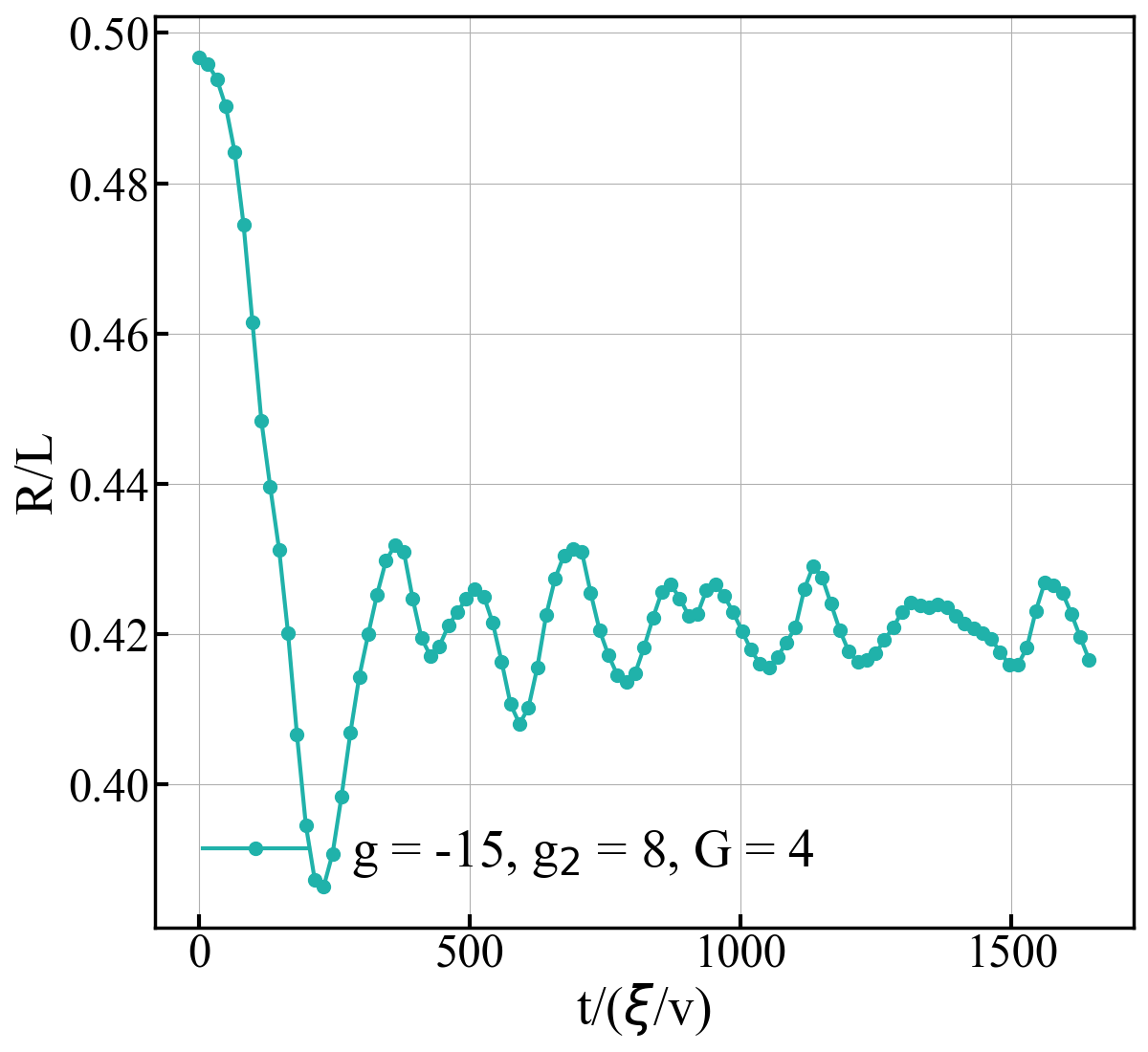} \put(-100,95){{\bf (e)}}
	\caption{Ten-level contour plots of $|\psi({\bf x},t)|^2$ from the cq-SGLPE~(\ref{eq:TSGLPE}) at $T = 0$, with initial and final states in columns 1 and 2, respectively, and $G=0$ [first row], $G=4$ [second row], $G=34$ [third row], and $G=98$ [fourth row] and the initial conditions given in the text. Column 3: $R/L$ versus the scaled time $t/(\xi/v)$ for (a) $G=0$, (b) $G=4$, (c) $G=34$, and (d) $G=98$. Row 5 shows the contour plots of $|\psi({\bf x},t)|^2$, for the same parameters as for row 2, but by using the cq-GPPE Eq.~(\ref{eq:TGPPE}). }	
	\label{fig:SGLPE_T0}
\end{figure*}

Our Gaussian-Ansatz study of $\mathcal{V}_{eff}$ suggests that we use the illustrative values $g=-15$ and $g_2=8$, when we go beyond this Ansatz to investigate the $G$-dependence of the axionic condensate by solving the cq-SGLPE~(\ref{eq:TSGLPE}). We first consider $T = 0$ and start with the initial condition {\bf IC1}, a nearly homogeneous distribution of $\psi$, and then we increase $G$ from zero to large values.  In Fig.~\ref{fig:SGLPE_T0}, we show ten-level contour plots of the spatial variation of $|\psi({\bf x},t)|^2$, at the initial time (column-1) and final time (column-2) for $G=0$ [first row of Fig.~\ref{fig:SGLPE_T0}], $G=4$ [second row of Fig.~\ref{fig:SGLPE_T0}], $G=34$ [third row of Fig.~\ref{fig:SGLPE_T0}], and $G=98$ [fourth row of Fig.~\ref{fig:SGLPE_T0}]. The intial conditions for the runs in rows 2, 3, and 4, are, respectively, the final configurations in rows 1, 2, and 3. In each row, the equilibrium configuration (column 2) for a given value of $G$, is a result of a balance between the  self-interactions and the repulsive quantum pressure. Note that, as $G$ increases, the equilibrium configuration goes from a tenuous axionic gas [row 1], to flat sheets [row 2] or \textit{Zeldovich pancakes}~\cite{1970A&A.....5...84Z}, cylindrical structures [row 3], and finally a spherical axionic condensate [row 4]. In column 3 of Fig.~\ref{fig:SGLPE_T0} we illustrate the evolution of the scaled radius of gyration $R/L$ as a function of the scaled time $t/(\xi/v)$ for (a) $G=0$, (b) $G=4$, (c) $G=34$, and (d) $G=98$. In row 5 we show ten-level contour plots of $|\psi({\bf x},t)|^2$, with the same parameters as for row 2, but which we obtain by using the cq-GPPE~(\ref{eq:TGPPE}).

\begin{figure}[!hbt]
\resizebox{\linewidth}{!}{
	\includegraphics[width=1\linewidth]{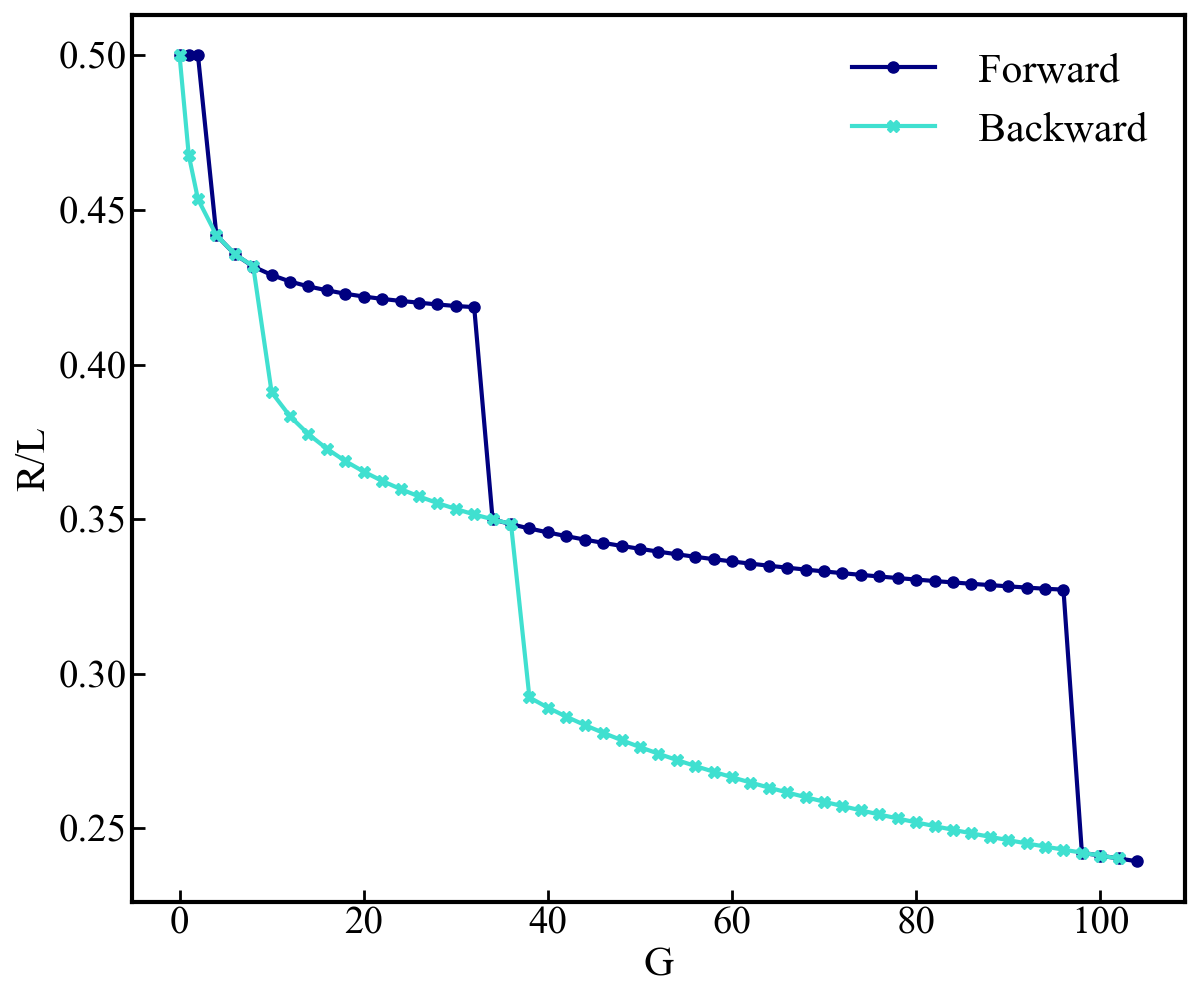}\put (-204,200) {\huge $\nearrow$}
	\put (-195,155) {\huge $\nearrow$}
	\put (-148,138) {\huge $\rightarrow$}
	\put (-33,88) {\huge $\uparrow$}
	\put(-100,180){(a)}
	\put(-182,210){\includegraphics[width= 0.2\linewidth]{figures/initia_G_0.png} }
	%\put(-173,168){(b)}
	\put(-173,158){\includegraphics[width= 0.2\linewidth]{figures/final_G_4.png} } 
	%\put(-128,162){(c)}
	\put(-128,105){\includegraphics[width= 0.2\linewidth]{figures/final_G_34.png} }
	%\put(-16,105){(d)}	
	\put(-55,105){\includegraphics[width= 0.2\linewidth]{figures/final_G_98.png} }
	%\includegraphics[width=1\linewidth]{figures/radius_vs_g2.png}
	%\put(-100,180){(b)}
	}
	\caption{Plot of the scaled radius of gyration $R/L$ versus the gravitational interaction parameter $G$ 
 for $g = -15$ and $g_2 = 8$ in the cq-GPPE; blue and green curves show, respectively, curves along which
 $G$ increases and decreases. } %The points $(a)$, $(b)$, $(c)$, and $(d)$ corresponds to the values of $G$ for which the density distribution $|\psi({\bf x},t)|^2$ is shown in Fig.\ref{fig:SGLPE_T0} for row1, row2, row3, and row4 respectively.}
	\label{fig:SGLPE_hys}
\end{figure}

To study the transitions between the $T=0$ equilibrium configurations shown in column 2, rows 1-4, in Fig.~\ref{fig:SGLPE_T0}, we increase (blue curve) and then decrease (green curve) $G$, as we show via the plot of $R/L$ versus $G$ in Fig.~\ref{fig:SGLPE_hys}, for the illustrative values  $g = -15$ and $g_2 = 8$ in the  cq-GPPE. As we increase $G$, there are three first-order transitions, first from a statistically homogeneous state to pancakes, then to a cylindrical configuration, and finally to a collapsed spherical object. At the transitions between these configurations, $R/L$ jumps discontinuously. Given that we change the values of $G$ at a finite rate, the metastability of these configurations makes the first-order jumps appear as hysteresis loops~\cite{rao1990magnetic}, in which the increasing-$G$ (blue curve) and decreasing-$G$ (green curve) scans yield different branches. It is interesting to speculate if this is an example of  the \textit{cosmological hysteresis} proposed in Ref.~\cite{PRD_cosmo_hysteresis}: ``a universe filled with scalar field exhibits cosmological hysteresis.''

%\begin{figure}
%	\includegraphics[width=0.9\linewidth]{figures/radius_vs_g2.png}
%	\caption{Plot of dimensionless radius of gyration $R/L$ vs the quintic nonlinearity parameter $g_2$ for $g = -10$ and $G = 50$. For the navy curve, the value of the parameter $g_2$ increases while it decreases for the cyan curve.} %The points $(a)$, $(b)$, $(c)$, and $(d)$ corresponds to the values of $G$ for which the density distribution $|\psi({\bf x},t)|^2$ is shown in Fig.\ref{fig:SGLPE_T0} for row1, row2, row3, and row4 respectively.}
%	\label{fig:SGLPE_g2_hys}
%\end{figure}

\subsection{Formation of axionic objects: $T > 0$}
\label{subsec:NonzeroT}

\begin{figure*}
	\resizebox{\linewidth}{!}{
		\includegraphics[width=0.25\linewidth]{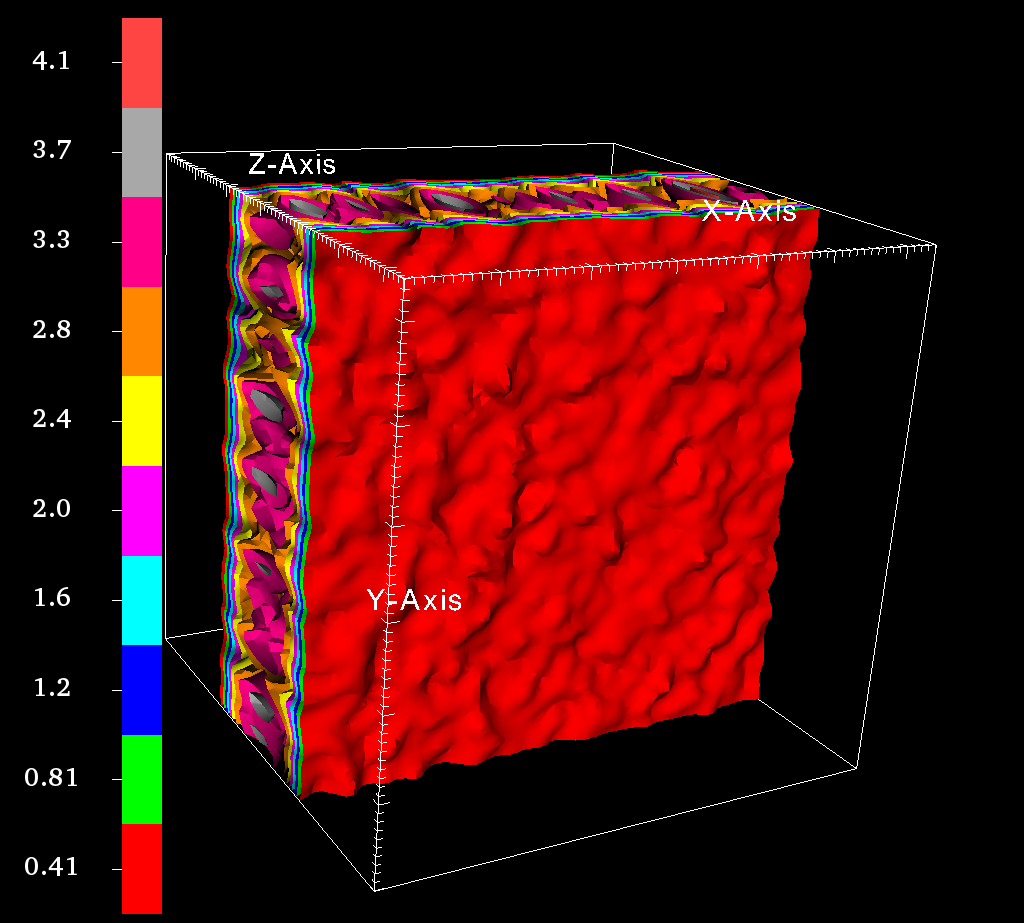}
		\put (-105,102){{\bf \color{white} (A)}}
		\includegraphics[width=0.25\linewidth]{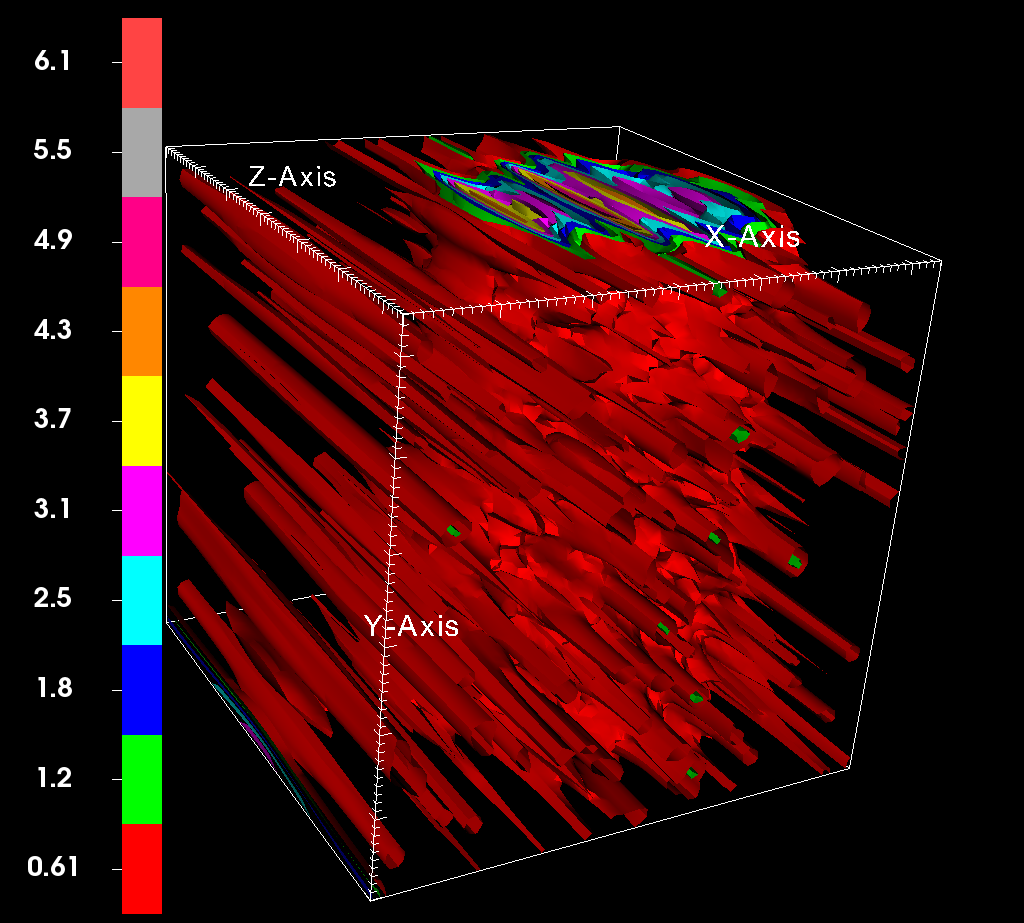} 
		\put (-105,102){{\bf \color{white}(B)}}
		\includegraphics[width=0.25\linewidth]{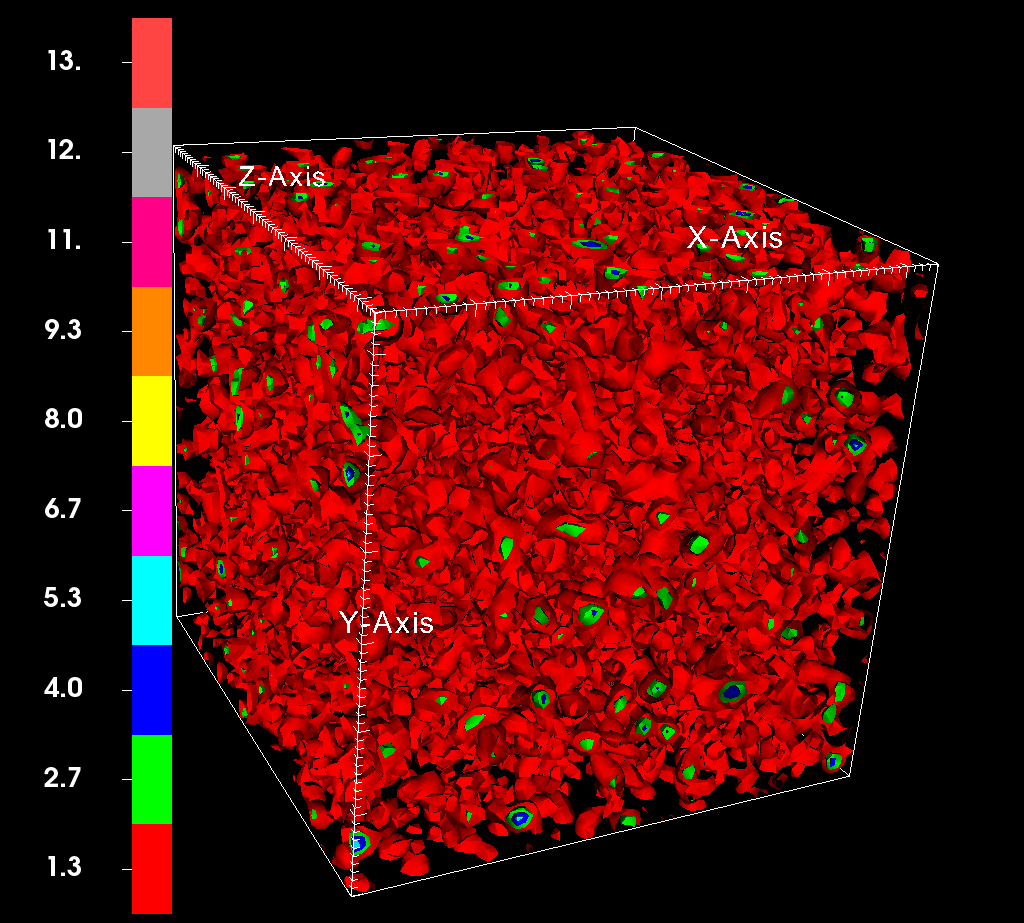}
		\put (-105,102){{\bf \color{white}(C)}}
		\includegraphics[width=0.25\linewidth]{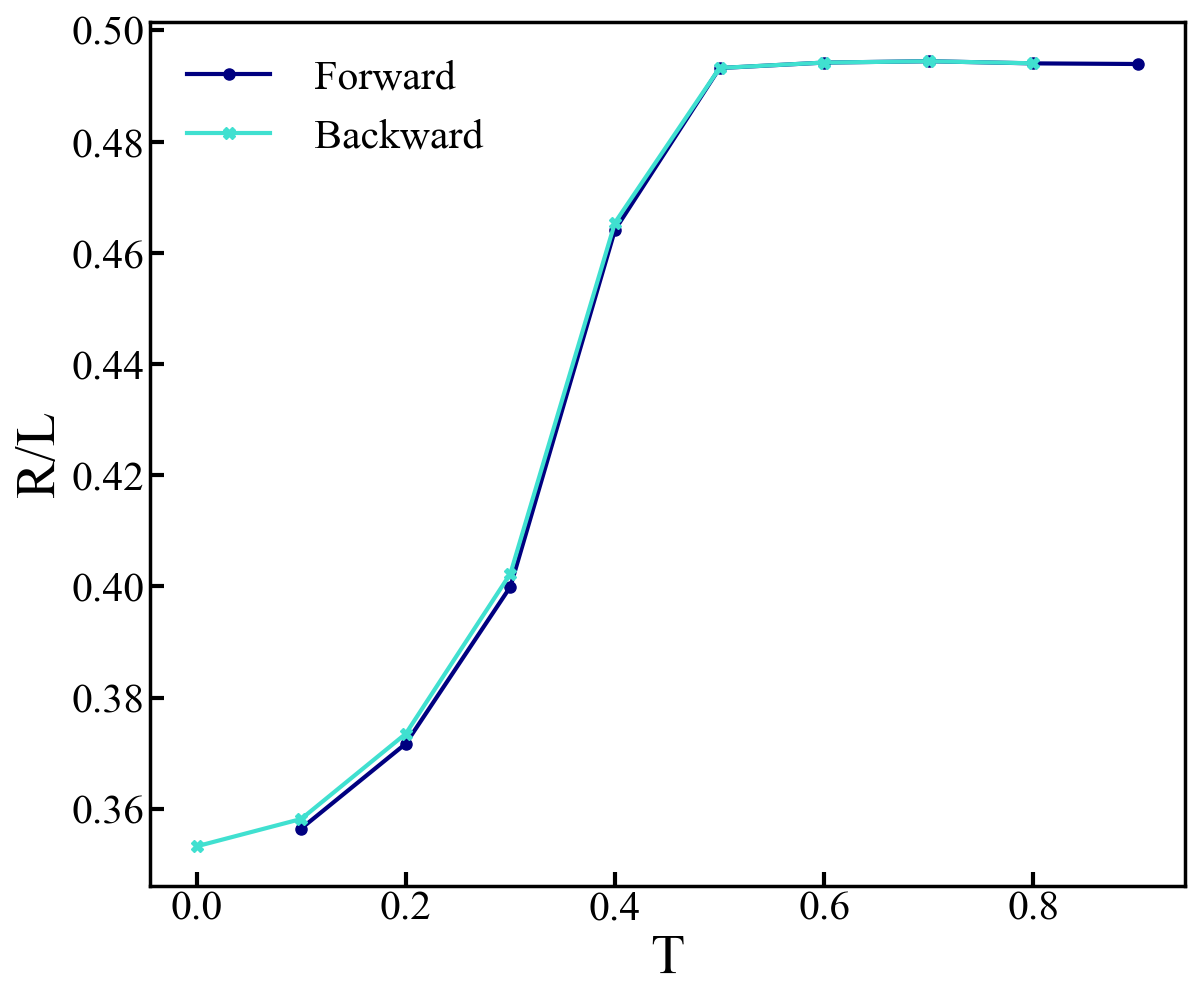}
		\put (-115,108){{\bf (D)}}
	\put (-100,17){ $ {\bf o} (A)$}
	\put (-79,40){ $ {\bf o} (B)$}
	\put (-25,97){ $ {\bf o} (C)$}}
	
	\resizebox{\linewidth}{!}{
		\includegraphics[width=0.25\linewidth]{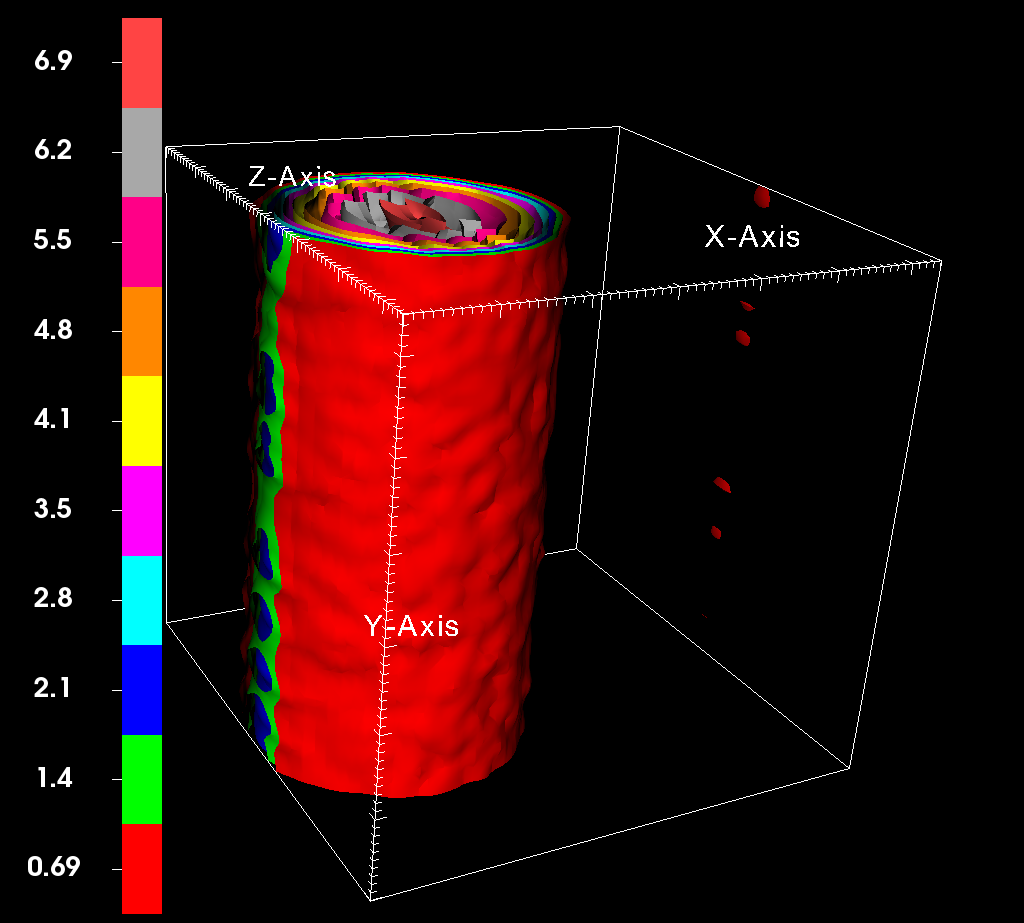}
        \put (-105,102){{\bf \color{white} (E)}}
		\includegraphics[width=0.25\linewidth]{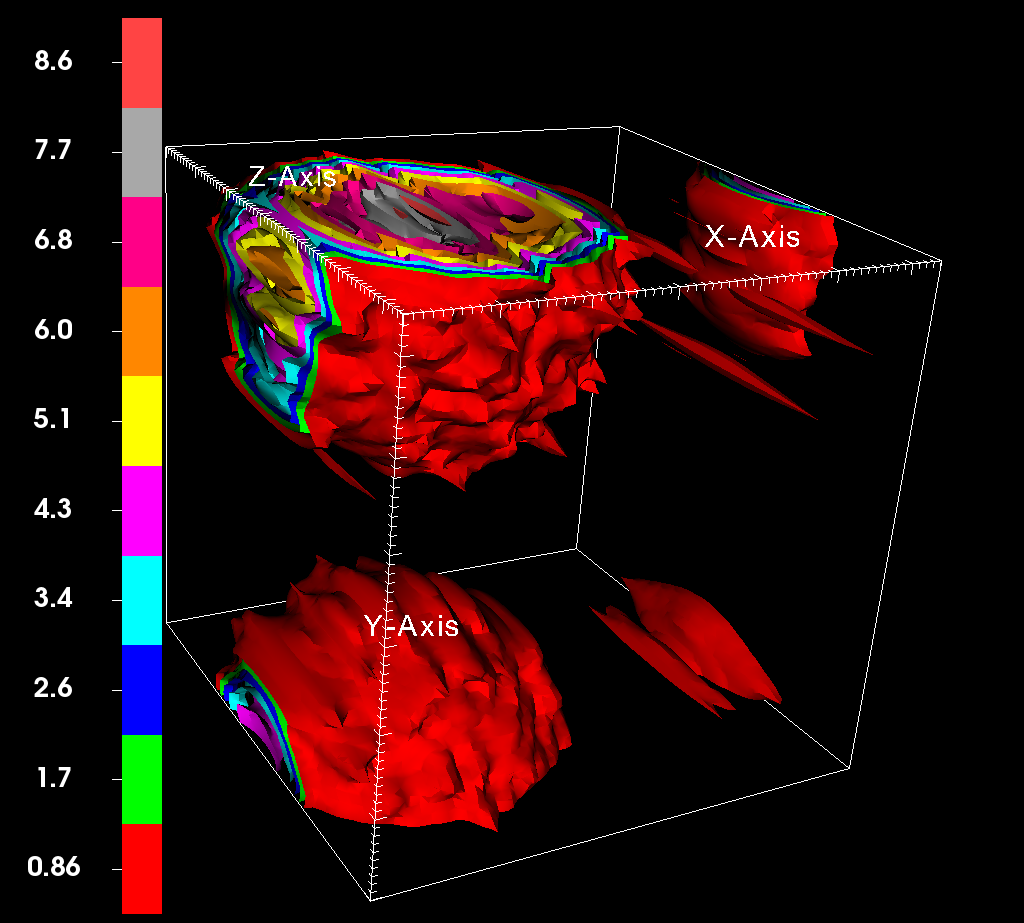}
        \put (-105,102){{\bf \color{white} (F)}}
		\includegraphics[width=0.25\linewidth]{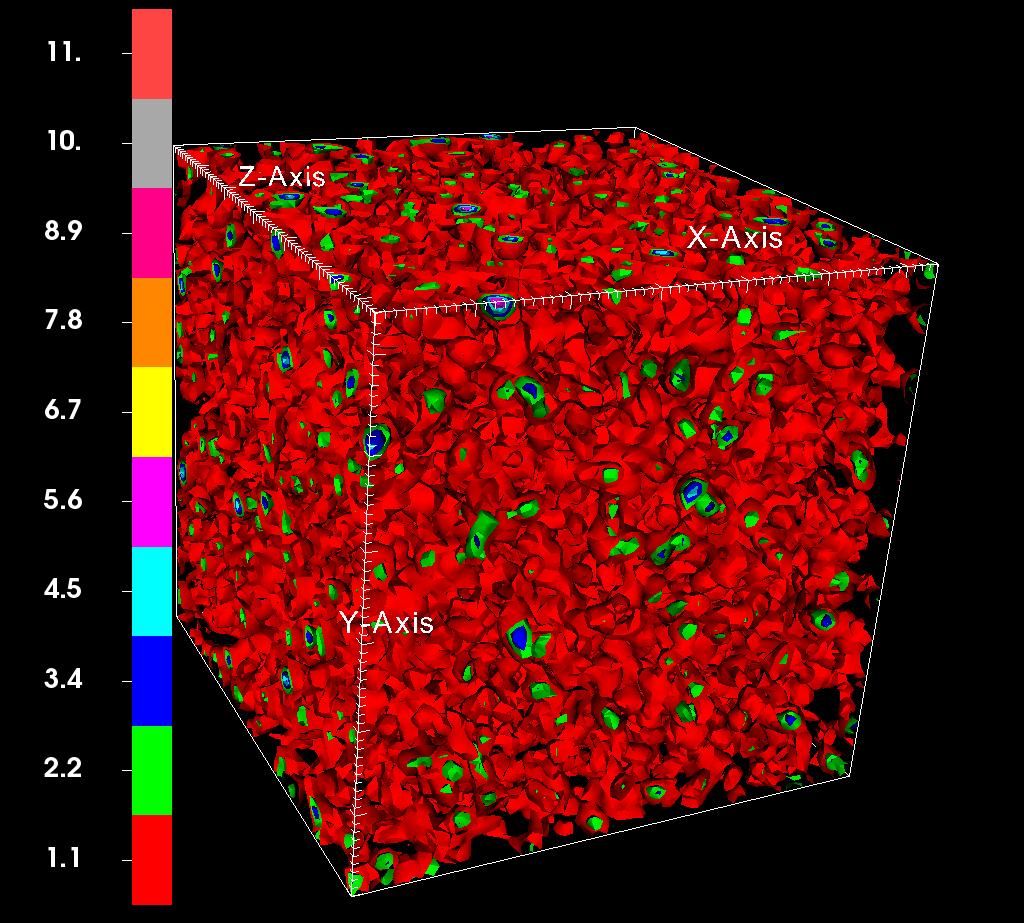}
        \put (-105,102){{\bf \color{white} (G)}}
		\includegraphics[width=0.25\linewidth]{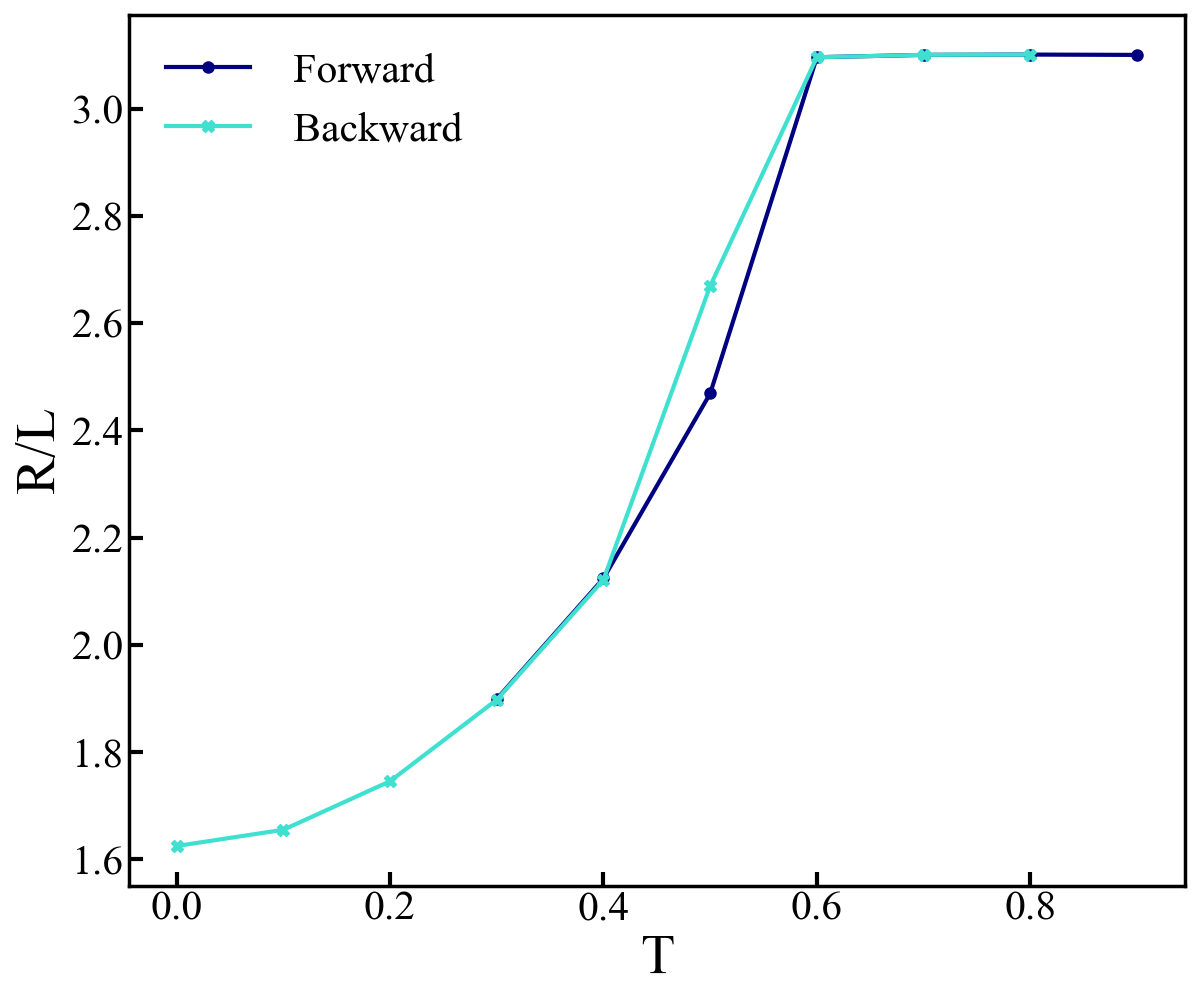}
        \put (-115,108){{\bf (H)}}
	\put (-100,16){ $ {\bf o} (E)$}
	\put (-79,28){ $ {\bf o} (F)$}
	\put (-25,98){ $ {\bf o} (G)$}}
	
	\resizebox{\linewidth}{!}{
		\includegraphics[width=0.25\linewidth]{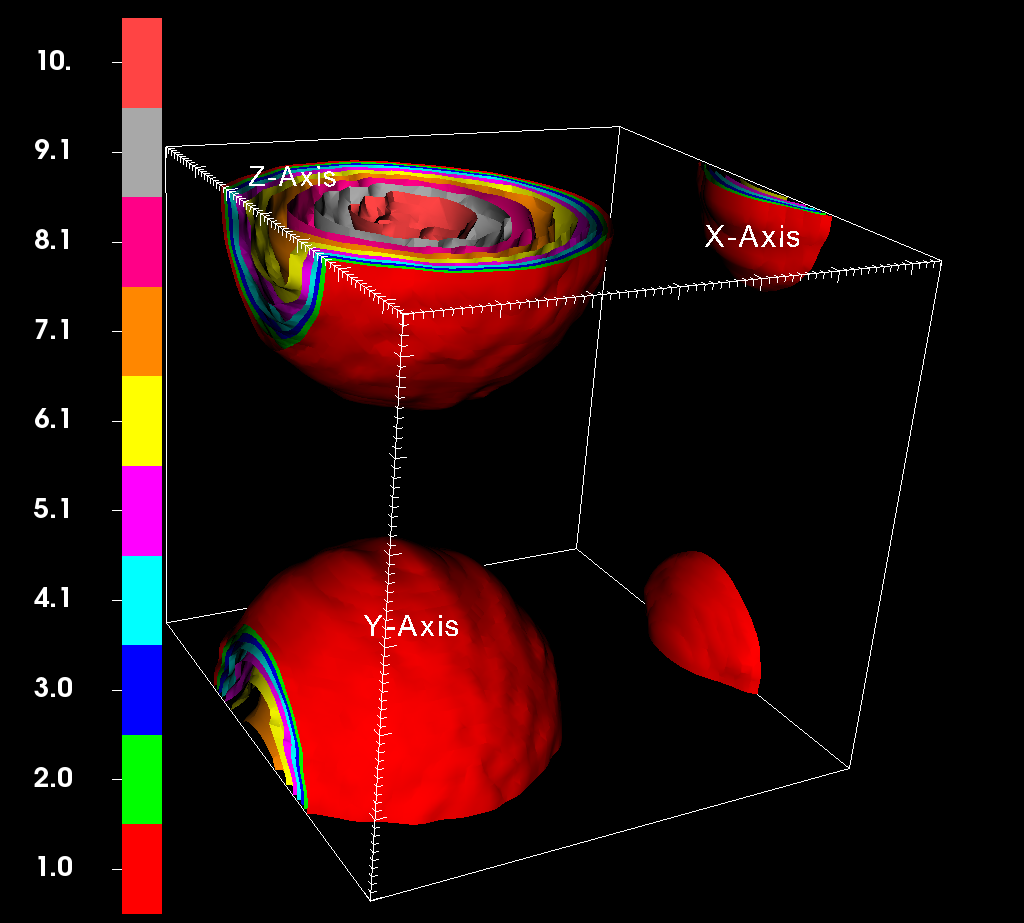} 
       \put (-105,102){{\bf \color{white} (I)}}
		\includegraphics[width=0.25\linewidth]{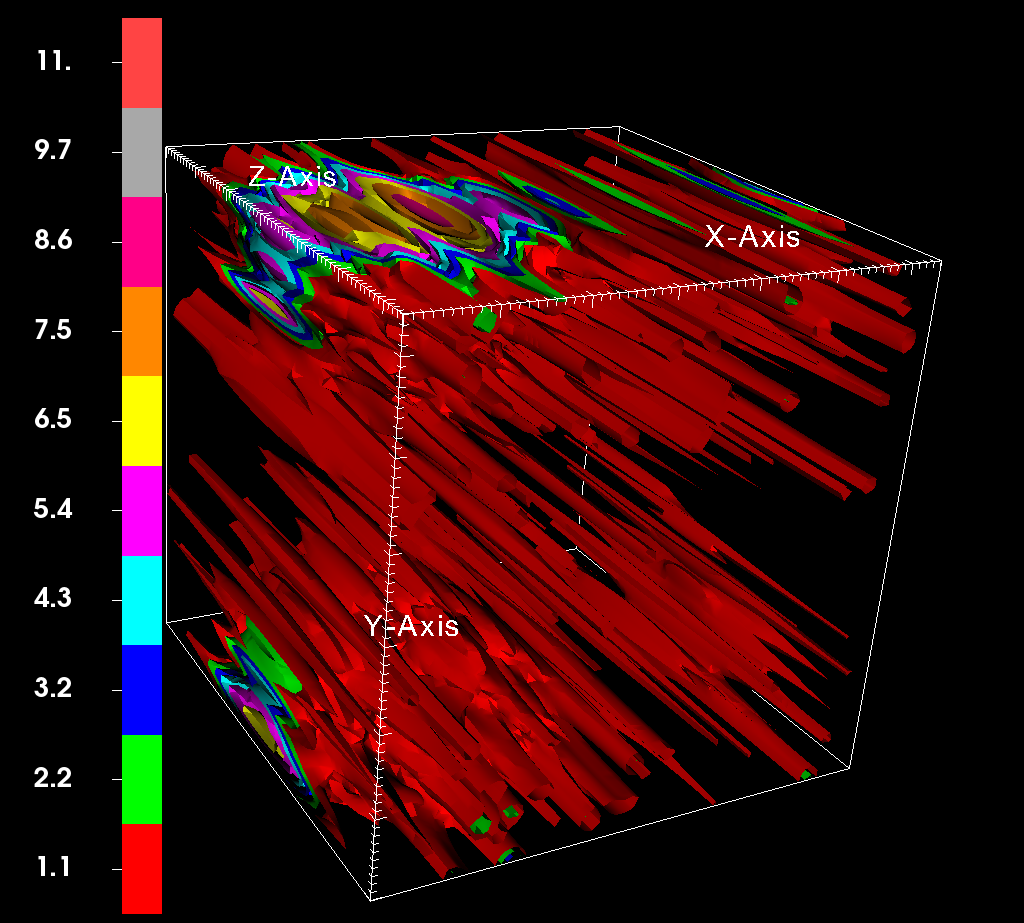}
       \put (-105,102){{\bf \color{white} (J)}}
		\includegraphics[width=0.25\linewidth]{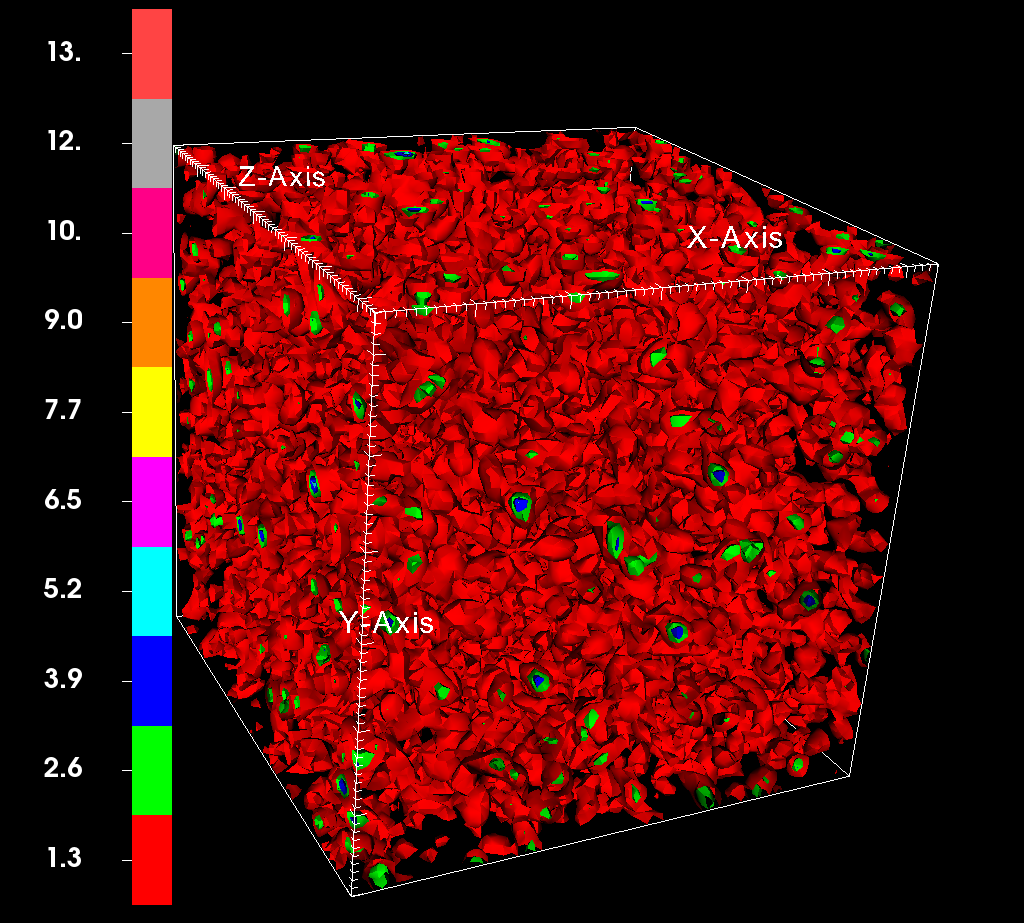}
       \put (-105,102){{\bf \color{white} (K)}}
       \includegraphics[width=0.25\linewidth]{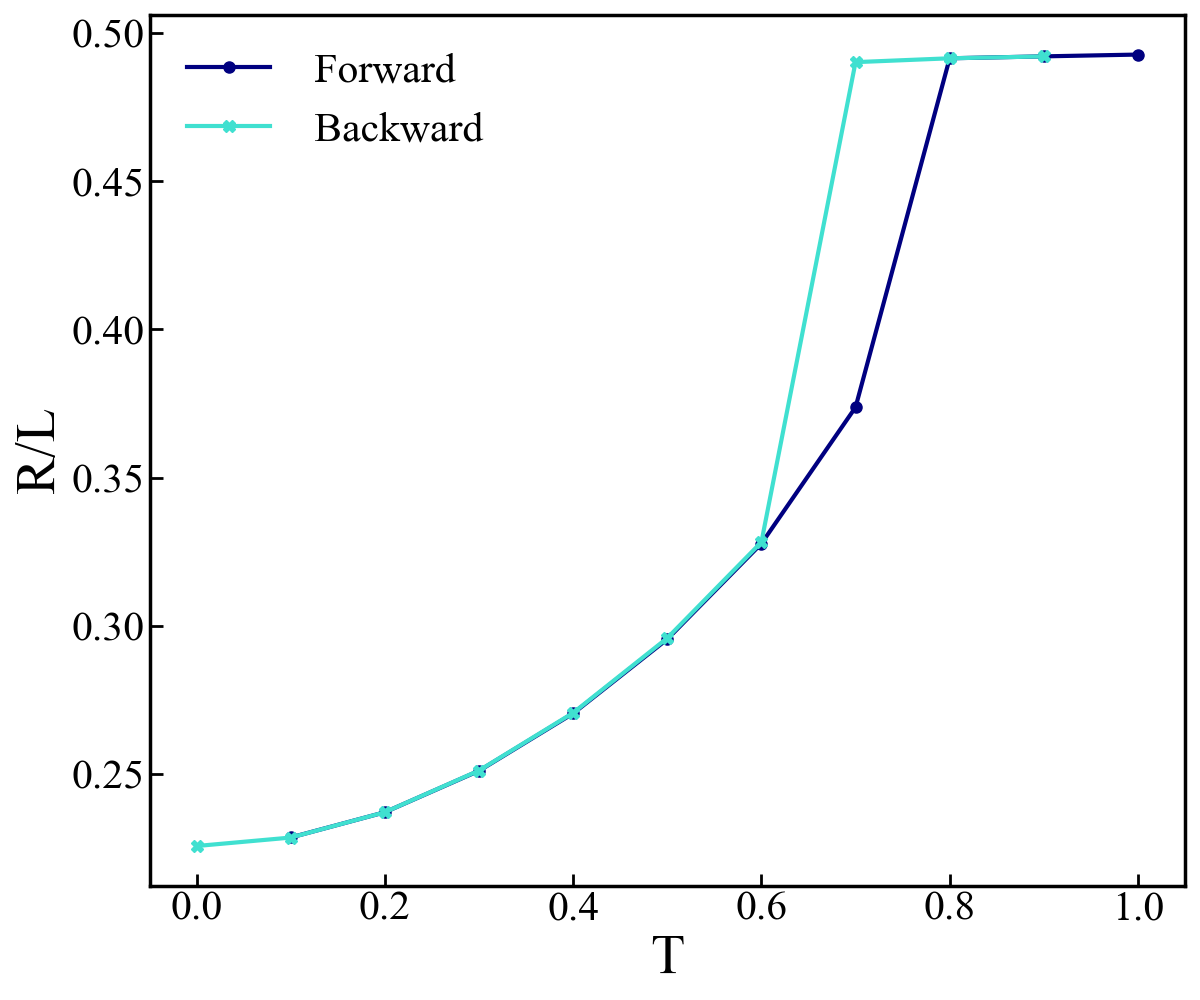}
     \put (-115,108){{\bf (L)}}
	\put (-102,15){ $ {\bf o} (I)$}
	\put (-42,60){ $ {\bf o} (J)$}
	\put (-22,98){ $ {\bf o} (K)$}}
	\caption{Columns 1-3: Ten level contour plots of $|\psi({\bf x},t)|^2$ at different temperatures as we heat last snapshots of Fig.\ref{fig:SGLPE_T0}. The three rows are arranged with increasing value of $G$; for $G = 30$ in row-1, $G = 70$ in row-2, and $G = 104$ in row-3. Column-4 shows the scaled radius of gyration, $R/L$ vs the temperature $T$ .
	%{\bf (x)} for pancakes of row-2 in Fig.\ref{fig:SGLPE_T0} and $g = -10, g2 = -6, G = 30 $, {\bf (H)} for cylinders of row-3 in Fig.\ref{fig:SGLPE_T0} and $g = -10, g2 = -6, G = 70 $, and {\bf (L)} for spheres of row-4 in Fig.\ref{fig:SGLPE_T0} and $g = -10, g2 = -6, G = 104 $. 
	The black curve is for the heating part of the cycle and the green one is for the cooling part. The points $(A)$, $(B)$, and $(C)$ in Fig.$(D)$ corresponds to density distributions in columns 1-3.}
	\label{fig:SGLPE_T}
\end{figure*}

We now study finite-temperature ($T>0$) effects, on the various structures obtained in Fig.~\ref{fig:SGLPE_T0}, by using the Fourier-truncated cq-GPPE~(\ref{eq:TGPPE}); we also construct the thermalized state directly by using the cq-SGLPE~(\ref{eq:TSGLPE}). 
Columns 1, 2, and 3 of Fig.~\ref{fig:SGLPE_T} show ten-level contour plots of $|\psi({\bf x},t)|^2$ at different representative temperatures, which increase from column 1 to 3. At $T=0$ we have pancake, cylindrical, and spherical structures at $G = 30,\,G = 70,$ and $G = 104$ in rows 1, 2, and 3, respectively. As we increase $T$ and move from column 1 to 3, we see that, in all the rows, each one 
of the condensed structures becomes a disordered tenuous axionic assembly. In column 4 of Fig.~\ref{fig:SGLPE_T0} we show plots of  the scaled radius of gyration $R/L$ versus the temperature $T$. We start with the density distributions illustrated in column 1 of Fig.~\ref{fig:SGLPE_T}. We then increase $T$, for each one of these initial conditions; we follow this by a cooling cycle until the system returns to the initial temperature. These heating and cooling cycles yield the hysteresis loops that we show in column 4 of Fig.~\ref{fig:SGLPE_T} (with black and blue lines for heating and cooling cycles, respectively). Although the loops are clearly visible, they are not as pronounced as their counterparts in Fig.~\ref{fig:SGLPE_hys}. 

\subsection{Rotational dynamics of a single axionic condensate }
\label{subsec:Rotation}

Quantized vortices appear when we rotate a superfluid with a sufficiently large angular speed $\Omega$. To obtain such quantized vortices in our self-gravitating system of axions, we solve the cq-SGLPE~(\ref{eq_SGLPE_rot}). First we use initial condition {\bf IC1} in Eq.~(\ref{eq_SGLPE_rot}), with $\Omega=0$; for the chosen set of parameters this yields a spherical collapsed object. We now use this collapsed object as the initial condition for Eq.~(\ref{eq_SGLPE_rot}) and slowly increase the angular speed $\Omega$. We use the final steady-state value for the field $\psi$, for a given value of $\Omega$, as the initial condition for the next value of $\Omega$.

\begin{figure*}
	\resizebox{\linewidth}{!}{
	 \begin{tikzpicture}
	 \node[anchor=south west,inner sep=0] at (0,0)
	{\includegraphics[width=0.5\linewidth]{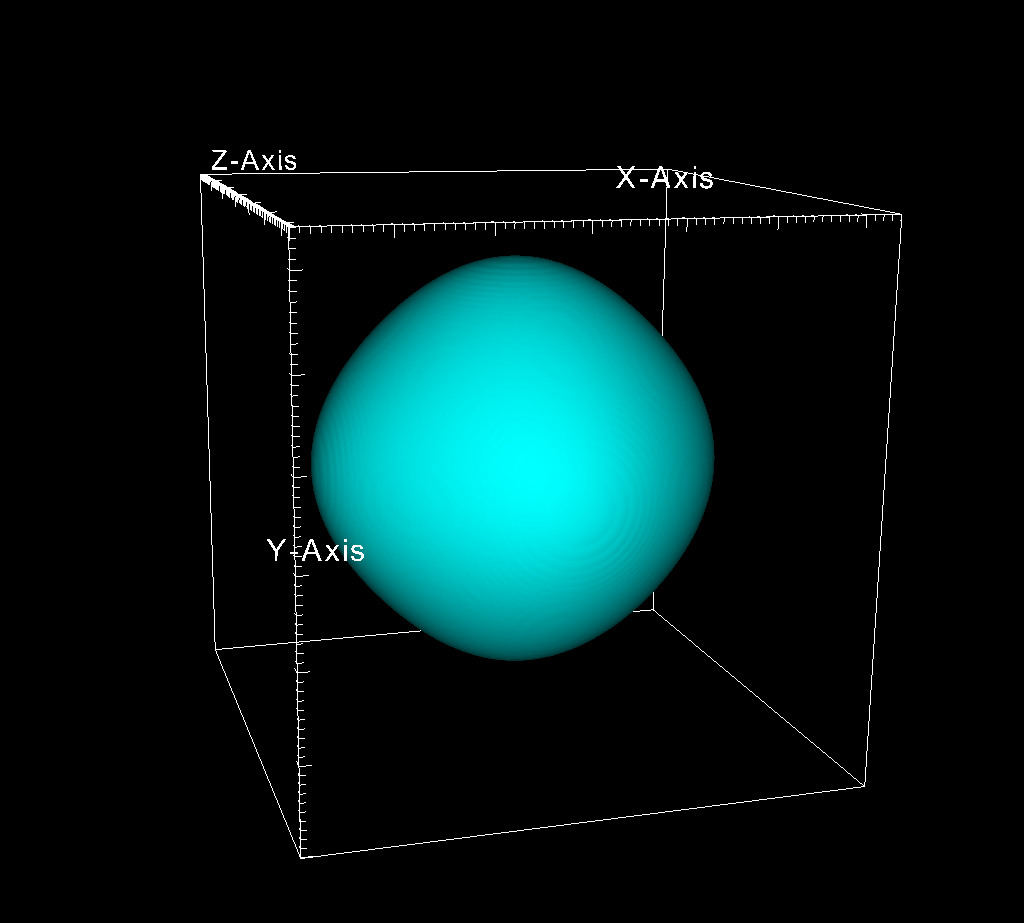}
		\put(-250,200){\color{white} \huge \bf  (a): $\Omega=3$}
	\includegraphics[width=0.5\linewidth]{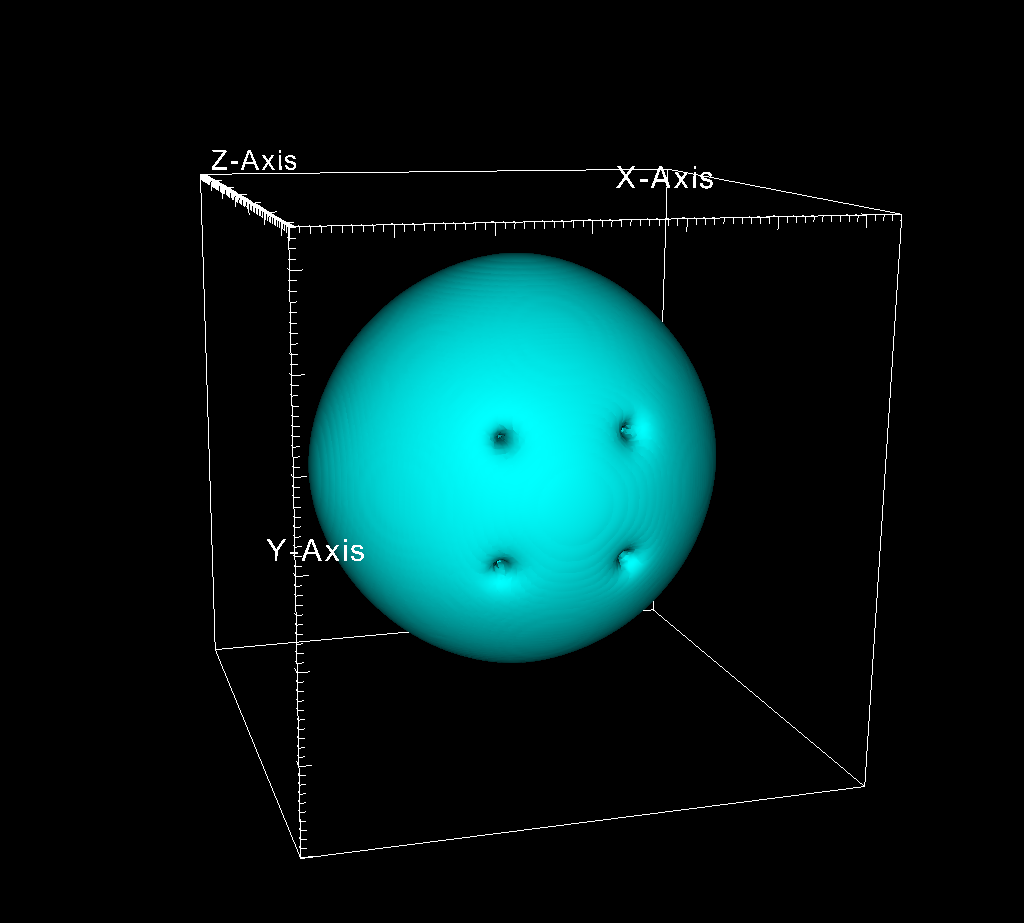}
    	\put(-250,200){\color{white} \huge \bf  (b): $\Omega=4$}
	\includegraphics[width=0.5\linewidth]{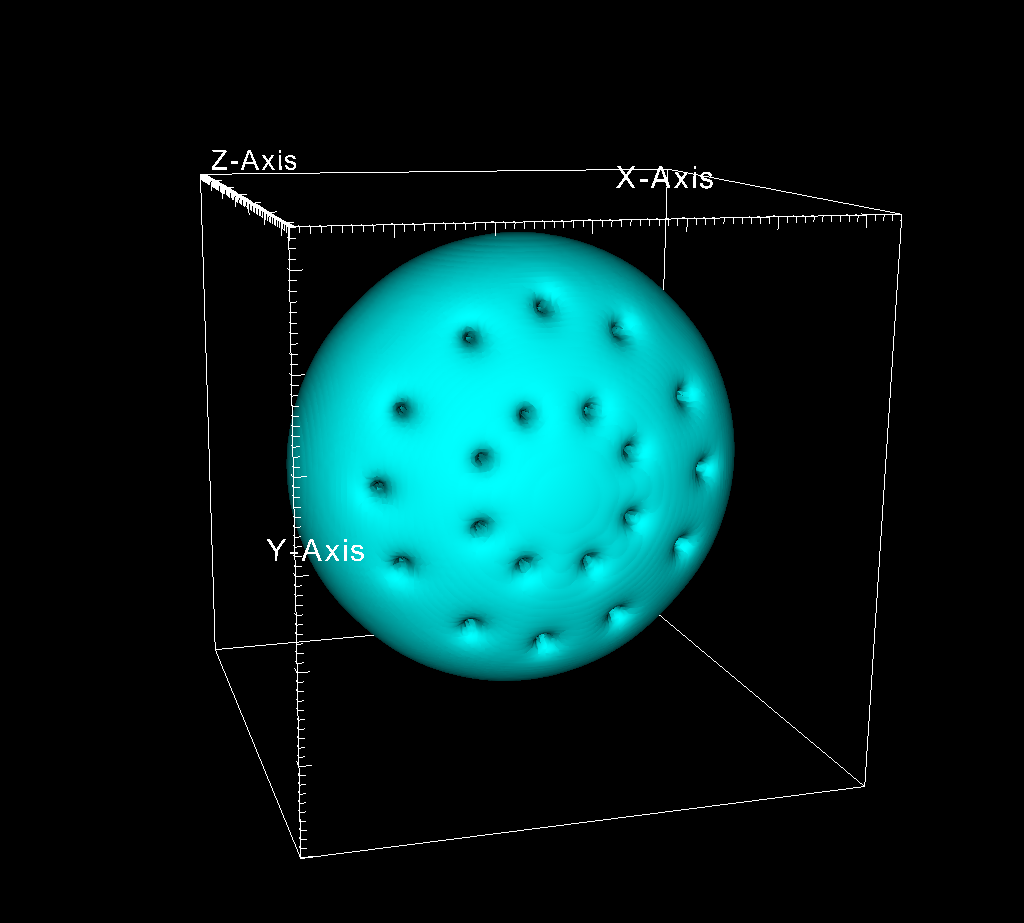}
	    \put(-250,200){\color{white} \huge \bf  (c): $\Omega=5$}
	\includegraphics[width=0.5\linewidth]{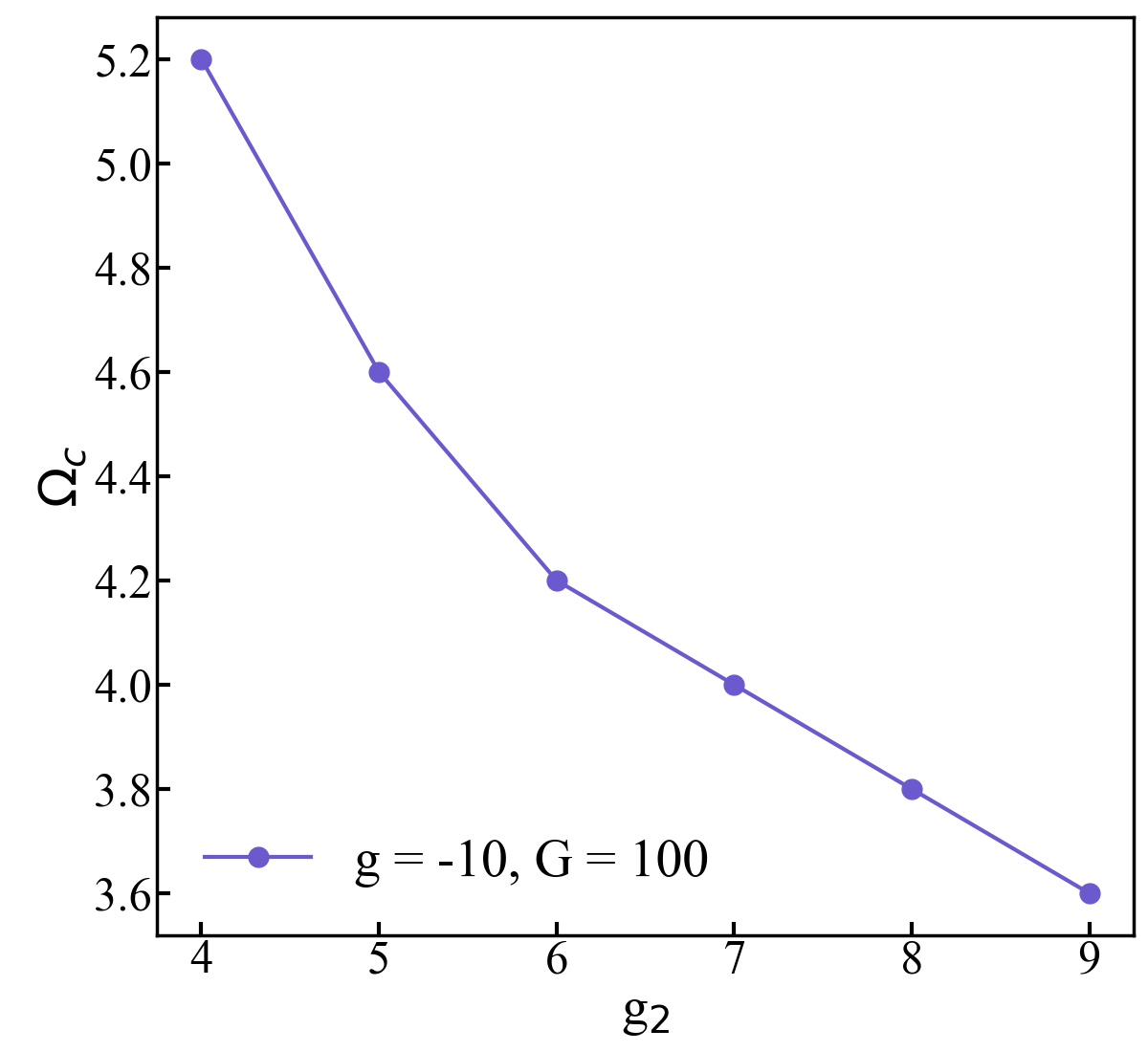}
	  \put(-200,200){ \huge \bf  (d)}};
    \draw[line width=1mm,green, ->] (4.2,4.2) -- (5,3) ;
    \draw [line width=1mm,yellow,->] (5.4,3.3) arc (0:-300:0.5);
    
    \draw[line width=1mm,green, ->] (13.5,4.2) -- (14.5,3) ;
    \draw [line width=1mm,yellow,->] (14.8,3.3) arc (0:-300:0.5);
    
    \draw[line width=1mm,green, ->] (22.8,4.2) -- (23.8,3) ;
    \draw [line width=1mm,yellow,->] (24,3.3) arc (0:-300:0.5);
	\end{tikzpicture}
}
	\caption{Contour plots of $|\psi({\bf x},t)|^2$, for a single rotating compact axionic object, which we obtain by solving the cq-SGLPE for $g = -15$, $g_2=8$, $G = 100$, and (a) $\Omega = 3$, (b) $\Omega = 4$, and (c) $\Omega = 5$. Vortices appear once $\Omega > \Omega_c$, a critical angular speed. The axis of rotation is indicated by the green arrow  (the $Z$-axis about which the axion condensate is rotated). (d) Plot of $\Omega_c$ versus $g_2$.} 
	\label{fig:SGLPE_rotation}
\end{figure*}

In Figs.~\ref{fig:SGLPE_rotation} (a)-(c), we present contour plots of $|\psi({\bf x},t)|^2$, for a single rotating compact axionic object, with the $(z)$ axis of rotation indicated by a green arrow. We obtain this configuration by solving the cq-SGLPE for $g = -15$, $g_2=8$, $G = 100$, and (a) $\Omega = 3$, (b) $\Omega = 4$, and (c) $\Omega = 5$. Note that vortices thread the collapsed object once $\Omega > \Omega_c \gtrsim 3$, where $\Omega_c$ is the critical angular speed required for the appearance of vortices. The number density of vortices increases as we increase $\Omega$ [cf. Figs.~\ref{fig:SGLPE_rotation} (b) and (c)]. Furthermore, we show 
in Fig.~\ref{fig:SGLPE_rotation} (d) that $\Omega_c$ decreases as $g_2$ increases. Our result is akin to that of Ref.~\cite{PRA_omega_c_vs_g}, for a trapped BEC without the $G$ and $g_2$ terms, where it is found that the critical angular speed decreases as the repulsive interaction $g$ between the bosons increases. 
%In our case of self-gravitating system of axions, parameter $g$ is positive and without a quintic term $g_2$, no vortices are possible. So here we study the dependence of the critical angular speed on the parameter $g_2$. We have shown critical angular speed $\Omega_c$ vs  parameter $g_2$ in Fig.\ref{fig:SGLPE_rotation}(d) and found it to decrease as we increase the parameter $g_2$. 
For our self-gravitating axionic system, with $g <0$ and $G$ held fixed, Fig.~\ref{fig:SGLPE_rotation} (d) suggests that, as $g_2\rightarrow 0$, the critical angular speed $\Omega_c$ becomes so  high that the system cannot support vortices.

\subsection{Rotational dynamics of binary axionic systems}
\label{subsec:Rotbinary}

We now investigate the dynamics of a rotating binary axionic system by using the following initial condition that consists of two spherical collapsed axionic objects, separated by a distance $d$, along the $y$-direction, and with equal and opposite initial velocities, ${\bf v_1} = (v,0,0)$, ${\bf v_2} = (-v,0,0)$, in the $x$-direction:
\begin{eqnarray}
\psi_b({\bf x},t=0) &=& f(|{\bf x - x_0}|)e^{i\bf v_1\cdot x}\nonumber \\ 
&+& f(|{\bf x + x_0}|)e^{i{\bf (v_2\cdot x} + \Delta \phi)}\,, 
\label{eq:binary_ini}
\end{eqnarray}
where ${\bf x_0} = (0,d/2,0)$ and $\Delta \phi$ is the relative phase between the two objects. We obtain the function $f(|\bf x|)$ by using initial condition {\bf IC2} and then the Newton method [cf. the discussion around Eq.~(\ref{eq:relaxation})], which converges rapidly to the stationary state of the cq-GPPE~(\ref{eq:GPPE}). We explore the dependence of the dynamics of such axionic binaries in the following two illustrative parameter regimes, PI and PII.

\subsubsection*{\bf{Parameters PI}}

%$ N_1 = N_2,  g=-0.5, g_2=0.001,  and G=2.0 $

\begin{figure*}[!hbt]
	\resizebox{\linewidth}{!}{
		\includegraphics[width=0.5\linewidth]{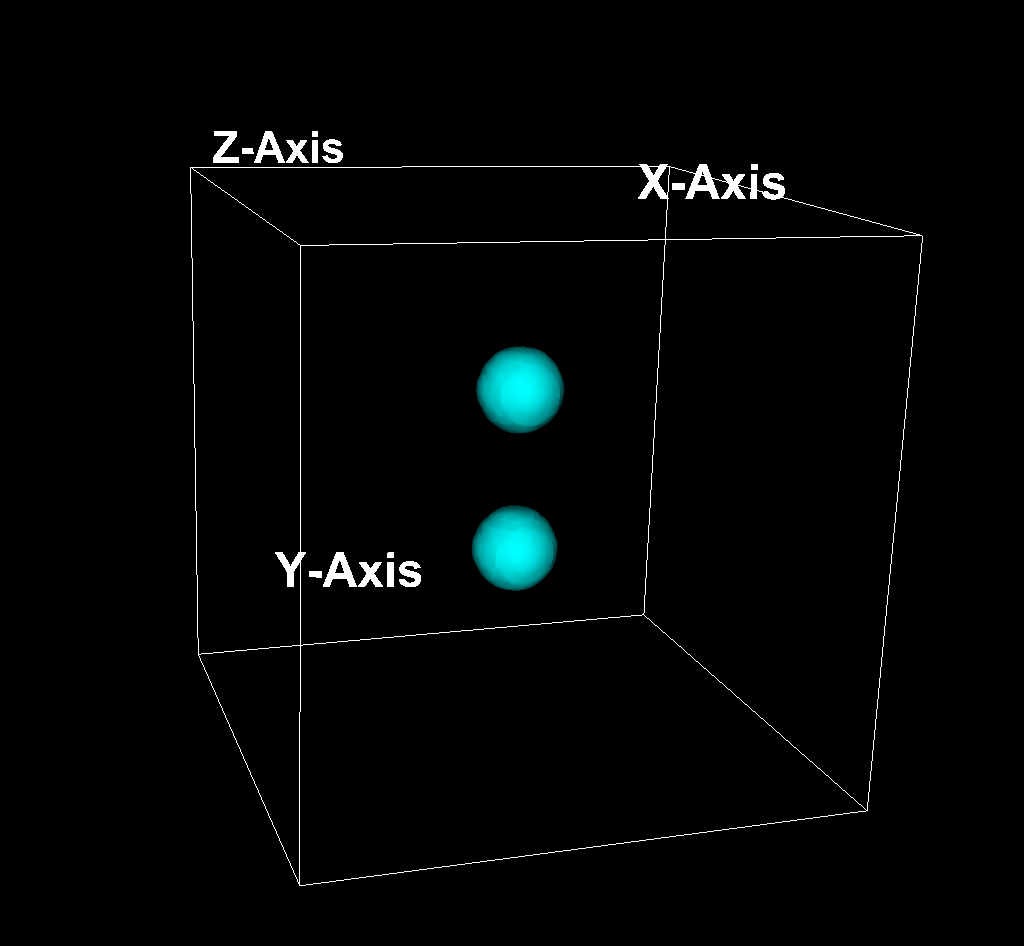}
		\put(-360,100){\huge $\Delta \phi=0$}
		\put(-290,100){\bf \huge  $\rightarrow$}
		\put(-150,210){\color{white}\huge t=0\;s}
		\includegraphics[width=0.5\linewidth]{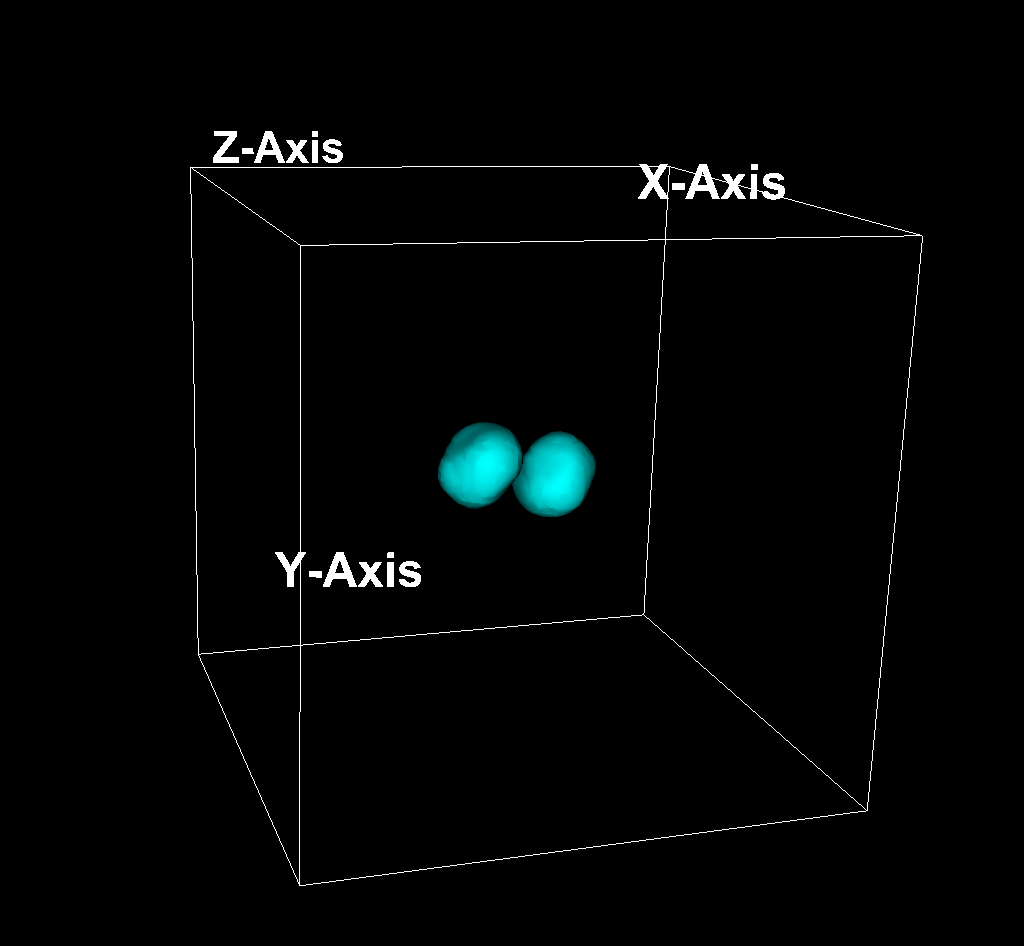}
		\put(-150,210){\color{white} \huge t=2\;s}
		\includegraphics[width=0.5\linewidth]{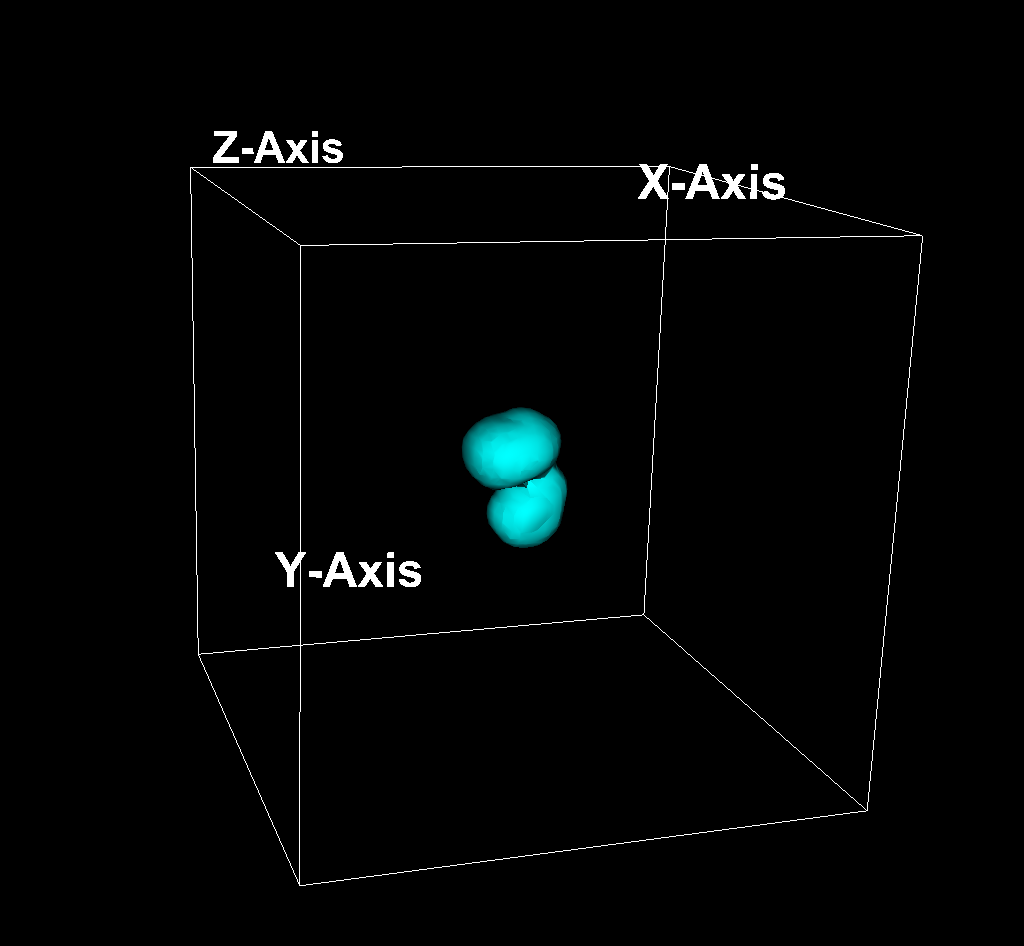}
		\put(-150,210){\color{white} \huge t=4\;s}
		%\includegraphics[width=0.5\linewidth]{figures/binary_star_0.5_0.001_2_0.5_in0006.png}
		%\put(-150,210){\color{white} \huge t=6\;s}
		\includegraphics[width=0.5\linewidth]{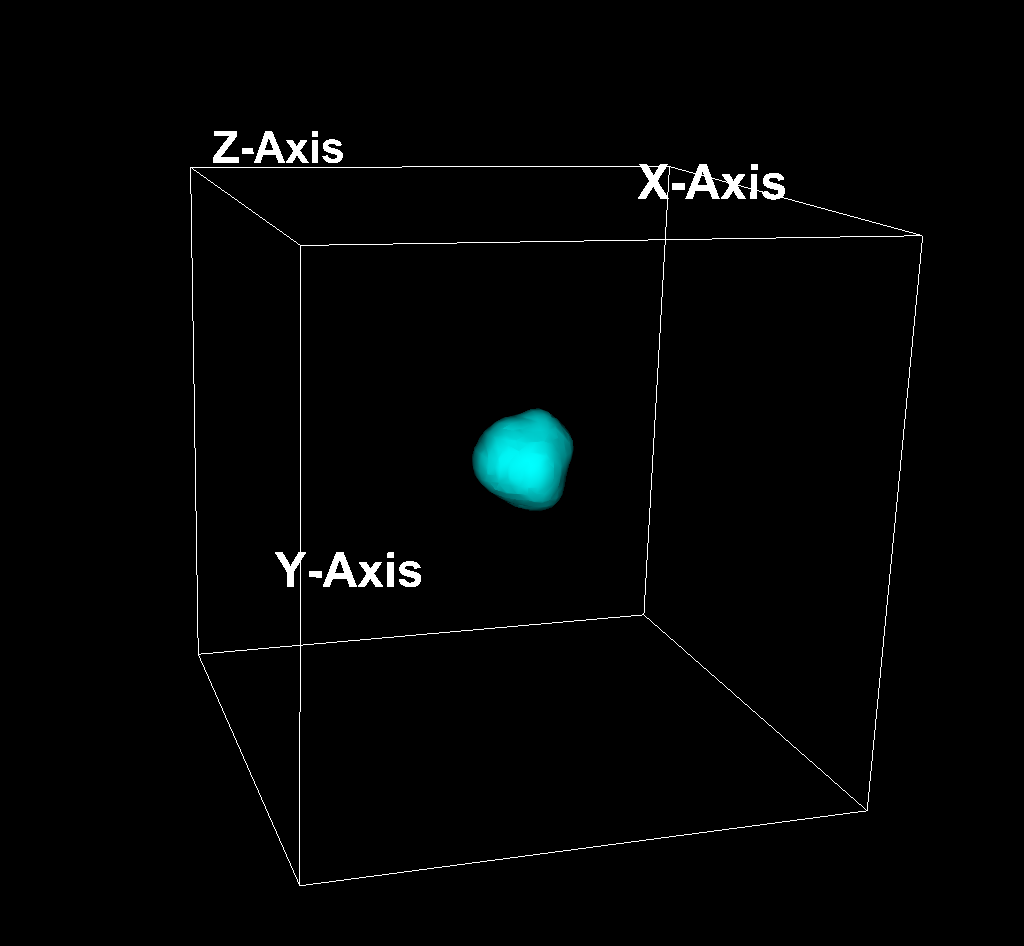}
		\put(-150,210){\color{white}\huge t=10\;s}
		\includegraphics[width=0.55\linewidth]{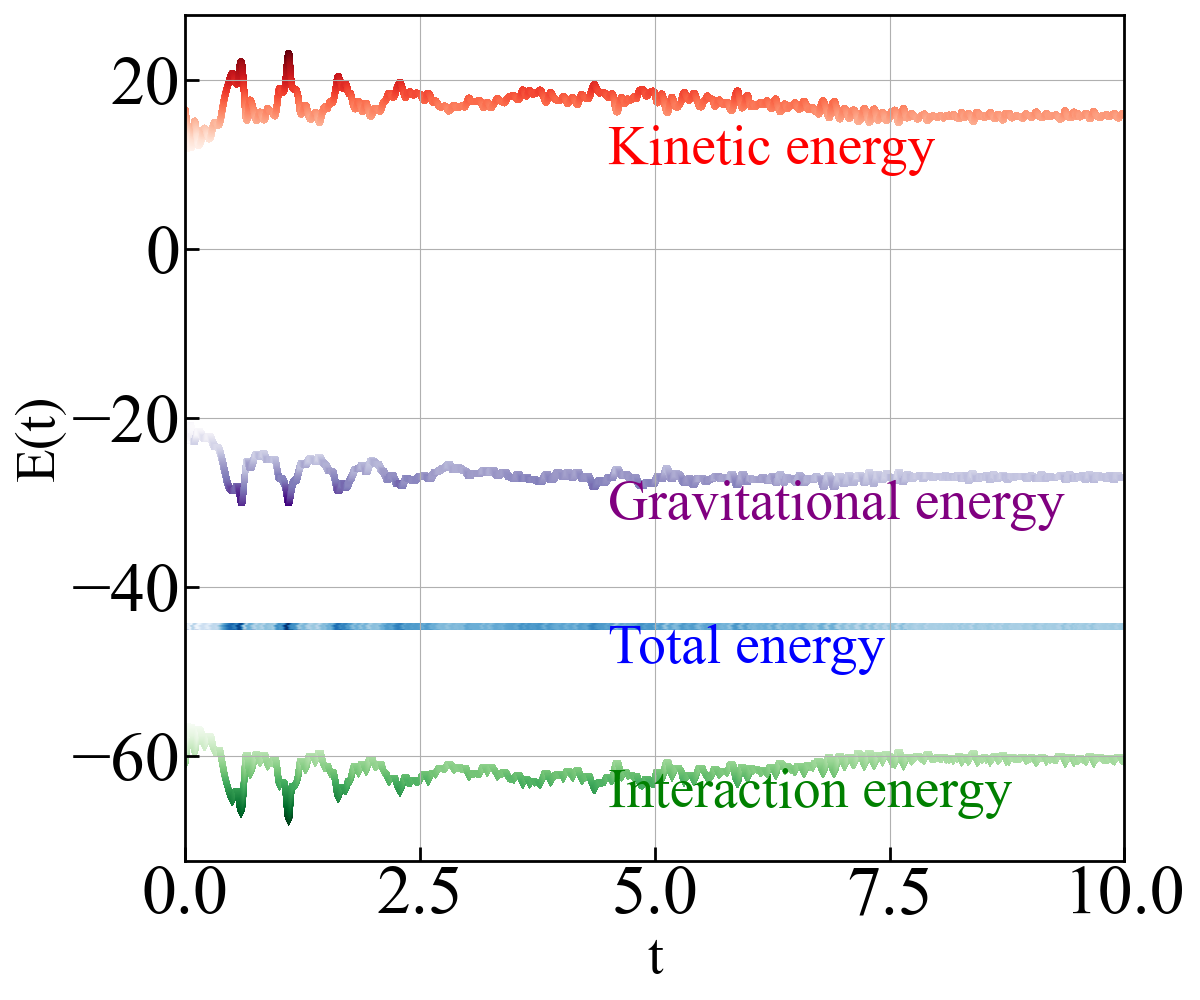}
	}
    
    \resizebox{\linewidth}{!}{
    	\includegraphics[width=0.5\linewidth]{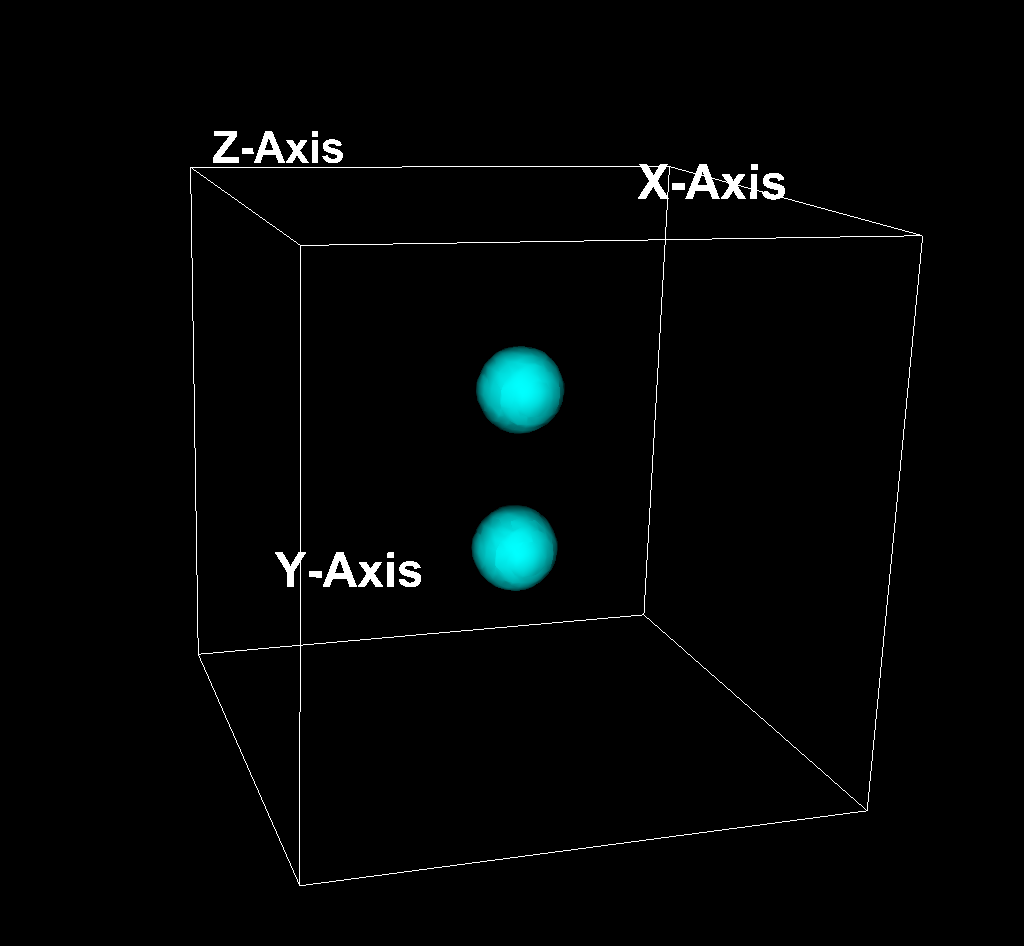}
    	\put(-360,100){\huge $\Delta \phi=\pi$}
    	\put(-290,100){\bf \huge  $\rightarrow$}
    	\put(-150,210){\color{white} \huge t=0\;s}
    	\includegraphics[width=0.5\linewidth]{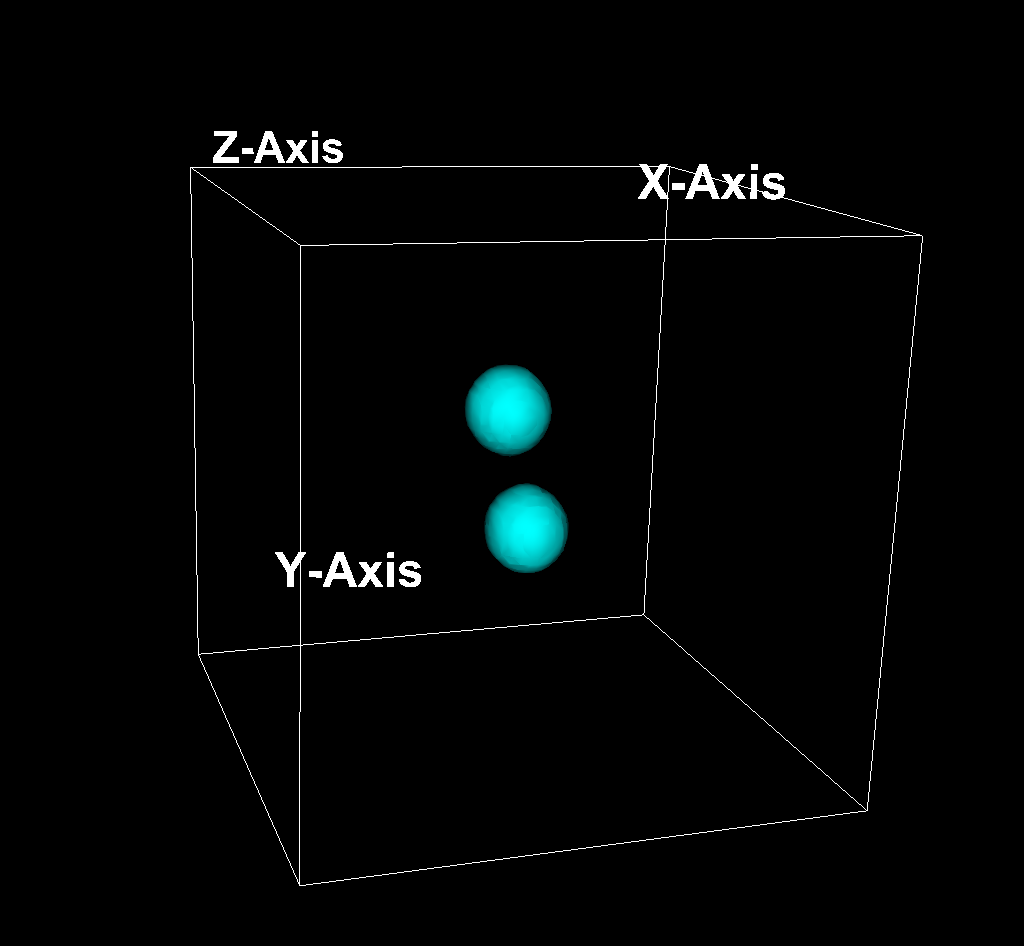}
    	\put(-150,210){\color{white} \huge t=0.3\;s}
    	\includegraphics[width=0.5\linewidth]{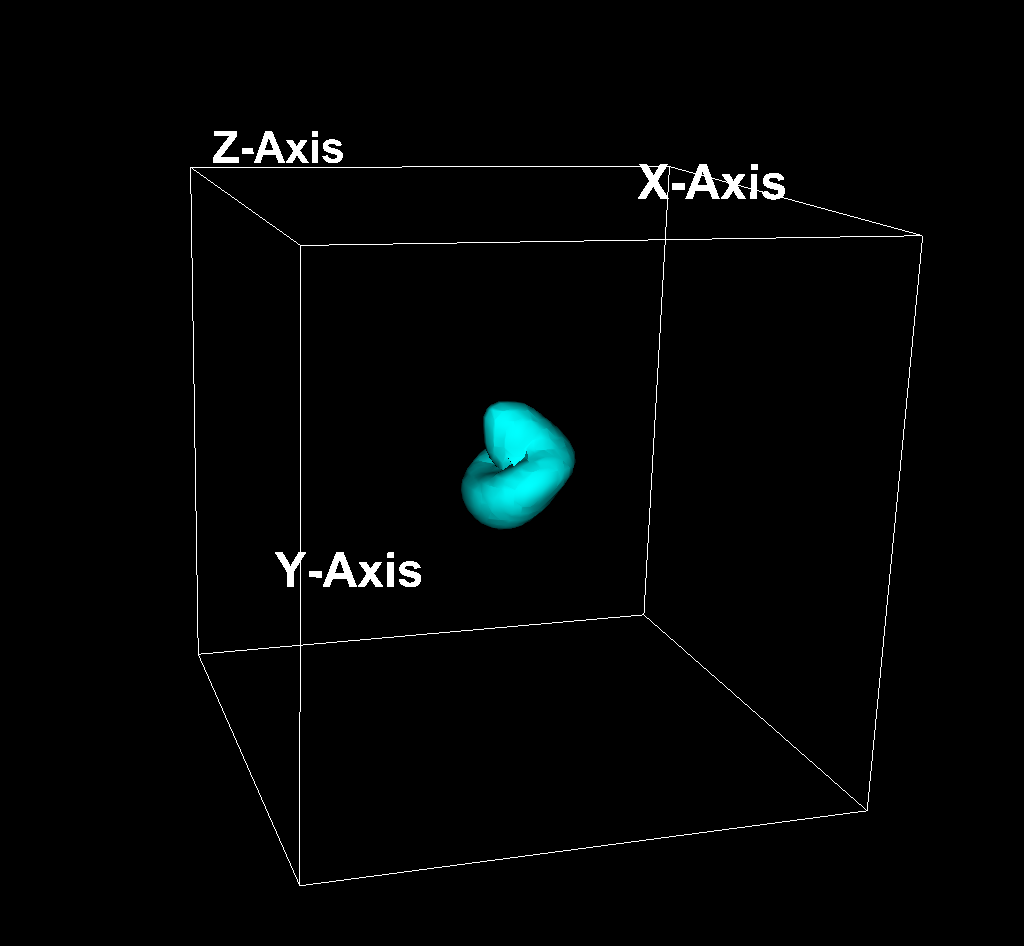}
    	\put(-150,210){\color{white} \huge t=0.6\;s}
    	%\includegraphics[width=0.5\linewidth]{figures/binary_star_0.5_0.001_2_out0003.png}
    	%\put(-150,210){\color{white} \huge t=0.9\;s}
    	\includegraphics[width=0.5\linewidth]{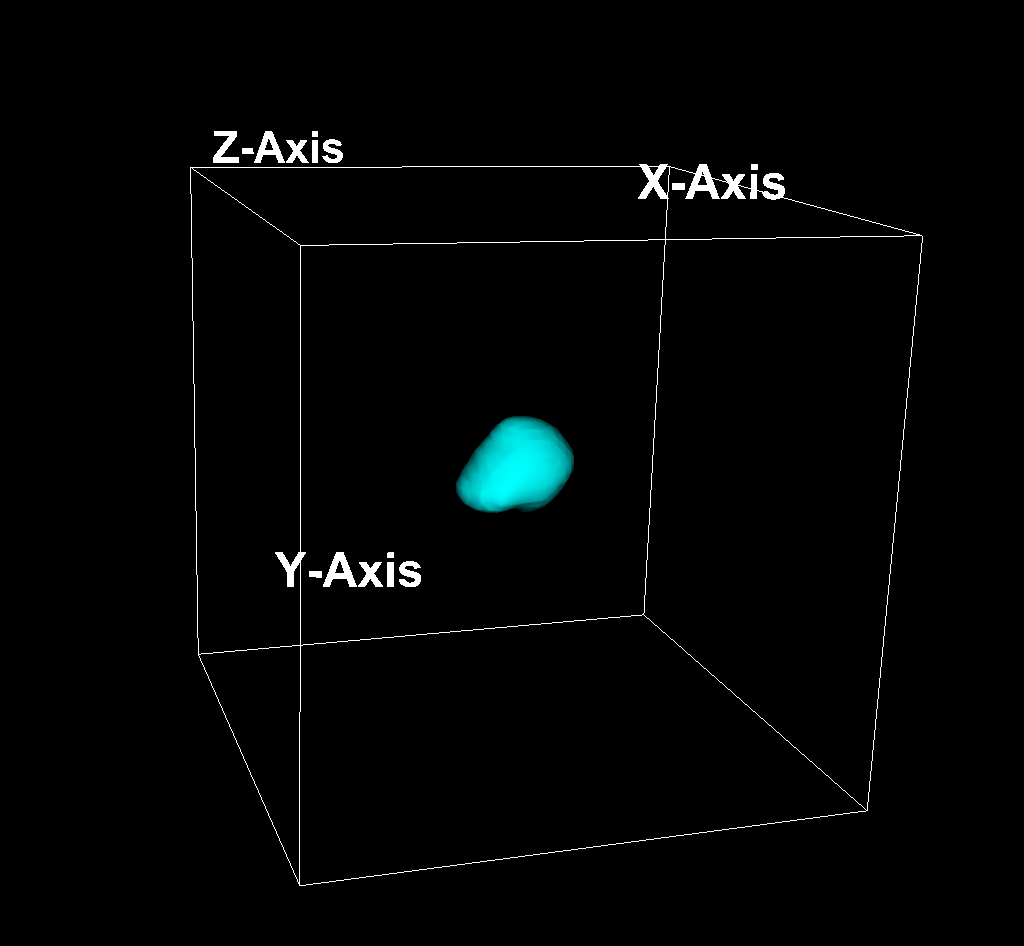}
    	\put(-150,210){\color{white} \huge t=1.2\;s}
    	\includegraphics[width=0.55\linewidth]{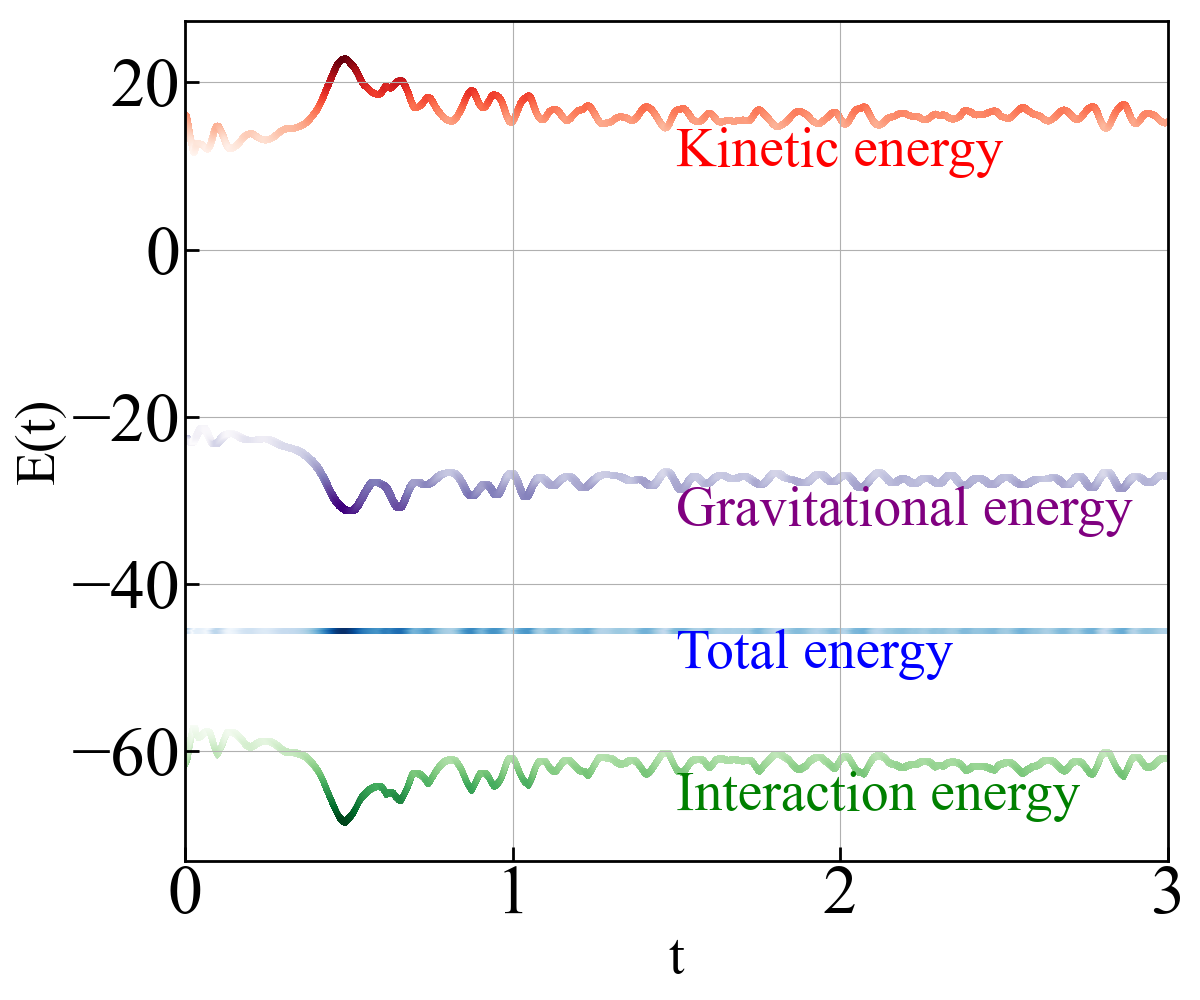}
    }
  
	\caption{Contour plots of $|\psi({\bf x},t)|^2$ for a rotating binary axion system, obtained by using the cq-GPPE for parameter set PI, i.e., $g = -0.5$, $g_2 = 0.001$, $G = 2.0$ and relative phase, $\Delta \phi =0$ (top row and $\Delta \phi = \pi$ (bottom row). Plots in the last column show the time evolution of the kinetic, gravitational, interaction, and total energies.} %The points $(a)$, $(b)$, $(c)$, and $(d)$ corresponds to the values of $G$ for which the density distribution $|\psi({\bf x},t)|^2$ is shown in Fig.\ref{fig:SGLPE_T0} for row1, row2, row3, and row4 respectively.}
	\label{fig:gpe_binary_star}
\end{figure*}

We first study the binary system in which the two axionic compact objects have the same mass, $N_1=N_2=N/2$, where $N$ is the number of bosons. We evolve Eq.~(\ref{eq:TGPPE}) in time, by starting with the initial condition of Eq.~(\ref{eq:binary_ini}), $v = 0.5$, and $  g=-0.5,\, g_2=0.001,\,$  and $G=2.0$. In the black panel of the first row of Fig.~\ref{fig:gpe_binary_star}, we show iso-surface plots of $|\psi({\bf x},t)|^2$ for the rotating binary system at different representative times and with $\Delta \phi = 0$. We observe clearly that the two components of the binary system approach each other initially and collide; then they bounce off of each other, but again come close together, albeit with a reduced separation; and finally these components merge after a large time.  

In the black panel of the second row of Fig.~\ref{fig:gpe_binary_star}, 
we show iso-surface plots of $|\psi({\bf x},t)|^2$ for the rotating binary system at different representative times and with $\Delta \phi = \pi$, i.e., with the two components of the binary system out of phase. We observe that the collapse of the two components of the 
binary system occurs in a very short time, compared to the collapse time for $\Delta \phi=0$ [cf. the time labels in the black panels of the first and second rows in Fig.~\ref{fig:gpe_binary_star}]. 

In the last column of Fig.~\ref{fig:gpe_binary_star}, we present plots versus time $t$ of the kinetic energy $E_{kq}$, the interaction energy $E_{int}$, the gravitational energy $E_G$, and the total energy $E$; the plot in the first (second) row is for $\Delta \phi=0$ ($\Delta \phi=\pi$). These plots lead to the following important results:

\begin{itemize}
    \item The temporal oscillations in these curves are associated with the bouncing of the two components of the binary system before their eventual merger. The differences in the time scales on the horizontal axes of top and bottom graphs confirm that the collapse of the binary system occurs more rapidly when $\Delta \phi=\pi$ than if $\Delta \phi=0$.
    \item The total energy $E < 0$ for both $\Delta \phi=0$ and $\Delta \phi=\pi$. This is similar to the result of Ref.~\cite{PRD_head_on_bin_star}, for the interaction between two BEC halos, which finds that the two halos collide and merge when $E<0$. 
    \item The rapid merger for $\Delta \phi=\pi$ in our cq-GPPE system suggests that there is a well in the potential energy,  which favors the formation of a bound state for our binary system. Studies of a rotating-binary system in the conventional Gross-Pitaveskii-Poisson equation (GPPE) with repulsive self interactions, i.e., $g > 0$ [but $g_2=0$] also show that the two components merge more easily when they are out-of-phase than if they are in-phase.

\end{itemize}
%It is possible to observe the presence of quantized vortices as the two objects merge and it is preferable to observe them when there is sufficient self-interaction between the bosons. We show the binary rotating system in Fig.\ref{fig:gpe_binary_star2} for value of $g_2$ that is five times larger than that of Fig.\ref{fig:gpe_binary_star}. It is observed that with increase in $g_2$, a vortex is formed when the two stars merge with $\Delta \phi=0$. 

\subsubsection*{\bf {Parameters PII}  } 

%$N_1 = N_2,  g=-0.5, g_2=0.005, G=2.0 $

Next we investigate the binary system in which the two axionic compact objects have the same mass, $N_1=N_2=N/2$, where $N$ is the number of bosons. We evolve Eq.~(\ref{eq:TGPPE}) in time, by starting with the initial condition of Eq.~(\ref{eq:binary_ini}), $v = 0.5$, and $ g=-0.5,\, g_2=0.005,\,$  and $G=2.0$, i.e., the same as the parameter set PI except for a five-fold increase in the axion-interaction strength $g_2$.

\begin{figure*}
	\resizebox{\linewidth}{!}{
			\includegraphics[width=0.5\linewidth]{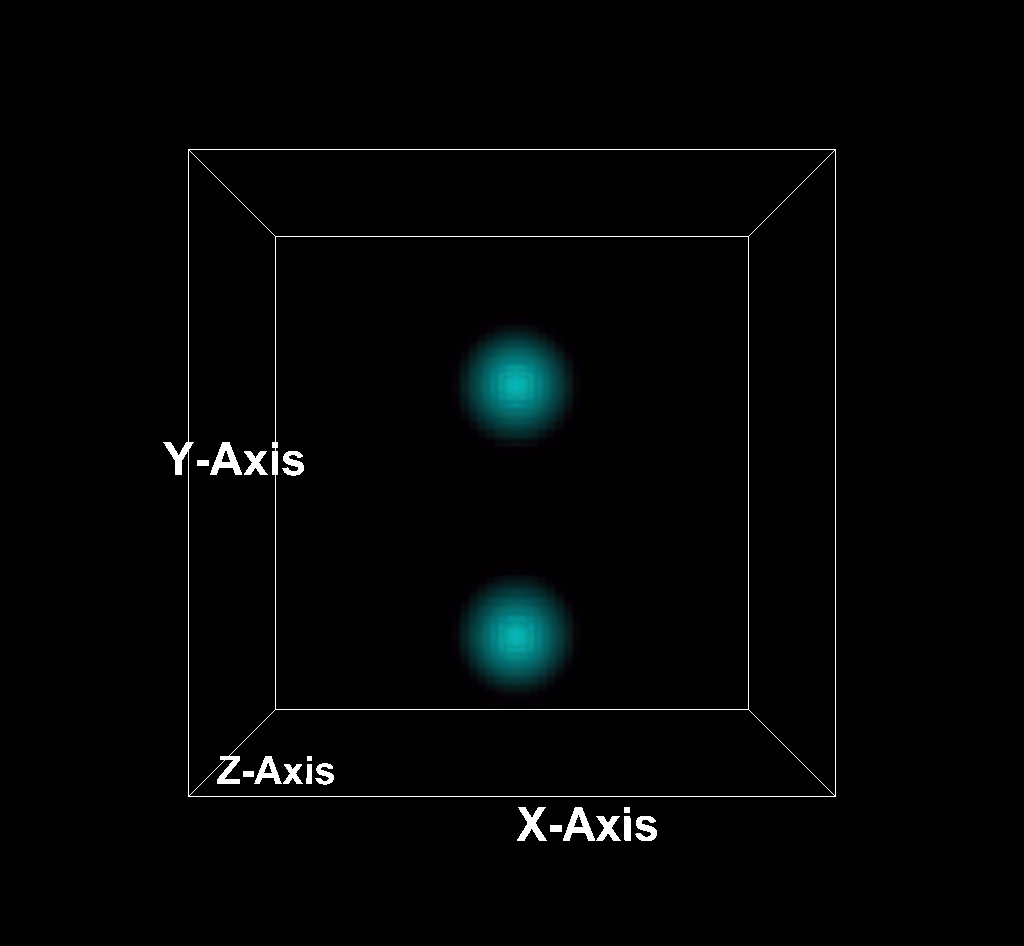}
			\put(-360,100){\huge $\Delta \phi=0$}
			\put(-290,100){\bf \huge  $\rightarrow$}
				\put(-150,210){\color{white} \huge t=0\;s}
			\includegraphics[width=0.5\linewidth]{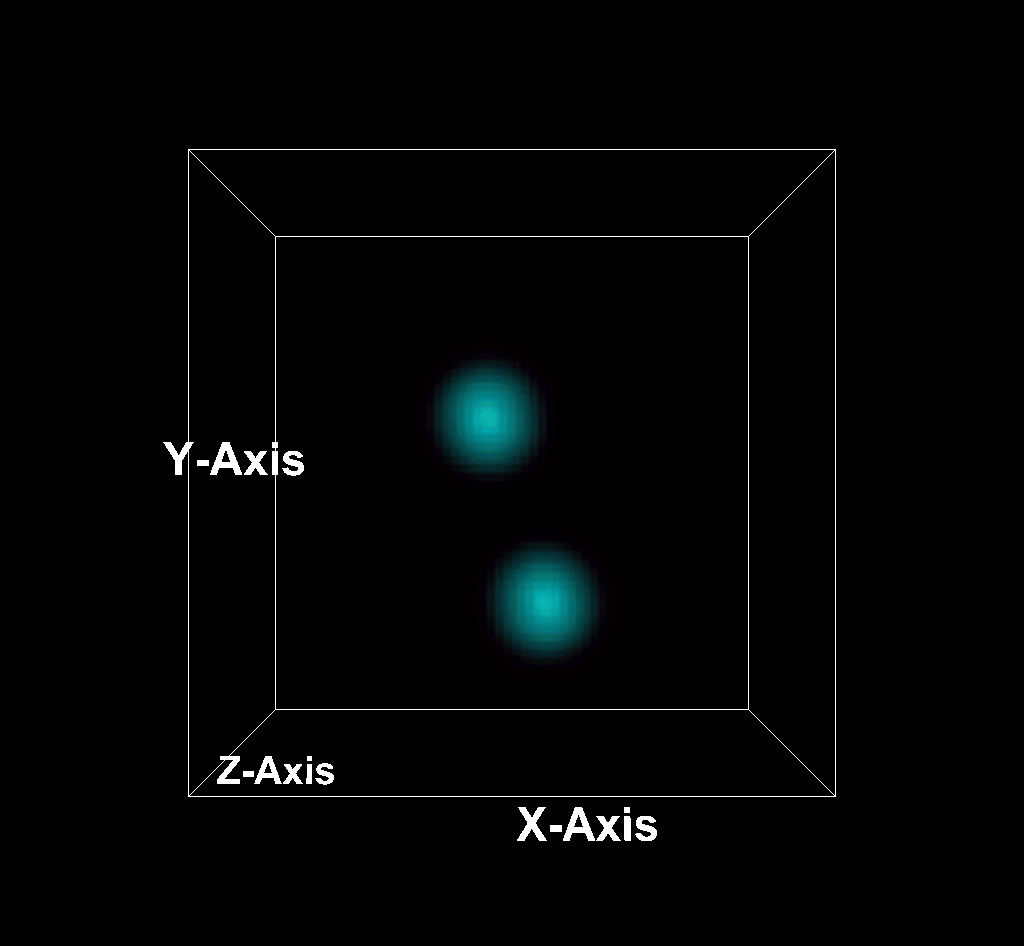}
				\put(-150,210){\color{white} \huge t=1.6\;s}
			\includegraphics[width=0.5\linewidth]{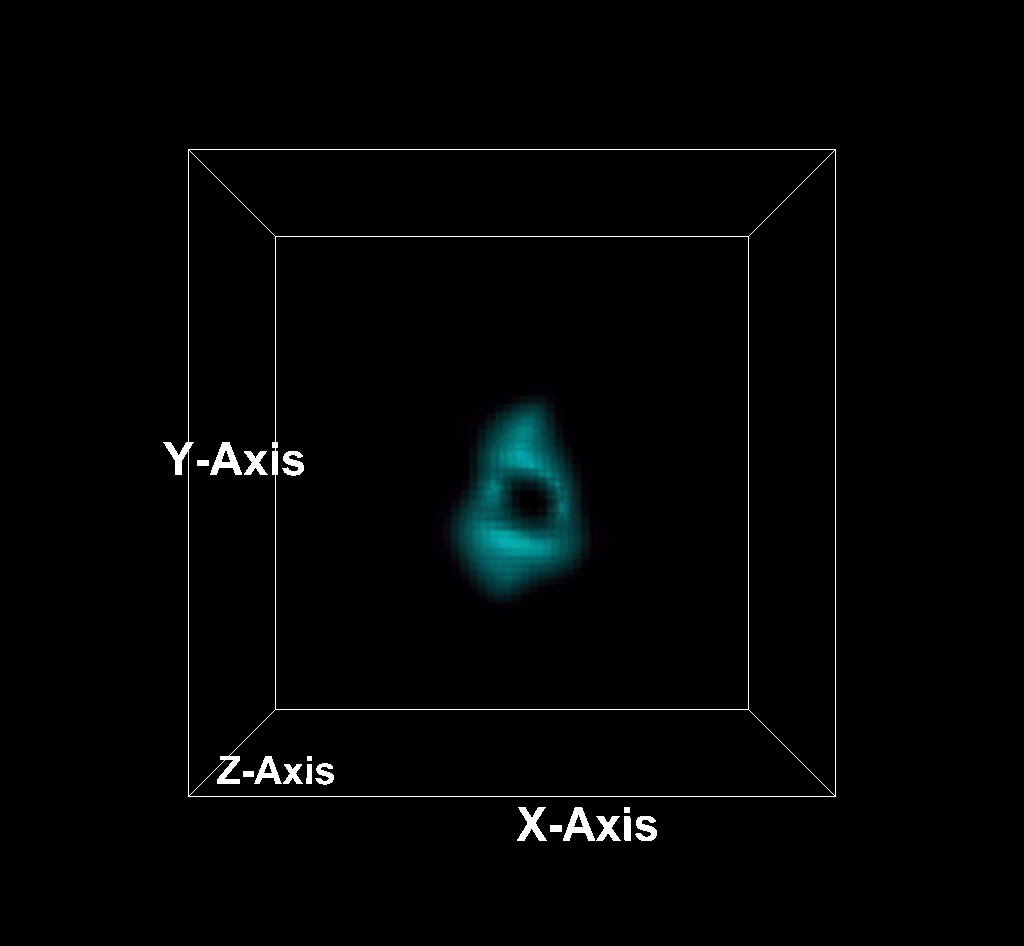}
				\put(-150,210){\color{white} \huge t=4.8\;s}
			%\includegraphics[width=0.5\linewidth]{figures/binary_star_0.5_0.005_2_in0008.png}
			%	\put(-150,210){\color{white} \huge t=6.4\;s}
			\includegraphics[width=0.5\linewidth]{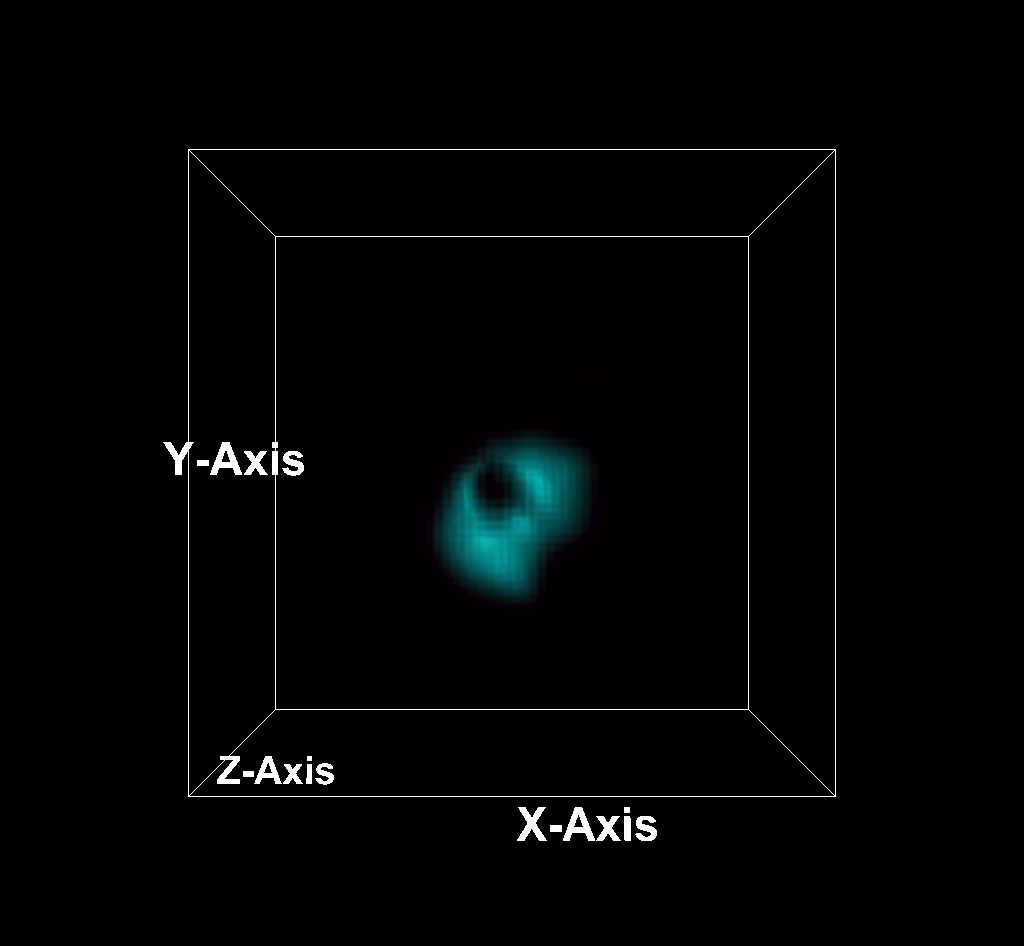}
				\put(-150,210){\color{white} \huge t=8.0\;s}
				\includegraphics[width=0.55\linewidth]{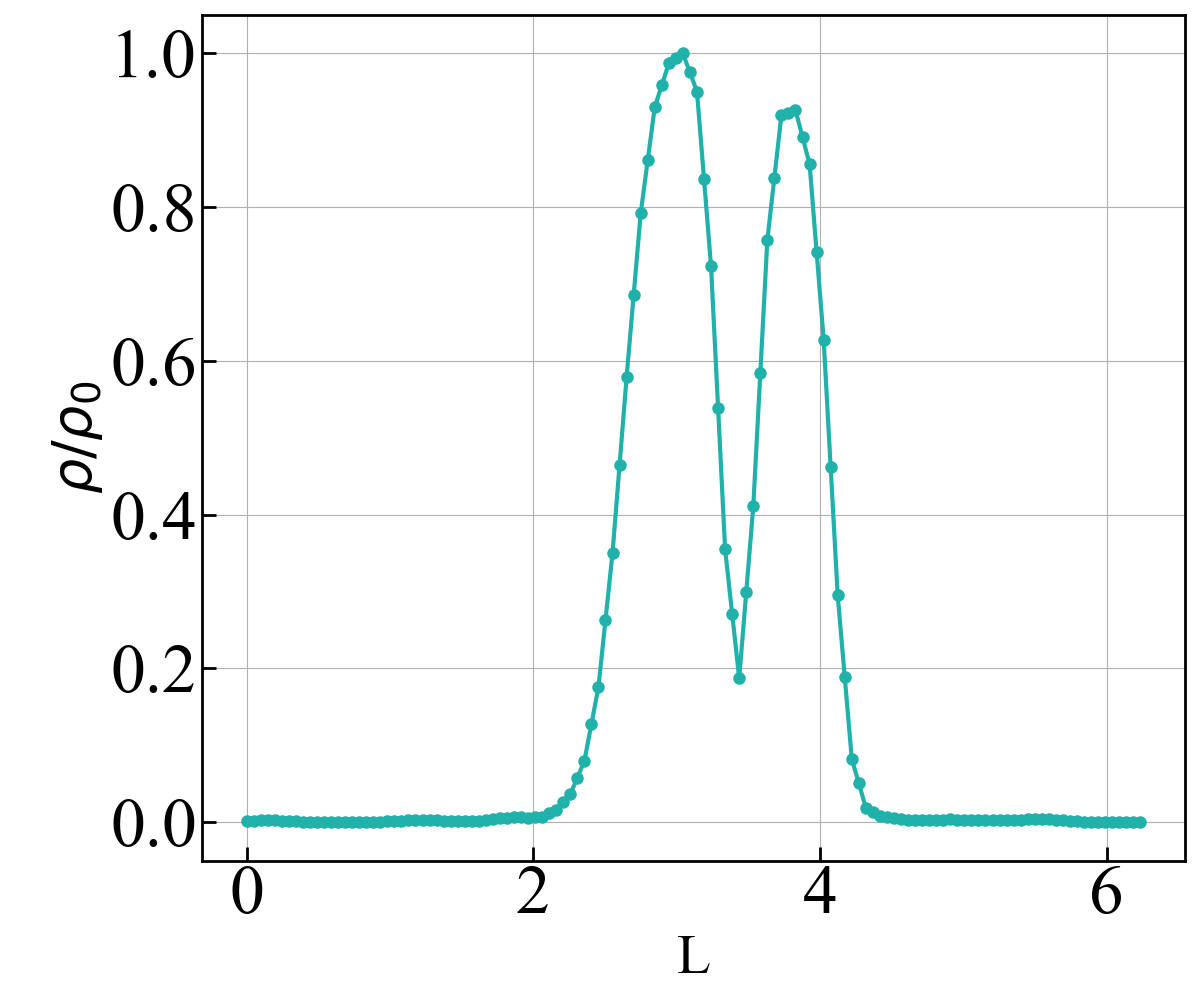}
	}
    \resizebox{\linewidth}{!}{
    	\includegraphics[width=0.5\linewidth]{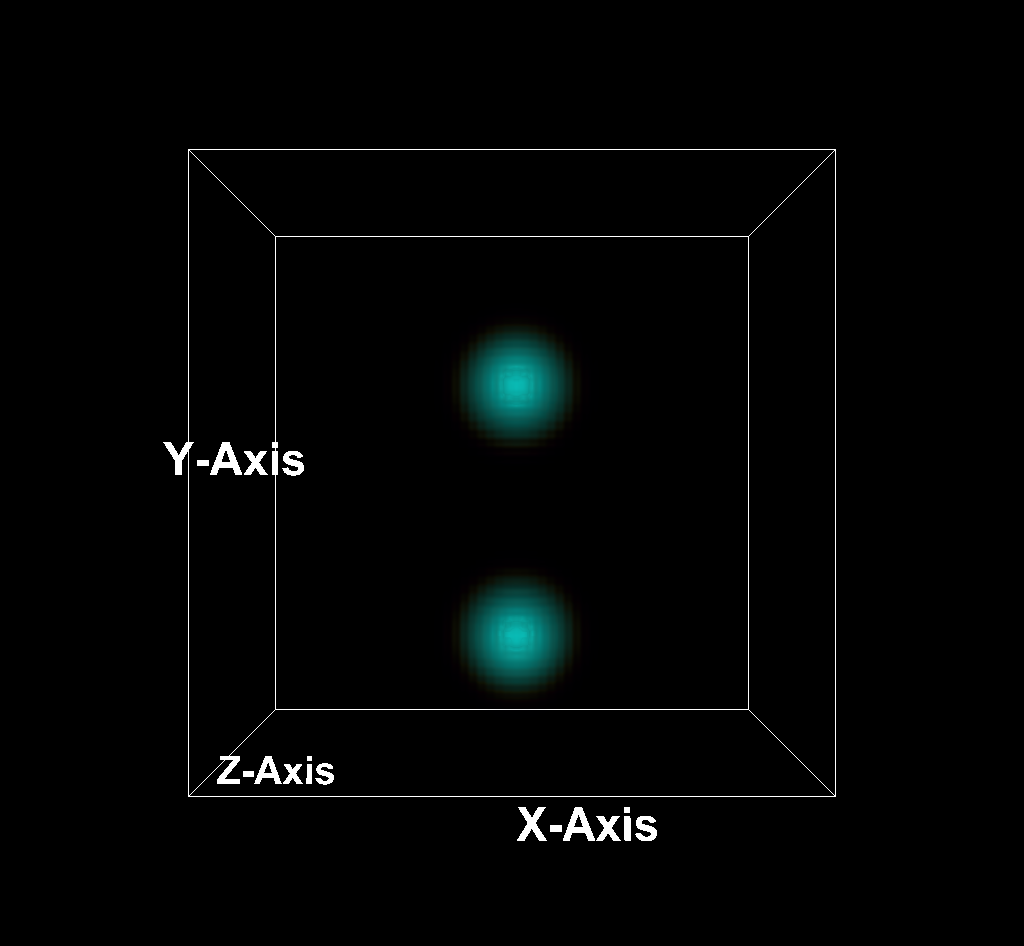}
    	\put(-360,100){\huge $\Delta \phi=\pi$}
    	\put(-290,100){\bf \huge  $\rightarrow$}
    		\put(-150,210){\color{white} \huge t=0\;s}
    	\includegraphics[width=0.5\linewidth]{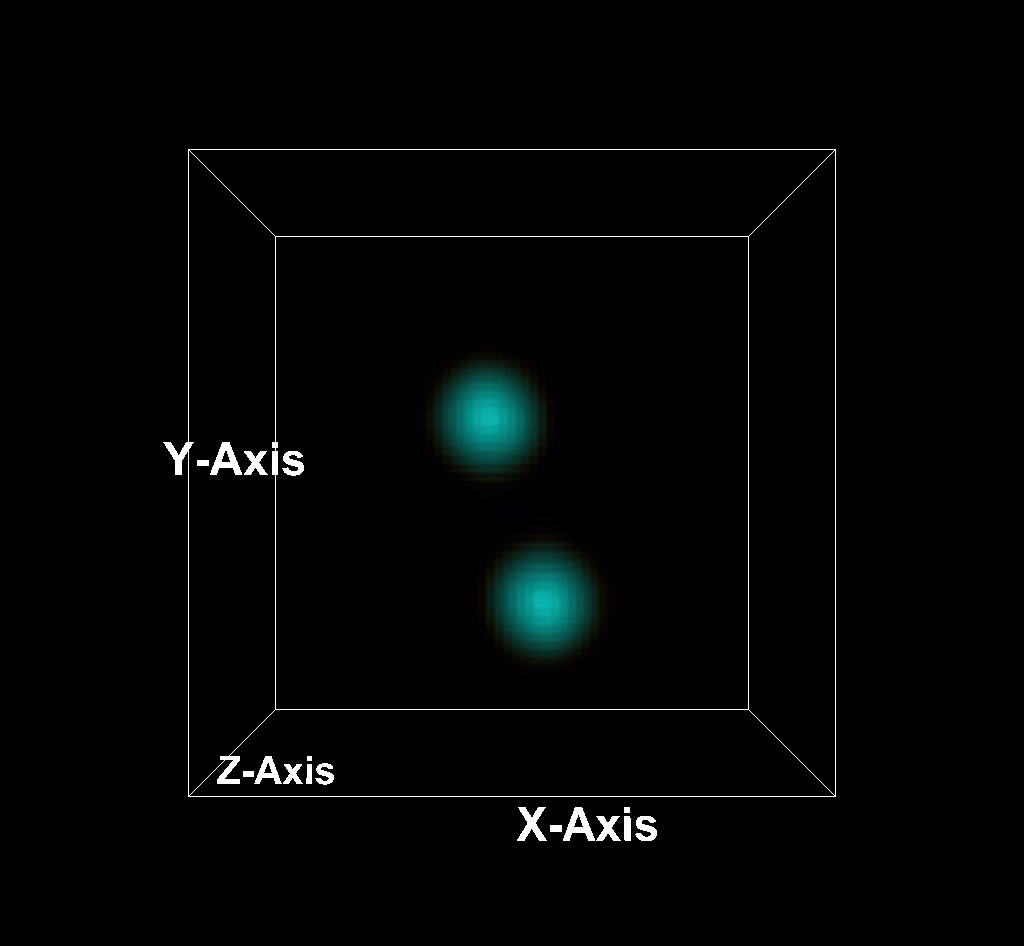}
    		\put(-150,210){\color{white} \huge t=0.8\;s}
    	\includegraphics[width=0.5\linewidth]{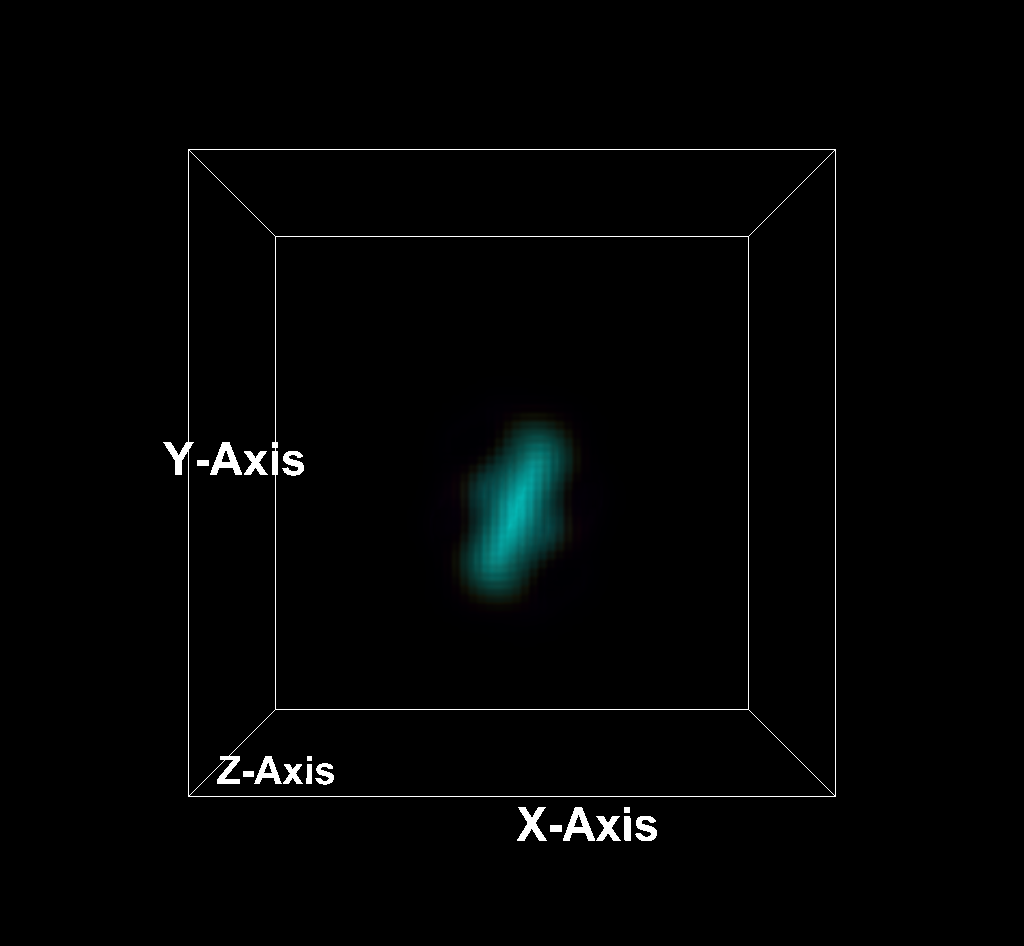}
    		\put(-150,210){\color{white} \huge t=2.4\;s}
    %	\includegraphics[width=0.5\linewidth]{figures/binary_star_0.5_0.005_2_out0007.png}
    %		\put(-150,210){\color{white} \huge t=5.6\;s}
    	\includegraphics[width=0.5\linewidth]{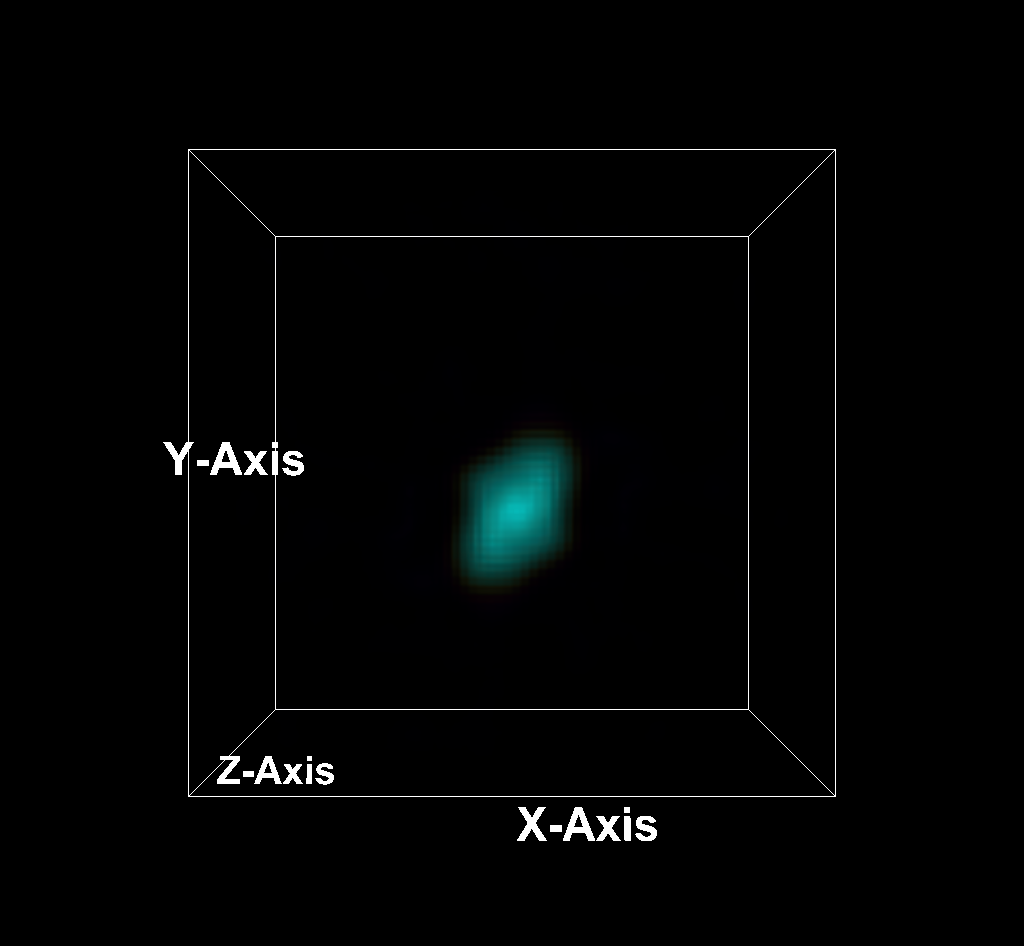}
    		\put(-150,210){\color{white} \huge t=7.2\;s}
    		\includegraphics[width=0.55\linewidth]{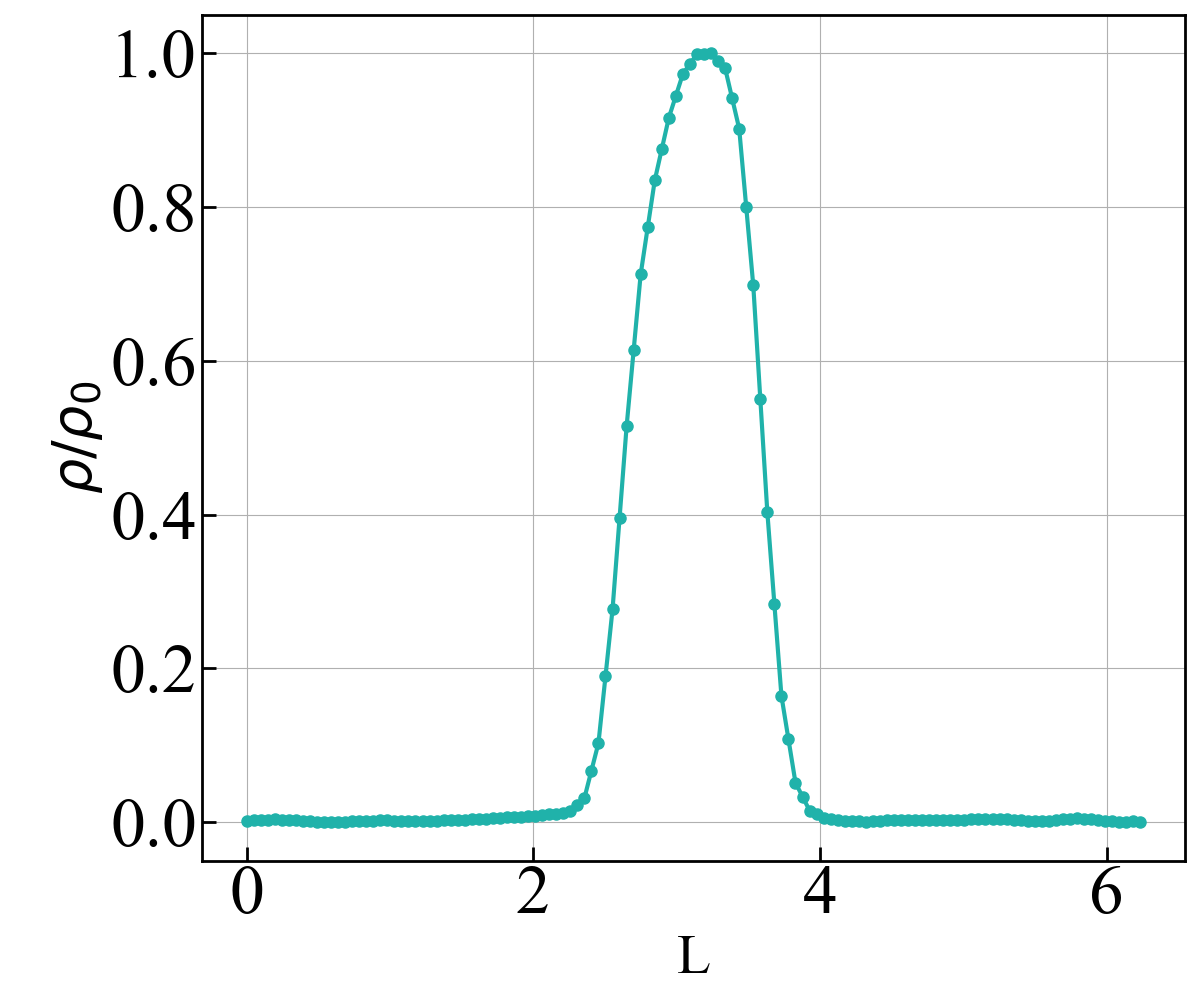}
    }
   \caption{Volume plots of $|\psi({\bf x},t)|^2$ for a rotating binary axionic system obtained by using the cq-GPPE for parameter set PII, i.e., $g = -0.5$, $g_2 = 0.005$, $G = 2.0$ and relative phase $\Delta \phi =0$ (first row) and $\Delta \phi = \pi$ (second row). The last column shows plots of the density variation along the line passing through the center of the last configuration of the collapsed, rotating axionic system.} %The points $(a)$, $(b)$, $(c)$, and $(d)$ corresponds to the values of $G$ for which the density distribution $|\psi({\bf x},t)|^2$ is shown in Fig.\ref{fig:SGLPE_T0} for row1, row2, row3, and row4 respectively.}
   \label{fig:gpe_binary_star2}
\end{figure*}

In Fig.~\ref{fig:gpe_binary_star2} we show volume plots of $|\psi({\bf x},t)|^2$ for the in-phase $\Delta \phi = 0$ (first row) and out-of-phase $\Delta \phi = \pi$ (second row) cases. In this instance too, the two objects merge more quickly in the second case than in the first. 

The two compact objects in this binary system have equal and opposite velocities initially. As the system evolves, the two objects merge and the single collapsed object rotates with a finite angular momentum. If this angular momentum is sufficiently high, it is possible to obtain quantized vortices, if the interaction strength 
$g_2$ is large. The volume plot in the first row of Fig.~\ref{fig:gpe_binary_star2} clearly shows a vortex. Such a
vortex can also be visualized by plotting the density variation along one direction for the last configuration of the collapsed, rotating
axionic object. We show such plots in the last column of Fig.~\ref{fig:gpe_binary_star2}. The large dip in the density, approximately at the middle, indicates a vortex when $\Delta \phi=0$; we do not observe such a dip if $\Delta \phi=\pi$.

\section{Conclusions}
\label{sec:conclusion}

We have shown how to use the cq-GPPE~(\ref{eq:TGPPE}) and the cq-SGLPE~(\ref{eq:TSGLPE}) to investigate the gravitational collapse of a tenuous axionic gas into a collapsed axionic condensate for $T \geq 0$. We have first presented analytical results at $T=0$, which use a Gaussian Ansatz for a spherically symmetric density profile~\cite{chavanis_2011} and suggest parameter regimes in which we might expect to find compact axionic condensates. We have gone beyond this Ansatz by using the cq-SGLPE~(\ref{eq:TSGLPE}) to investigate the $G$-dependence of the axionic condensate at $T = 0$, and shown that, as $G$ increases, the equilibrium configuration goes from a tenuous axionic gas, to flat sheets or \textit{Zeldovich pancakes}~\cite{1970A&A.....5...84Z}, cylindrical structures, and finally a spherical axion condensate [see Fig.~\ref{fig:SGLPE_T0}]. By varying $G$, we have shown that there are first-order phase transitions as the system goes from one of these structures to the next one, as we see clearly by the hysteresis loops in Fig.~\ref{fig:SGLPE_hys}. We have then examined these states and the transitions between these states via the Fourier truncated cq-GPPE~(\ref{eq:TGPPE}) and also by obtaining the thermalized $T > 0$ states [see Fig.~\ref{fig:SGLPE_T}] from the cq-SGLPE~(\ref{eq:TSGLPE}); transitions between these states yield thermally driven first-order phase transitions and their associated hysteresis loops. Finally, we have discussed how our cq-GPPE approach can be used to follow the spatiotemporal evolution of a rotating axion condensate and also a rotating binary-axion-condensate system; in particular, we have examined, in the former, the emergence of vortices at large angular speeds $\Omega$ and, in the latter, the rich dynamics of the mergers of the components of this binary system, which can yield vortices in the process of merging. 

Our work goes beyond earlier studies~\cite{PRR_akverma2021,verma2022rotating,PRD_rotating_binary,axion_merger_iop} that use the conventional cubic GPPE, which is appropriate for boson condensates that are not axionic. The latter require the inclusion of the 
quintic term in the cq-GPPE, which we have studied in detail here. We also note that the cq-GPPE arises naturally in the Taylor expansion of instantonic potentials of axions~\cite{PRD_cq_chavanis}. In future work we will explore axionic generalizations of Ref.~\cite{verma2022rotating} and explore relations, if any, of the emergence of vortices in a recent study of the mergers of black holes and saturons~\cite{dvali2023vortex}.  If dark matter indeed consists of BECs, then dark-matter galactic halos and axionic or bosonic stars should be capable of generating quantized vortices, because of the tidal torques from the surrounding matter as studied; for instance, Refs.~\cite{vortices_BEC} and ~\cite{vortices_BEC_2} study the formation and effects of vortices on the rotation curves of spiral galaxies; their results are in agreement with observations obtained from the Andromeda Galaxy and suggest the existence of sub-structures on these curves, which agree with the observations on some spiral galaxies. Our studies could find applications in such astrophysical settings.

\section*{Acknowledgments}
We thank the Indo-french Centre for Applied Mathematics (IFCAM), the Science and Engineering Research Board (SERB), and the National Supercomputing Mission (NSM), India for support, and the Supercomputer Education and Research Centre (IISc) for computational resources.

\bibliographystyle{ieeetr}

\bibliography{reference.bib}

\section{Appendix}

We use the dimensionless form of the cubic-quintic Gross-Pitaevskii-Poisson equation (cq-GPPE), which we obtain by setting $\hbar=1$, and $m=1$. Here we discuss the units relevant for diffrerent astrophysical settings. For $\hbar=1$, we have
\begin{eqnarray}
1\frac{([M] kg) \cdot ([L] m)^2}{([T] s)} = 1.054 \times 10^{-34} \; \frac{m^2 \cdot kg}{s} \,,
\label{eq:unit_hbar}
\end{eqnarray}

where $[L]$, $[T]$, and $[M]$ are the units of length, time, and mass, respectively. We now calculate the astrophysically relevant mass and time scales (depending on the object of interest).

\begin{itemize}
    \item { If the cq-GPPE~(\ref{eq:GPPE}) describes a dark-matter halo, we consider ultra-light axions of mass $m\simeq 10^{-23} eV/c^2$, which fixes the unit of mass. Therefore, choosing $m=1$ amounts to using}
\begin{eqnarray}
1 ([M ]kg) = 10^{-23} \; \frac{eV}{c^2}\,, \nonumber \\
\left[ M\right]  = 1.78\times 10^{-59}\; kg\,.
\label{eq:unit_mass_halo}
\end{eqnarray}
By using Eqs.~(\ref{eq:unit_mass_halo}) and (\ref{eq:unit_hbar}) we get
\begin{eqnarray}
\left[ T\right] = 1.69 \times 10^{-25} \left[ L\right]^2\; s\,;
\end{eqnarray}
and we can choose the unit of length to be $1\; kpc \simeq 3\times 10^{19} \; m$ \cite{GPPE_units}. Therefore, for dark-matter haloes
\begin{eqnarray}
\left[ T\right] &=& 1.52 \times 10^{14} \; s \simeq 4.8 \times 10^6\; yr = 4.8 \; Myr \,, \nonumber \\
\left[ M\right]  &=& 1.78\times 10^{-59}\; kg\,.
\end{eqnarray}

With these units of length $[L]$, mass $[M]$, and time $[T]$ for dark-matter haloes, our simulation box is of size $(2\pi \times 2\pi \times 2\pi) \; kpc^3$ and the time step of $0.000001$ is equivalent to $dt=4.8\; yrs$.

\item {If the cq-GPPE~(\ref{eq:GPPE}) describes an axionic star, we consider axions of mass $m\simeq 10^{-4} eV/c^2$, which fixes the unit of mass. Therefore, choosing $m=1$ amounts to using}
\begin{eqnarray}
1 ([M ]kg) = 10^{-4} \; \frac{eV}{c^2}\,, \nonumber \\
\left[ M\right]  = 1.78\times 10^{-40}\; kg\,.
\label{eq:unit_mass_halo_axion}
\end{eqnarray}
By using Eqs.~(\ref{eq:unit_mass_halo_axion}) and (\ref{eq:unit_hbar}) we obtain

\begin{eqnarray}
\left[ T\right] = 1.69 \times 10^{-6} \left[ L\right]^2\; s\,.
\end{eqnarray}

If we choose the unit of length to be $1\; km $ then, for axionic stars,
\begin{eqnarray}
\left[ T\right] &=& 1.69 \ s \\
\left[ M\right]  &=& 1.78\times 10^{-40}\; kg\,.
\end{eqnarray}

With these units of length $[L]$, mass $[M]$, and time $[T]$ for axionic stars, our simulation box is of size $(2\pi \times 2\pi \times 2\pi) \; km^3$ and the time step of $0.000001$ is equivalent to $dt=1.69\; \mu s$.

\end{itemize}

\end{document}